\documentclass[%
superscriptaddress,
nofootinbib,
 amsmath,amssymb,
 aps,
 pra,
 10pt,
]{revtex4-2}

\usepackage{uniinput}
\usepackage[english]{babel}

\usepackage{graphicx} 
\usepackage{dcolumn} 
\usepackage{bm}
\usepackage{bbm}
\usepackage{bbold}
\usepackage{mathtools}

\usepackage{array}
\usepackage{siunitx} 

\usepackage{listings}

\newtheorem{proposition}{Proposition}

\newcommand{\epsdiamond}{\varepsilon_{\diamond}}
\newcommand{\epstotal}{\varepsilon_{\mathrm{total}}}
\newcommand{\rzchannel}{\mathcal{Z}_{\theta}}
\newcommand{\sign}{\mathrm{sign}}
\newcommand{\Tr}{\operatorname{Tr}}

\usepackage{tikz}
\usepackage[compat=0.6]{yquant}
\usetikzlibrary{quotes}

\usepackage{xcolor}
\definecolor{riverlane_dark_green}{RGB}{0, 59, 64}
\definecolor{riverlane_green}{RGB}{0, 111, 98}
\definecolor{riverlane_light_green}{RGB}{0, 150, 143}
\definecolor{riverlane_orange}{RGB}{255, 117, 0}
\definecolor{riverlane_red}{RGB}{220, 68, 5}
\definecolor{riverlane_pink}{RGB}{207, 111, 127}
\usepackage{hyperref}
\usepackage[capitalise]{cleveref}
\hypersetup{
  colorlinks   = true, 
  urlcolor     = riverlane_green, 
  linkcolor    = riverlane_green, 
  citecolor   = riverlane_green  
}

\begin{document}

\title{More efficient Clifford+T synthesis for small-angle rotations and application to Trotterization}

\author{Marius Bothe$^*$}
\affiliation{Riverlane, Cambridge, CB2 3BZ, UK}
\affiliation{Astex Pharmaceuticals, 436 Cambridge Science Park, Cambridge, CB4 0QA, UK}
\author{Christoph Sünderhauf$^\dagger$}
\affiliation{Riverlane, Cambridge, CB2 3BZ, UK}
\author{Michael J. Witham}
\affiliation{Riverlane, Cambridge, CB2 3BZ, UK}
\author{Earl T. Campbell}
\affiliation{Riverlane, Cambridge, CB2 3BZ, UK}
\affiliation{School of Mathematical and Physical Sciences, University of Sheffield, Sheffield, S3 7RH, UK}
\author{Nick S. Blunt$^\ddagger$}
\affiliation{Riverlane, Cambridge, CB2 3BZ, UK}

\date{\today}

\begin{abstract}
Clifford+T synthesis of rotation gates is an important routine in fault-tolerant quantum compilation. While Clifford+T synthesis is scalable, it has a high overhead of tens of T gates per rotation in practice, translating to high resource estimates for many fault-tolerant algorithms. However, these well-known results, including those using probabilistic mixtures [Quantum 7, 1208 (2023)], are independent of the rotation angle $\theta$, requiring $O(\log1/δ)$ T gates. We show that it is possible to do much better for small angles, reducing the T cost to $\tilde{O}(θ^2/δ)$, and returning to existing $O(\log1/δ)$ results in the worst case. This is particularly important since many algorithms, such as Trotterization, are dominated by small-angle rotations. Further, we perform a detailed theoretical and numerical study of quasi-probabilities, which can further reduce the total T cost of large circuits by orders of magnitude with only a small overhead in sample complexity. We also develop a scheme based on quasi-probability mixtures of Clifford+T fallback channels. We derive new $\theta$-dependent formulas that can be used for resource estimation of fault-tolerant quantum algorithms. As an application of our results, we show that the gate cost of Trotterization circuits compiled to a Clifford+T gate set is constant in the small Trotter step size limit, and can be reduced by orders of magnitude even for large step sizes. The cost of fault-tolerant Trotterization for a variety of applications should be re-examined in light of these results. Our work dispels the widely-stated claim that Clifford+T rotation synthesis has a high cost independent of $\theta$, and further develops a scalable quasi-probability method for rotation synthesis. We also expect our results to bring forward useful early fault-tolerant quantum computing by reducing required magic state resources.
\end{abstract}

\maketitle

\def\thefootnote{$^*$}\footnotetext{\href{marius.bothe@riverlane.com}{marius.bothe@riverlane.com}; Equal contribution}\def\thefootnote{\arabic{footnote}}
\def\thefootnote{$\dagger$}\footnotetext{\href{christoph.sunderhauf}{christoph.sunderhauf@riverlane.com}; Equal contribution}\def\thefootnote{\arabic{footnote}}
\def\thefootnote{$\ddagger$}\footnotetext{\href{nick.blunt@riverlane.com}{nick.blunt@riverlane.com}; Equal contribution}\def\thefootnote{\arabic{footnote}}

\newpage

\tableofcontents

\section{Introduction}
\label{sec:intro}

Compilation of quantum circuits into a Clifford+T gate set is an important aspect of fault-tolerant quantum computing. Within a quantum error correcting code, arbitrary-angle rotation gates cannot be directly implemented with high fault distance, and instead it is common to work with a finite set of gates that can be implemented with a low logical error rate. The most common such gate set to work with is a set of Clifford gates together with the T gate. Under the surface code, Clifford gates can be implemented either transversally or by well-established techniques such as lattice surgery \cite{horsman2012surface,fowler2018low,litinski2019game,chamberland2022circuit}, while T gates can be implemented by teleporting a magic state \cite{gottesman1999demonstrating,bravyi2005universal}, which can be prepared to high accuracy, for example using various magic state distillation \cite{bravyi2005universal,bravyi2012magic} or cultivation protocols \cite{chamberland2020very,gidney2024magic}.

On the one hand, the task of approximating single-qubit rotation gates in the Clifford+T gate set has been well studied, and information theoretic bounds on the performance of such algorithms is known. Under unitary synthesis, the lower bound on the number of required T gates required is $3 \log_2(1/\varepsilon)$ \cite{harrow2002efficient,ross2016optimal}. This cost can be reduced by a number of schemes, notably fallback schemes which use an ancilla and measurement, and mixing schemes which take a probabilistic mixture of Clifford+T channels \cite{Campbell2016, Hastings2017}. Such mixing schemes essentially convert coherent errors into incoherent errors, reducing errors from $\varepsilon$ to $\mathcal{O}(\varepsilon^2)$, and give an average T gate count of $1.5 \log_2(1/\varepsilon)+C$.

However, these formulas for T gate counts are independent of the rotation angle, while the true cost for implementing specific rotation angles may be lower. The mixing approach of Campbell \cite{Campbell2016} and Hastings \cite{Hastings2017} is particularly powerful here because, if the rotation gate is close to a gate which can be implemented cheaply, then the mixture will be dominated by that gate, and the average synthesis cost will be reduced accordingly. An important example of this is that of small-angle rotations, which are close to the identity. Indeed, many quantum algorithms, such as Trotterization \cite{lloyd1996universal,childs2021theory}, can be dominated by extremely small-angle rotations. In the fault-tolerant setting, Trotterization is regarded as being a particularly expensive quantum algorithm, estimated to have a much higher cost for studying large-scale problems of interest compared to more modern quantum simulation methods \cite{reiher2017, lee2021, blunt_2022}. These resource estimates are based on formulas that estimate the cost of implementing each rotation gate independently of the angle size.

In this work, we perform a detailed numerical and theoretical study of mixed approximations for Clifford+T rotation synthesis in the small-angle regime. We obtain results in two settings: first using a probabilistic mixtures of Clifford+T channels, which builds on existing work \cite{Campbell2016, Hastings2017, Kliunchnikov_shorter2023}; and second by developing a scheme based on quasi-probability mixtures of Clifford+T channels, which is significantly less well explored, but has been considered in \cite{Koczor2024,khitrin2026unbiased}. We show that, in either approach, the average cost of Clifford+T synthesis of small-angle rotations is substantially reduced compared to current angle-independent formulas. In addition to obtaining numerical data to demonstrate this, we derive formulas which can be used for resource estimation for fault-tolerant quantum algorithms. Angle-dependent data and formulas for resource estimation are obtained for both ancilla-free and fallback schemes.

As an application of our results, we perform resource estimation of Trotterization applied to chemistry Hamiltonians, with application to pentacene and iron porphyrin molecules. We show that, when compiled to a Clifford+T gate set, the gate cost of Trotterization is independent of the Trotter step size, $t$, in the limit of small $t$. Even for large Trotter step sizes, the cost of Trotterization can be reduced by multiple orders of magnitude compared to standard resource estimation techniques, dependent on factors such as the total evolution time. The reduction we find is significant enough that we argue the cost of fault-tolerant Trotterization should be reassessed for a variety of applications under these new results. The fact that the cost of Trotterization becomes independent of the step size for small $t$ is also valuable because it lessens the importance of accurately estimating Trotter error (necessary to choose the optimal step size), which is a notoriously difficult and widely-studied task \cite{childs2021theory, Su2021, Zhao2022, Yi2022, Mizuta2025, Blunt2025, baysmidt2026}. We perform resource estimation using both probability and quasi-probability mixtures of Clifford+T channels, and show that the gate cost can be significantly lower in the quasi-probability method. However, the probability method can be applied to any quantum circuit, while the quasi-probability method is primarily of use for estimating expectation values, so both approaches are valuable.

We also believe that our results will be important for early fault-tolerant (FT) quantum computing. In particular, the quasi-probability approach allows one to reduce Clifford+T depth for increased sample overhead in a straightforward manner, allowing the use of shorter Clifford+T sequences for circuits where the total rotation angle is not too large, which we expect will be valuable for many early error-corrected experiments. Our results also correct an established narrative in the literature, that Clifford+T rotation synthesis remains expensive in the small-angle case. This has been one motivation for development of alternative partially fault-tolerant techniques \cite{Toshio2025, Akahoshi2025, Chung2026, Kanasugi2026, Toshio2026} which have benefits in the small-angle regime. These alternative techniques remain highly compelling, but our results will allow a fair comparison between the two methodologies. In the context of Trotterization our results have parallels with existing quantum simulation methods with small-angle savings, such as qDRIFT \cite{Campbell2019}, recent partially-randomized quantum simulation methods \cite{Gunther2025}, and TE-PAI \cite{Kiumi2025}, and we discuss differences between these approaches. We emphasize that the Clifford+T results presented in this paper have the important benefit that they both achieve small-angle savings and also match state-of-the-art Clifford+T results in the worst case. Moreover, we emphasize that this reduction is achieved \emph{automatically} (without considering quantum simulation methods beyond deterministic product formulas) by simply performing compilation to a Clifford+T gate set with a mixed approximation.

\section{A preliminary example}
\label{sec:introductory_example}

Let us begin with an introductory example for motivation and to set up the problem. In this paper we are concerned with implementing the rotation gate
\begin{equation}
    R_Z(\theta) = e^{i \theta Z}
\end{equation}
and its channel $\rzchannel$. It is well known (see e.g. Ref. \cite{Luthra2025}) that $\rzchannel$ can be represented as a linear combination of channels (LCC) using \emph{only} Cliffords,
\begin{equation}
    \rzchannel = c_I \mathcal{I} + c_S \mathcal{S}^{\dagger} + c_Z \mathcal{Z},
    \label{eq:exact_rot_clifford}
\end{equation}
where  $\mathcal{I}$,  $\mathcal{S}$ and $\mathcal{Z}$ are channels corresponding to $I$, $S$ and $Z$ unitaries. However, we cannot claim that this implements the channel $\rzchannel$ because the coefficients $\{c_I, c_S, c_Z \}$ do not form a probability distribution that can be sampled from. In particular, the L1 norm of this LCC,
\begin{equation}
    \lambda = \sum_i |c_i|,
    \label{eq:l1_norm}
\end{equation}
can be calculated as (assuming $0 \le \theta \le \pi/4$)
\begin{equation}
    \lambda = \sin(2\theta) + \cos(2\theta),
    \label{eq:introductory example I S^dag Z channel}
\end{equation}
and we have $\lambda > 1$ for $ 0 < \theta < \pi/4$. Quasi-probability distributions will be introduced in \Cref{sec:quasi_prob}, but for a circuit with $N$ such rotations the sampling overhead factor to estimate an expectation value in this approach is $\lambda^{2N}$. This exponential sampling overhead means that this approach is not scalable. However, it also gives us a clue; for small $\theta$ we have $\lambda \approx 1 + 2\theta$ and so if $\theta \lesssim 1/2N$ then this overhead is acceptable, and we need not look for more expensive representations of $\rzchannel$ using $T$ gates.

A second way to approach the problem is to approximate $\rzchannel$ with a channel $\tilde{\rzchannel}$, that can be decomposed as a strict probability distribution but with an error channel $\mathcal{E}$. In particular, using only Clifford channels we can write, for example,
\begin{equation}
    \tilde{\rzchannel} := \mathcal{E}(\mathcal{Z}_{\theta}) = p_I \mathcal{I} + p_S \mathcal{S}^{\dagger},
    \label{eq:approx_rot_clifford}
\end{equation}
where $\{ p_I, p_S \}$ form a probability distribution and therefore, unlike \cref{eq:exact_rot_clifford}, this can be implemented directly. We can assess the diamond norm error \cite{watrous2018theory} of this approximation,
\begin{equation}
    \epsdiamond =  \lVert \tilde{\rzchannel}- \rzchannel \rVert_{\diamond} =  \lVert \mathcal{E} - \mathcal{I} \rVert_{\diamond}.
    \label{eq:diamond_norm_error}
\end{equation}
The mixture in \cref{eq:approx_rot_clifford} is a special case of the mixed diagonal approximation of Kliuchnikov \emph{et al.} \cite{Kliunchnikov_shorter2023}. In this case $I$ is an ``under-rotation'' while $S^{\dagger}$ is an ``over-rotation'' for the desired $R_z(\theta)$. Using results from Ref.~\cite{Kliunchnikov_shorter2023} and in the small-$\theta$ limit one can calculate $\epsdiamond \approx 2 \theta$. Again, if the circuit contains $N$ such rotations, and assuming we target $\varepsilon$ total error, then this is acceptable provided $\theta \lesssim \varepsilon/2N$, similarly to the quasi-probability example above.

From this example, it is clear that the angle-independent formulas to estimate T gate counts per rotation, widely used for resource estimation, can be incorrect for small angles. Of course, using only Clifford gates to represent rotations channels is not scalable beyond trivial problems. Nonetheless, by taking mixtures over Clifford+T channels, this observation can be made scalable while maintaining significant reductions at small angles. In particular, in the case of quasi-probability mixtures we can write
\begin{equation}
    \rzchannel = \sum_i c_i \mathcal{U}_i,
    \label{eq:general_quasi_prob_lcc}
\end{equation}
or in the case of probability mixtures
\begin{equation}
    \tilde{\rzchannel} := \mathcal{E}(\rzchannel) = \sum_i p_i \mathcal{U}_i,
    \label{eq:general_prob_lcc}
\end{equation}
where $\{ \mathcal{U}_i \}$ are channels corresponding to Clifford+T operators, including the identity. The key observation is that, for small $\theta$, the identity term will have $c_I \approx 1$ while $|c_i| \ll 1$ for all other terms. Therefore, when compiling the circuit, most rotations will be replaced by the identity, and the Clifford+T gate counts will be significantly reduced. By considering LCC's that contain Clifford+T sequences of varying length, we can formulate this observation such that it is scalable to the limit of large circuits, returning to existing state-of-the-art results in the worst case.

\section{Summary of results}
\label{sec:summary}

In this paper we present a rotation synthesis approach using quasi-probability mixtures of Clifford+T channels, \cref{eq:general_quasi_prob_lcc}, and study their behavior in the small-angle regime. This is first performed for ancilla-free channels and then extended to fallback channels. We also provide data and derive a connection with probabilistic mixtures of Clifford+T channels, \cref{eq:general_prob_lcc}, corresponding to mixed diagonal (ancilla-free) and mixed fallback approximations from Kliuchnikov \emph{et al.} \cite{Kliunchnikov_shorter2023}; however, unlike Ref.~\cite{Kliunchnikov_shorter2023} we establish the behavior of these approximations in the small-angle regime. In this section we summarize our main results.

For the quasi-probability scheme, results are presented in terms of $\lambda-1$, where $\lambda$ is the L1 norm of the quasi-probability distribution, defined in \cref{eq:l1_norm}. For the probability scheme, results are presented in terms of $\epsdiamond$ defined in \cref{eq:diamond_norm_error}. As we shall see, equal values of $\lambda-1$ and $\epsdiamond$ result in essentially equal T counts, and so we define a variable $\delta$ which can be set equal to either $\lambda-1$ or $\epsdiamond$ depending on the scheme.

\subsection{Ancilla-free scheme}
\label{sec:summary_ancilla_free}

\begin{figure}
    \centering
    \includegraphics[width=1.0\linewidth]{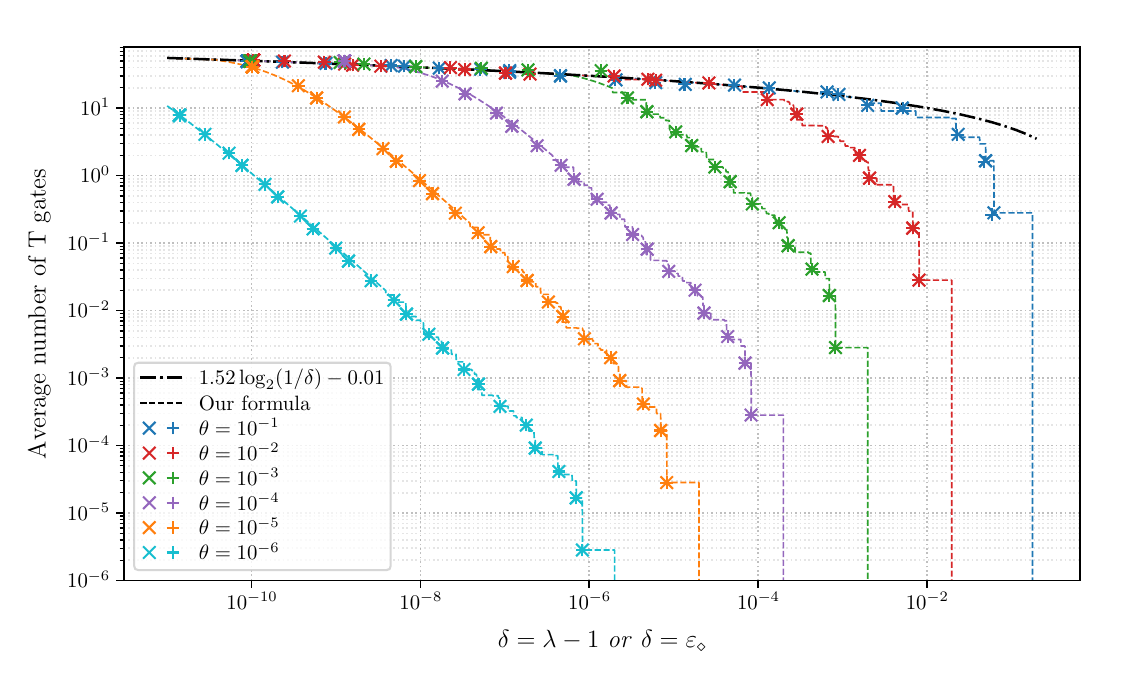}
    \caption{Average T gate counts to implement a rotation channel $\rzchannel$ in the ancilla-free scheme for two different formulations of our results, first in terms of quasi-probability mixtures with L1 norm $\lambda$ (results marked as ``$\times$''), and second in terms of strict probability mixtures with diamond norm error $\epsdiamond$ (results marked as ``$+$''). The black dashed-dotted line indicates the previous angle-independent formula for the mixed diagonal approximation from Kliuchnikov \emph{et al} \cite{Kliunchnikov_shorter2023}. Data points represent the obtained results for different angles, $\theta$, for the quasi-probability and quantum channel approaches. We see that, while results for all angles eventually converge to the angle-independent formula, for small $\theta$ there is very significant reduction in the average T count, dependent on $\delta$. We also see that equal values of $\lambda-1$ and $\epsdiamond$ result in essentially equal average T gate counts. Indeed, the respective data points lie almost precisely on top of each other, so they often appear together as a $\mathrlap{\times}+$ data point. The dashed lines indicate results from our angle-dependent formula, which can be used for resource estimation. A Python implementation of the full formula is given in \Cref{app:costing script}.}
    \label{fig:av_t_count_per_rot}
\end{figure}

Our results for the ancilla-free case are summarized in \Cref{fig:av_t_count_per_rot}, showing data for both quasi-probability mixtures (``$\times$'' data points) and probability mixtures (``$+$'' data points). The black dash-dotted line indicates the previous angle-independent formula for the mixed diagonal approximation \cite{Kliunchnikov_shorter2023},
\begin{equation}
    T_{\textrm{mixed-diagonal}}(\delta) = 1.52 \log_2\Big( \frac{1}{\delta} \Big) - 0.01.
    \label{eq:mixed_diagonal_theta_independent}
\end{equation}
The plotted data is obtained by varying the maximum number of T gates in the Clifford+T sequences used for each mixture. As can be seen, the average T gate count for a given $\delta$ can be substantially below that estimated by $T_{\textrm{mixed-diagonal}}(\delta)$, dependent on $\theta$. Our approach involves finding ``under-rotation'' and ``over-rotation'' Clifford+T sequences. Roughly speaking, the under-rotation corresponds to an angle smaller than the target $\theta$, while the over-rotation corresponds to an angle greater than $\theta$; a more precise definition is given in \Cref{sec:under_and_over_rot}. The savings at small $\theta$ ultimately come about because the under-rotation can be chosen to be the identity, which has zero cost. Therefore, smaller rotation angles can use fewer Clifford and T gates on average. However, crucially, for sufficiently small $\delta$, the cost of implementing any rotation gate is seen to always return to the worst case result with scaling $\mathcal{O}(\log_2(1/\delta))$, and therefore the approach remains scalable in the limit of large circuits.

By fixing the under-rotation to the identity, we obtain a formula for the average T gate count as a function of both $\theta$ and $\delta$. The formula has two parts: first, a ``staircase'' that corresponds to optimal over-rotations calculated by brute force enumeration of all Clifford+T sequences up to a maximum length (we obtain solutions up to $35$ T gates); and second, a general formula that is derived (\emph{not} fitted) and can be applied for T gate counts beyond the staircase. The general formula is
\begin{equation}
T_{\textrm{small-angle}}(\theta, \delta) = \frac{3θ}{α+2φ_0}\log_2\frac{12}{(α-φ_0)^2(α+2φ_0)},
\label{eq:asymptotic_summary}
\end{equation}
\begin{equation}
    \alpha = \frac{\delta}{2\theta} + \theta,
    \;\;\;\;\;\;\;
    φ_0 \approx \max\left(α-\frac{α}{\ln(K/α)},θ\right)
\end{equation}
with $K:=(\frac{2\sqrt{2e^3}}{3})^{2/3} \approx 2.61$.

Solutions for optimal Clifford+T over-rotations up to $35$ T gates are presented in Tables~ \ref{tab:staircase app} and \ref{tab:staircase gate sequences app} in \Cref{app:matsumoto_amano}. Note that the first two over-rotations in the staircase correspond to $S^{\dagger}$ and $T^{\dagger}$; the special case of $S^{\dagger}$ was described in the preliminary example, \Cref{sec:introductory_example}. The derived formula is valid for $0 < \theta  \le \pi/8$. Other rotations $R_z(\theta)$ can always be first mapped to this range by applying Cliffords. For example, $R_z(-\theta) = X R_z(\theta) X$, so it follows that small negative angles, $-\theta$, have the same T gate cost as small positive angles, $\theta$.

The formula \cref{eq:asymptotic_summary} was derived by considering a quasi-probability mixture approach, rather than the probability mixture approach of Ref.~\cite{Kliunchnikov_shorter2023}. However, from the data in \Cref{fig:av_t_count_per_rot} we see that $\lambda-1$ in the quasi-probability scheme and $\epsdiamond$ in the probabilistic scheme are close to equal in almost all cases. We further prove this closeness in \Cref{app:comparison kliuchnikov}. Therefore, our resource estimation formulas can be used with either the probability or quasi-probability approach by simply setting $\delta = \epsdiamond$ or $\delta = \lambda - 1$, respectively. Despite this correspondence, we will show that using quasi-probability distributions can significantly reduce the gate cost of implementing algorithms such as Trotterization.

For $\delta \lesssim \theta^2$ the identity is no longer the optimal under-rotation. Instead, there are more accurate Clifford+T sequences that can be used as the under-rotation instead of the identity. The formula \Cref{eq:asymptotic_summary} therefore no longer matches the data for $\delta \lesssim \theta^2$; the data in \Cref{fig:av_t_count_per_rot} does not fix the under-rotation to the identity, but allows the optimal Clifford+T sequence found to be used. Since both the under- and over-rotations are Clifford+T sequences in this regime, there is no longer a small-angle saving, and data returns to the angle-independent formula, \cref{eq:mixed_diagonal_theta_independent}. The following formula accounts for both regimes:
\begin{equation}
    T(\theta,\delta) = \min(T_{\textrm{small-angle}}(\theta, \delta), \, T_{\textrm{mixed-diagonal}}(\delta)).
    \label{eq:min_of_small_and_worst}
\end{equation}
The result plotted as ``Our formula'' in \cref{fig:av_t_count_per_rot} is given as a Python script in \Cref{app:costing script}. This combines the staircase solutions, Tables~ \ref{tab:staircase app} and \ref{tab:staircase gate sequences app}, the analytic small-angle formula, \cref{eq:asymptotic_summary}, and worst-case formula, \cref{eq:mixed_diagonal_theta_independent}, and is seen to match the data well in all regimes.

\subsection{Fallback scheme}
\label{sec:summary_fallback}

We also consider fallback channels, which use Clifford+T unitaries together with a projective measurement via an ancilla. When the projective measurement fails, a fallback step is performed using the mixed diagonal approximation, as in the mixed fallback approximation of Ref.~\cite{Kliunchnikov_shorter2023}.
Data for this scheme is presented in \Cref{fig:av_t_count_per_rot_fallback} in \Cref{sec:mixed_fallback_scheme}, again showing data for both quasi-probability mixtures (``$\times$'' data points) and probability mixtures (``$+$'' data points). Data points are obtained by varying the maximum T gate count in the Clifford+T sequences for the projective step. The black dashed line indicates the previous angle-independent formula for the mixed fallback approximation \cite{Kliunchnikov_shorter2023},
\begin{equation}
    T_{\textrm{mixed-fallback}}(\delta) = 0.53 \log_2\Big( \frac{1}{\delta} \Big) + 4.86.
    \label{eq:mixed_fallback_theta_independent}
\end{equation}
We again derive an angle-dependent formula for the case where the under-rotation is the identity. We obtain
\begin{equation}
    T_{\textrm{small-angle-fallback}}(\theta, \delta) = T_{\textrm{small-angle}}(\theta, 2\delta)
\end{equation}
where $f_{\textrm{small-angle}}(\theta, \delta)$ is defined as in \cref{eq:asymptotic_summary}. The fallback case is modified by substituting $\delta \rightarrow 2 \delta$, which roughly halves the average T gate count for a given $\delta$ compared to the ancilla-free scheme.

As for the ancilla-free case, for $\delta \lesssim \theta^2$ the optimal under-rotation is no longer the identity, but rather a Clifford+T fallback channel. In this case there is no small-angle saving to be made, and results return to the angle-independent formula, therefore
\begin{equation}
    T(\theta,\delta) = \min(T_{\textrm{small-angle-fallback}}(\theta, \delta), \, T_{\textrm{mixed-fallback}}(\delta))
    \label{eq:min_of_small_and_worst_fallback}
\end{equation}
is accurate for resource estimation. The result ``Our formula'' in \Cref{fig:av_t_count_per_rot_fallback} uses this result, and also includes exact over-rotations corresponding to $S^{\dagger}$ and $T^{\dagger}$, as in the ancilla-free case.

\subsection{Resource estimates for Trotterized quantum dynamics}
\label{sec:trotter_summary}

\begin{figure}
    \centering
    \includegraphics[width=0.7\linewidth]{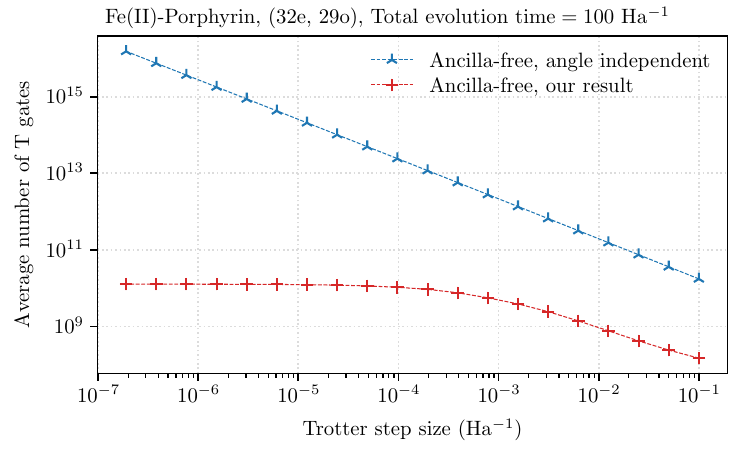}
    \caption{Average T gate cost to implement a first-order Trotterization circuit using the quasi-probability Clifford+T approach, as a function of Trotter step size. The system is the Fe(II)-Porphyrin complex in a (32e,29o) active space and using localized orbitals. The total evolution time is fixed to $T = 100$ Ha$^{-1}$. The ancilla-free Clifford+T synthesis scheme is used. Results in blue show the costing results using the previous angle-independent formula, while the results in red used our angle-dependent formula. The cost of Trotterization becomes independent of the step size in the limit of small steps sizes, while the cost is reduced by two orders of magnitude even for large step sizes. Note that this is the same figure as \Cref{fig:quasi_prob_comparison}(c) in \Cref{sec:trotter_results}.}
    \label{fig:trotter_summary}
\end{figure}

The derived angle-dependent formulas can be used in resource estimation studies. While the formulas are naturally more complicated than the angle-independent formulas, they are simple to implement; see \Cref{app:costing script} for a Python implementation of the ancilla-free formula.

As an example, \Cref{fig:trotter_summary} shows the average T gate count to implement a first-order Trotter circuit with evolution time $T = 100$ Ha$^{-1}$ for a chemistry example, a Fe(II)-Porphyrin complex in a (32e,29o) active space; for further details of the Hamiltonian, see \Cref{sec:chemistry_hamiltonians}. Results use the quasi-probability ancilla-free synthesis scheme. For each rotation we take $\delta$ such that the total sampling overhead for the full Trotter circuit is approximately $e^2 \approx 7.39$, which is a small sampling overhead that does not increase with circuit size; see \Cref{sec:assigning_delta} for further details.

The results plotted in blue perform costing using the angle-independent formula, \cref{eq:mixed_diagonal_theta_independent}, while the results plotted in red use our angle-dependent formula. For large step sizes, $t$, (i.e., $t=0.1$ Ha$^{-1}$) the average T gate count per Trotter circuit is reduced by roughly two orders of magnitude by using the angle-dependent formula. In the limit of small step sizes the T gate cost is independent of $t$. This result is also straightforward to derive directly from \cref{eq:asymptotic_summary}. This shows that properly assessing the cost of mixed Clifford+T synthesis in the small-angle regime can have a significant impact on the cost of fault-tolerant quantum simulation. Further Trotterization results are presented in \Cref{sec:trotter}.

\section{Ancilla-free scheme}
\label{sec:ancilla_free_scheme}

\subsection{Quasi-probability distributions to calculate expectation values}
\label{sec:quasi_prob}

We begin by providing an introduction to quasi-probability distributions and their use in calculating expectation values of observables from quantum circuits containing rotation gates. Given $R_z(\theta) = e^{i \theta Z}$ and its channel $\rzchannel(\rho) = e^{i \theta Z} \rho e^{-i \theta Z}$, we wish to write $\rzchannel$ as a linear combination of channels (LCC)
\begin{equation}
    \rzchannel = \sum_i c_i \mathcal{U}_i.
    \label{eq:rzchannel_lcc}
\end{equation}
In particular, we consider quasi-probability mixtures of channels, for which
\begin{equation}
    \lambda = \sum_i |c_i|
\end{equation}
may be greater than $1$. Such an LCC cannot be implemented directly. Instead, we may rewrite this LCC as
\begin{equation}
    \rzchannel = \lambda \sum_i \sign(c_i) \, p_i \mathcal{U}_i,
\end{equation}
where $p_i = |c_i|/\lambda$ form a probability distribution that can be sampled from. In particular, if we wish to estimate the expectation value of an observable, $O$, then
\begin{equation}
    \mathrm{Tr}[O \rzchannel(\rho)] = \lambda \sum_i \sign(c_i) \, p_i \mathrm{Tr}[O \mathcal{U}_i(\rho)].
\end{equation}
This can be achieved by sampling quantum circuits to estimate $\mathrm{Tr}[O \mathcal{U}_i(\rho)]$ with probability $p_i$, and including the factors $\lambda \, \sign(c_i)$ in post processing when constructing the final expectation value estimate. Thereby, the quasi-probability approach provides an unbiased estimator for expectation values. If we wish to obtain estimate an expectation value to precision $\varepsilon_\text{sample}$ with probability $1-\eta$, then according to Hoeffding's inequality, we need to take $N_\text{sample}$ samples, where
\begin{equation}
    N_\text{sample} = 2 \frac{\lambda^2}{\varepsilon_\text{sample}^2} \ln \Big( \frac{2}{\eta} \Big).
\end{equation}
If we estimated the expectation value without the quasi-probability scheme then the result would be the same but with $\lambda=1$. Hence, $\lambda^2$ can be considered to be the sampling overhead of the quasi-probability distribution. We therefore wish to construct LCC's with $\delta = \lambda-1$ as close to zero as possible. We note that, if we were to run a circuit with $N$ such rotations, then the total sampling overhead would be $\lambda^{2N}$. In order to avoid a blow up in sampling overhead with increasing $N$, we need $\delta = \mathcal{O}(1/N)$. In particular, setting $\delta = 1/N$ gives
\begin{equation}
    \lambda^{2N} = \Big(1 + \frac{1}{N} \Big)^{2N},
\end{equation}
which converges to the constant $e^2 \approx 7.39$ for large $N$.

\subsection{LCC coefficients for a rotation channel}

\subsubsection{General setup}
\label{sec:general_setup}

\label{sec:analytic coefficients}

In the initial stages of this work we used a numerical approach to construct and optimize linear combinations of Clifford+T channels; this approach is described in \cref{app:cvx}. While useful for investigation, we ultimately found this approach to be limited by numerical issues that restrict its applicability. In this section we will instead derive analytic coefficients for a linear combination of channels (LCC) representation of a rotation channel, using a basis of Clifford+T channels, resulting in a scalable approach.

First, note that any $SU(2)$ unitary\footnote{As overall phase is unphysical, we can also use more general unitaries. However, the analysis requires the twirling proposition for unitaries in the form of \cref{eq:su2_def_2}.} (a $2 \times 2$ unitary with determinant $1$) can be written as
\begin{equation}
U = 
\begin{pmatrix}
u & -v^* \\
v & u^*
\end{pmatrix},\ u=re^{iθ}.
\label{eq:su2_def_2}
\end{equation}
We want to use unitaries of this form for the LCC, but it will be valuable to twirl these unitaries. Campbell twirled with respect to $Z$ \cite{Campbell2016}, while Kliuchnikov \emph{et al.} twirl with respect to $Z$ and $S$ \cite{Kliunchnikov_shorter2023}, which we follow in our approach. In particular, we define the twirl
\begin{equation}
    \mathcal{T}_U (\rho) = \frac{1}{4} \sum_{V \in \{I, S, S^{\dagger}, Z\}} (VUV^{\dagger}) \rho (V U^{\dagger} V^{\dagger}).
\end{equation}
For a unitary $U$ as defined in \cref{eq:su2_def_2} a short derivation shows (Proposition D.1, Ref.~\cite{Kliunchnikov_shorter2023})
\begin{equation}
    \mathcal{T}_U (\rho)  = \mathcal{P}_{r,\theta}(\rho) - \frac{ir^2\sin(2\theta)}{2} (\rho Z - Z \rho),
    \label{eq:twirl}
\end{equation}
where $\mathcal{P}_{r,\theta}(\rho)$ is a Pauli channel,
\begin{equation}
    \mathcal{P}_{r,\theta}(\rho) = r^2 \cos^2(\theta) \rho  + \frac{1-r^2}{2} (X \rho X + Y \rho Y) + r^2 \sin^2(\theta) Z \rho Z.
\end{equation}
Therefore, the twirl $\mathcal{T}_U (\rho)$ gives a Pauli channel except for $Z(\cdot)I - I(\cdot)Z$ cross terms.

The LCC of the twirls of a given set of unitaries $\{U_i\}$ with quasi-probabilities $\{c_i\}$ can now be explicitly calculated by combining the preceding two equations with the quantum channel $\mathcal{Z}_\theta(ρ) = \cos^2θρ + \sin^2θZρZ -i\sin(2θ)/2\,(ρZ-Zρ)$,
\begin{align} \label{eq:channel coeffients}
\sum_i c_i \mathcal{T}_{U_i}(ρ)
    ={}& e^{iθZ} ρ e^{-iθZ} \\ \nonumber
    & + \left(-\cos^2θ + \sum_i c_ir_i^2\cos^2φ_i\right) \rho \\ \nonumber
    & + \left(\sum_ic_i\frac{1-r_i^2}{2}\right)(XρX +YρY) \\ \nonumber
    & + \left(-\sin^2θ + \sum_i c_ir_i^2\sin^2φ_i\right)ZρZ\\ \nonumber
    & + \left(i\frac{\sin2θ}{2} -\sum_i c_i i\frac{r_i^2\sin2φ_i}{2}\right)(ρZ-Zρ).
\end{align}

Our strategy will be to choose $\{c_i\}$ such that the cross terms $(\rho Z - Z \rho)$ vanish. With such coefficients we have that the exact rotation channel is an LCC over (twirls of) the chosen unitaries $\{U_i\}$, as well as Pauli channels:
\begin{equation}
    \mathcal{Z_\theta}(\rho) = \sum_i c_i\mathcal{T}_{U_i}(\rho) + c_I \rho + c_X X \rho X + c_Y Y \rho Y + c_Z Z \rho Z.
\end{equation}
We may therefore identify
\begin{equation}
    c_I = \cos^2θ - \sum_i c_i r_i^2 \cos^2 φ_i,
\end{equation}
\begin{equation}
    c_X = c_Y = \sum_i c_i\frac{r_i^2 - 1}{2},
\end{equation}
\begin{equation}
    c_Z = \sin^2θ - \sum_i c_i r_i^2 \sin^2 φ_i,
\end{equation}
and we also define
\begin{equation}
    c_{\times} = \sum_i c_i i\frac{r_i^2\sin2φ_i}{2} - i\frac{\sin2θ}{2}
\end{equation}
as the coefficient for the cross-terms, which we will enforce to be $0$.

This LCC has a sample complexity\footnote{If one of the unitaries $U_i$ is a Pauli operator or the identity, its coefficient combines with that of the corresponding Pauli operator in the computation of $λ$. For example, if $U_1=\mathbb{1}$, then $|c_1|+|c_I|$ can be replaced by $|c_1+c_I|$.}
\begin{equation}
    λ =\sum_i |c_i| + |c_I| + 2|c_X| + |c_Z|,
    \label{eq:lambda channel coefficients}
\end{equation}
where we have used $c_X$ = $c_Y$. The Pauli channels do not contribute any T gates, such that the average T count of this protocol is
\begin{equation}
    T = \sum_i \frac{|c_i|}{λ}\cdot\text{T-count}(U_i).
    \label{eq:av_t_count_formula}
\end{equation}

\subsubsection{Taking an over-rotation and under-rotation}
\label{sec:under_and_over_rot}

The work of Ref.~\cite{Kliunchnikov_shorter2023}, following the original proposal by Campbell \cite{Campbell2016} and Hastings \cite{Hastings2017}, considers probabilistic mixtures over Clifford+T channels. These approaches primarily aim to find just two Clifford+T unitaries to mix: an \emph{under-rotation}, $U_1$, and an \emph{over-rotation}, $U_2$, defined
\begin{equation}
U_1 = 
\begin{pmatrix}
r_1 e^{i \varphi_1} & -v_1^* \\
v_1 & r_1 e^{-i \varphi_1}
\end{pmatrix},
\;\;\;\;\;\;\;\;\;
U_2 = 
\begin{pmatrix}
r_2 e^{i \varphi_2} & -v_2^* \\
v_2 & r_2 e^{-i \varphi_2}
\end{pmatrix},
\label{eq:under_and_over_rotation_def}
\end{equation}
such that $\varphi_1< \theta < \varphi_2$. By taking probabilistic mixtures over the corresponding channels (and their twirls), it is possible to convert coherent errors of size $\delta = \epsdiamond$ into incoherent errors of size $\mathcal{O}(\epsdiamond^2)$, which roughly halves the T gate count required to achieve a given accuracy, yielding the mixed-diagonal approximation formula, \cref{eq:mixed_diagonal_theta_independent}. To achieve this T count, Ref.~\cite{Kliunchnikov_shorter2023} defines regions in the complex plane, and $U_1$, $U_2$ must be found such that their top-left entries $u_1 = r_1 e^{i \varphi_1}$ and $u_2 = r_2 e^{i \varphi_2}$ are in the corresponding regions (see \Cref{fig:mixed_diagonal_u_region}).

In this work, we propose using quasi-probabilities.
Inspired by the mixed approach, we consider the case of a quasi-probability mixture using two unitaries. In order to cancel the cross-term in \cref{eq:channel coeffients}, we choose the quasi-probabilities as
\begin{align}
 c_1 &= \frac{r_2^2 \sin(2 \varphi_2) - \sin(2 \theta)}{r^2_2 \sin(2 \varphi_2) - r_1^2 \sin(2 \varphi_1)}, \nonumber \\
    c_2 &= 1-c_1=\frac{-r_1^2 \sin(2 \varphi_1) + \sin(2\theta)}{r^2_2 \sin(2 \varphi_2) - r_1^2 \sin(2 \varphi_1)}.
    \label{eq:quasiprobabilities 2 rotations}
\end{align}
Note that this is a slightly different choice of coefficients compared to the probabilities taken in Ref.~\cite{Kliunchnikov_shorter2023}; we provide a theoretical comparison of the two choices in \Cref{app:comparison kliuchnikov}, in addition to numerical comparison, showing that the two choices give near identical results in practice.

In the next section we describe how $U_1$ and $U_2$ may be generated in practice and give numerical results.

\subsection{Numerical results}
\label{sec:numerical_results_ancilla_free}

\begin{figure}
    \centering
    \includegraphics[width=0.4\linewidth]{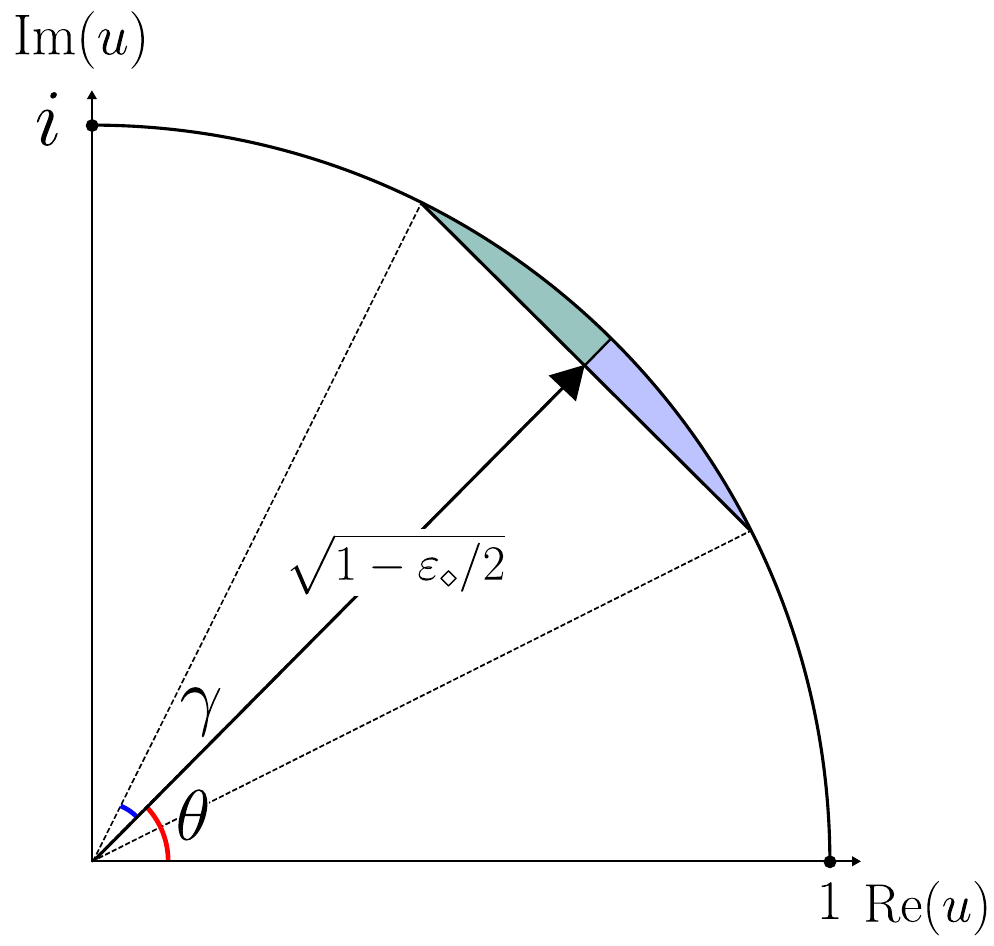}
    \caption{Regions of the complex plane for selecting the under- and over-rotation in the mixed diagonal approximation of Ref.~\cite{Kliunchnikov_shorter2023}. The target angle is $\theta$, and $\gamma = \arcsin(\sqrt{\epsdiamond/2})$. Selecting $u_1$ (for the under-rotation) in the blue region and $u_2$ (for the over-rotation) in the green region ensures a diamond norm error within $\epsdiamond$.}
    \label{fig:mixed_diagonal_u_region}
\end{figure}

In this section we use the approach of \Cref{sec:under_and_over_rot} to demonstrate the cost of the quasi-probability LCC approach in practice, and also show comparison to the approach using proper probabilities.

We generate $U_1$ and $U_2$ using the Ross-Selinger algorithm, as implemented in pygridsynth \cite{pygridsynth_github, pygridsynth_paper}. This allows generation of Clifford+T sequences with $u_1 = r_1 e^{i\varphi_1}$ and $u_2 = r_2 e^{i\varphi_2}$ in a given region of the complex plane. For each rotation angle, $\theta$, we select $u_1$ and $u_2$ in the blue and green regions of \Cref{fig:mixed_diagonal_u_region}, for a given $\epsdiamond$. These are the regions of Ref.~\cite{Kliunchnikov_shorter2023} which solve the mixed diagonal approximation to within diamond norm error $\epsdiamond$. Therefore, by varying $\epsdiamond$ we can generate under- and over-rotations of varying quality. We modified pygridsynth to obtain solutions in the blue and green regions in \Cref{fig:mixed_diagonal_u_region}.

Results are presented in \Cref{fig:av_t_count_per_rot} for angles from $\theta = 10^{-6}$ to $\theta = 10^{-1}$, and using a range of solutions of varying accuracy for $U_1$ and $U_2$. For each angle $\theta$, we generate the set of $U_1$ and $U_2$ as follows. We scanned over a range of $\epsdiamond$ values from $10^{-10}$ to $\sim 0.5$. For each even number of T gates we saved the under- and over-rotation. We also saved the $S^{\dagger}$ and $T^{\dagger}$ as optimal over-rotations with $0$ and $1$ T gates. Then, for each data point in the plot, we fixed a maximum T gate count, and took the best under- and over-rotation up to that T gate count from the saved set. For our quasi-probability approach the average T gate count is calculated using \cref{eq:av_t_count_formula}, and the L1 norm $\lambda$ is calculating using \cref{eq:lambda channel coefficients}, with the coefficients for the under- and over-rotation chosen as in \cref{eq:quasiprobabilities 2 rotations}. The results from this quasi-probability method are plotted as ``$\times$'' data points. We also show the results of the mixed diagonal approximation of Ref.~\cite{Kliunchnikov_shorter2023} using the same under- and over-rotations, and where the value on the $x$-axis is the diamond norm error, plotted as ``$+$'' data points.

\Cref{fig:av_t_count_per_rot} shows two distinct regions of behavior. For all rotation angles, for sufficiently low $\delta$ results return to the angle-independent result obtained in Ref.~\cite{Kliunchnikov_shorter2023}, which is \cref{eq:mixed_diagonal_theta_independent}. However, we also see a distinct second region of behavior, where the average number of T gates required to achieve a given $\delta$ is significantly below that predicted by the angle-independent formula. The size of this region of advantage grows with decreasing angle $\theta$. We see that the true average T gate count has significant $\theta$ dependence in this region.

The explanation for the angle-dependent and angle-independent regions is simple. At small $\theta$ the identity can be used as an accurate under-rotation, becoming more accurate as $\theta$ is decreased. For small $\theta$, almost all weight in the LCC will be on the identity channel, and the average T gate count is significantly reduced; when the circuit is compiled, most rotations will be implemented as the identity. As the over-rotation is implemented with increasing T gate counts, the over-rotation (and its twirls) become more accurate. The weight on the over-rotations in the LCC will therefore increase, while the weight on the identity decreases, hence the average T gate count increases (and $\delta$ decreases, because a more accurate basis is used for the LCC). As higher T gate counts are considered, eventually the identity will no longer be the best under-rotation compared to other Clifford+T unitaries available. At this point, it is optimal to not use the identity as the under-rotation, but to instead use the best Clifford+T unitary. There will be no small-angle savings here, and the average T count is angle-independent, governed by the existing formula, \cref{eq:mixed_diagonal_theta_independent}, which the data is seen to match well. We illustrate these two regimes in \Cref{fig:schematic}(a).

The fact that the average T gate count returns to $\mathcal{O}(\log(1/\delta))$ for all angles in the worst case is important, because it means that the protocol remains scalable to the limit of large circuits. Meanwhile, we see that the true cost can be much lower for small angles, dependent on $\delta$. We emphasize that, while probability and quasi-probability sampling techniques are often associated with NISQ or early fault-tolerant protocols, this protocol is entirely scalable to large-scale fault-tolerant circuits, just as for unitary Clifford+T synthesis algorithms. While the reductions in average T gate count in \Cref{fig:av_t_count_per_rot} look significant, one must remember that for circuits with a large number of rotations, $\delta$ for each rotation must be sufficiently small. Therefore, one must assess the improvement from these results for practical examples of interest. We will assess the reduction in T gate counts for Trotterized dynamics in \Cref{sec:trotter}.

\Cref{fig:av_t_count_per_rot} clearly shows that $\lambda-1$ in the quasi-probability scheme and $\epsdiamond$ in the probability scheme are extremely similar. In \Cref{app:comparison kliuchnikov} we provide a theoretical derivation of this result. However, this does \emph{not} imply that the two schemes are always equivalent. Indeed, the quasi-probability scheme can allow significantly lower resources: \Cref{app:prob_vs_quasi_prob} shows that, for equal total error, $λ-1$ can typically be taken orders of magnitudes larger than $\epsdiamond$.

\section{Quasi-probabilities for small angles}
\label{sec:small angles}

In our results in \Cref{fig:av_t_count_per_rot} we saw significant reductions in the average T gate count in the range $θ^2 \lesssim λ-1$. This is the range where is it optimal to set the under-rotation to the identity, with 0 T gate cost.
Here, we will perform such an analysis of quasi-probabilities with the under-rotation fixed as the identity. \Cref{prop:identity} in \Cref{app:comparison kliuchnikov} shows that our small-angle quasi-probability results closely match the outcome of using probabilities, when $ε_\diamond$ is identified with $λ-1$.

We will build on the setup in subsections~\ref{sec:general_setup} and \ref{sec:under_and_over_rot}, but specializing to the case where the under-rotation is fixed as the identity, $U_1 = \mathbb{1}$.
In this case \cref{eq:channel coeffients} can be modified as follows. Since the identity is already accounted for the in the $c_I$ term, the sum over $c_i$ only needs to include the over-rotation. We denote the coefficient for the over-rotation as $p$, and the upper-left element of the over-rotation as $u=re^{iφ}$. The coefficients for the LCC then become
\begin{equation}
    c_I = \cos^2θ - p r^2 \cos^2 φ,
\end{equation}
\begin{equation}
    c_X = c_Y = p\frac{r^2 - 1}{2},
\end{equation}
\begin{equation}
    c_Z = \sin^2θ - p r^2 \sin^2 φ,
\end{equation}
\begin{equation}
    c_{\times} = i \Big(p \frac{r^2\sin2φ}{2} - \frac{\sin2θ}{2} \Big).
\end{equation}

The quasi-probability of the over-rotation $p$ is fixed by the condition $c_\times=0$:
\begin{equation}
    p = \frac{\sin2θ}{r^2\sin2φ} = \frac{\sin2θ}{2xy},
    \label{eq:probability}
\end{equation}
where we have defined $x:=r\cosφ$ and $y:=r\sinφ$.

The sample complexity $λ$ is the sum of all coefficient absolute values. 
In the following we assume without loss of generality that $0 < \theta < \varphi < \pi/4$.
This range restriction leads to definite signs of the coefficients:
\begin{align}
    λ &= |p| + |c_I| + 2|c_X| + |c_Z| \\
      &= p + c_I - 2c_X - c_Z \\
      &= \tanα \sin2θ + \cos2θ ,
      \label{eq:lambda}
\end{align}
with the definition
\begin{equation}
    \tanα := \frac{2/r^2-1-\cos2φ}{\sin2φ} = \frac{1-x^2}{xy}. \label{eq:tanalpha}
\end{equation}
The result \cref{eq:lambda} can be obtained from the coefficients above by a short derivation\footnote{Note that with \cref{eq:probability}, $c_Z=\sin^2θ - \sin^2φ\sin2θ/\sin2φ = \sin^2θ\,(1-\tanφ/\tanθ)<0.$}. The angle $α$ has a natural interpretation: For fixed $λ$, the maximum possible over-rotation is $φ=α$, which is achieved when $r=1$ (see \Cref{fig:quasiregion} for a visual representation).

The best over-rotation for a given target angle $θ$ and maximum desired sample complexity $λ_\text{max}$ should be chosen such that $λ\leqλ_\text{max}$ and having lowest possible average T count, which is the T count of the over-rotation times its probability $p / \lambda$. As $\lambda - 1 \ll 1$, we will occasionally ignore the $1/λ \approx 1$ normalization of the quasi-probability distribution.

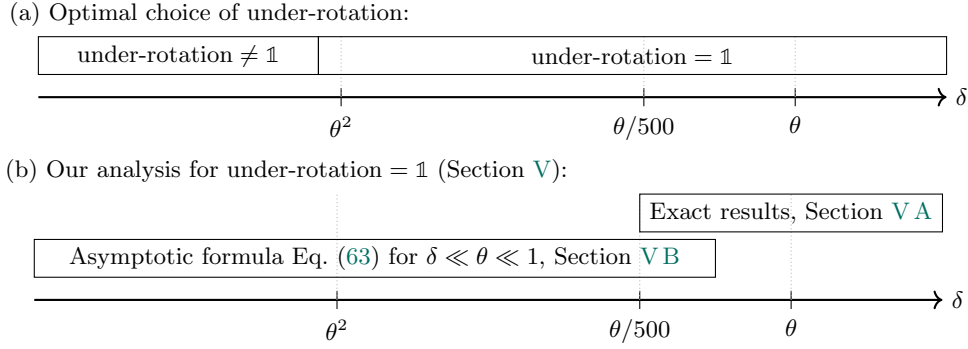
\begin{figure}

\begin{tikzpicture}
\draw[thick,->] (0,0) -- (12,0) node[right] {$\delta$};
\foreach \x/\lbl in {
    4/{$\theta^2$},
    8/{$\theta/500$},
    10/{$\theta$}
}{
    \draw[thin] (\x,0.12) -- (\x,-0.12);
    \node[below=4pt] at (\x,0) {\lbl};
    \draw[densely dotted,gray!55] (\x, 0) -- (\x, 0.8);
}
\node[anchor=west] at (-0.5, 1.1) {(a) Optimal choice of under-rotation:};
\draw (0,0.3) rectangle (3.7, 0.8) node[pos=.5] {under-rotation${}\neq\mathbb{1}$};
\draw (3.7,0.3) rectangle (12, 0.8) node[pos=.5] {under-rotation${}=\mathbb{1}$};
\end{tikzpicture}
\vspace{3ex}
\begin{tikzpicture}
\draw[thick,->] (0,0) -- (12,0) node[right] {$\delta$};
\foreach \x/\lbl in {
    4/{$\theta^2$},
    8/{$\theta/500$},
    10/{$\theta$}
}{
    \draw[thin] (\x,0.12) -- (\x,-0.12);
    \node[below=4pt] at (\x,0) {\lbl};
    \draw[densely dotted,gray!55] (\x, 0) -- (\x, 1.4);
}
\node[anchor=west] at (-0.5, 1.7) {(b) Our analysis for under-rotation${}=\mathbb{1}$ (\Cref{sec:small angles}):};
\draw (0, 0.3) rectangle (9, 0.8) node[pos=.5] {Asymptotic formula \cref{eq:asymptotic} for $δ\llθ\ll1$, \Cref{sec:asymptotic}};
\draw (8, 0.9) rectangle (12, 1.4) node[pos=.5] {Exact results, \Cref{sec:matsumoto_amano}};
\end{tikzpicture}
\vspace*{-5ex}
\caption{Schematic overview of methods for various regimes of $θ$ and $δ=λ-1$. (a) shows the preferable method in each regime. In the main plot, \Cref{fig:av_t_count_per_rot}, this crossover is visible at $δ$ just below $θ^2$. (b) summarizes our analysis in \Cref{sec:small angles} of the method with the under-rotation fixed as the identity. Results are approximately the same for $δ=ε_\diamond$, see \Cref{app:comparison kliuchnikov}.
\label{fig:schematic}} 
\end{figure}

The next subsections are summarized in \Cref{fig:schematic}(b):
First, we show how optimal over-rotations can be found by exhaustive search over all Clifford+T sequences up to a maximum T count (\Cref{sec:matsumoto_amano}). This is effective for $δ=λ-1 \gtrsim θ/500,$ after which the exponential cost makes exhaustive search prohibitively expensive. \Cref{tab:staircase} lists the optimal over-rotations which can directly be used in other work.
Secondly, we present an analytic formula estimating the average T count by performing an asymptotic analysis for $λ-1\llθ\ll1$ (\Cref{sec:asymptotic}).
In \Cref{app:gridsynth}, we also explain an adapted gridsynth algorithm that can determine good over-rotations.

\subsection{Exact results: Exhaustive search with the Matsumoto-Amano normal form}
\label{sec:matsumoto_amano}

We perform an exhaustive search over all possible Clifford+T sequences up to a maximal T count to find optimal over-rotations. Optimal over-rotations are those optimal according to the following partial order: An over-rotation with same or lower $λ$ and same or lower average T count is preferable, as it is more widely applicable to a larger range of target sample complexities $λ$ at same (or lower) cost.

In fact, optimal over-rotations can be characterized in a $θ$-independent fashion. Instead of $p$, \cref{eq:probability}, and $λ$, \cref{eq:lambda}, one can consider and extremize the quantities
\begin{equation}
    p/\sin2θ,\ \tanα = (λ-\cos2θ)/\sin2θ.
\end{equation}
In this way, the results in \Cref{tab:staircase} of our exhaustive search show optimal over-rotations for arbitrary angles\footnote{Subject to the requirement that the over-rotation angle $φ>θ$. Yet, as we argue in \Cref{app:matsumoto_amano}, this inequality will typically be true in all regimes of interest.} up to $λ-1 \gtrsim 0.01\,θ$. \Cref{app:matsumoto_amano} shows further results up to $λ-1\gtrsim0.002\,θ$. The main plot (\Cref{fig:av_t_count_per_rot}) shows these results in that regime of $λ-1$ as piecewise constant ``staircases'', owing to the discrete nature of unitaries with small $T$ count.

\begin{table}
\begin{tabular}{cc|cccl}
\multicolumn{2}{l|}{} & \multicolumn{4}{l}{Over-rotation unitary with $u=re^{iφ}$} \\
{$\tan \alpha$} & {$(p/\sin2θ)\cdot\text{T-count}$} & {T-count} & {$1-r$} & {$\varphi$} & {gate sequence} \\
\hline
\num{1.00e+00} & \num{0.00e+00} & 0 & \num{0.00e+00} & \num{7.85e-01} & ISZ \\
\num{4.14e-01} & \num{1.41e+00} & 1 & \num{0.00e+00} & \num{3.93e-01} & TSZ \\
\num{3.51e-01} & \num{8.36e+00} & 4 & \num{1.08e-02} & \num{2.55e-01} & ISHTHTSHTSHTHZ \\
\num{3.13e-01} & \num{1.49e+01} & 8 & \num{2.68e-03} & \num{2.85e-01} & ISHTSHTSHTHTHTHTHTSHTSHSI \\
\num{2.12e-01} & \num{1.87e+01} & 7 & \num{1.57e-03} & \num{1.93e-01} & IHTSHTHTHTSHTSHTSHTSHX \\
\num{1.99e-01} & \num{3.54e+01} & 12 & \num{2.01e-03} & \num{1.74e-01} & ISHTSHTHTHTHTHTSHTSHTSHTSHTSHTSHTHX \\
\num{1.38e-01} & \num{3.67e+01} & 9 & \num{7.86e-04} & \num{1.24e-01} & IHTHTSHTSHTHTSHTHTSHTSHTHSI \\
\num{1.15e-01} & \num{4.58e+01} & 10 & \num{2.30e-04} & \num{1.10e-01} & ISHTSHTSHTSHTHTSHTHTHTHTHTSHX \\
\num{1.11e-01} & \num{5.96e+01} & 13 & \num{3.37e-05} & \num{1.10e-01} & ISHTSHTSHTSHTHTHTSHTHTSHTHTHTSHTSHTSHZ \\
\end{tabular}
\caption{$θ$-independent optimal over-rotations, found by exhaustive search (\Cref{sec:matsumoto_amano}) up to T count 13. For given target rotation angle $θ$ and target sample complexity $λ_\text{max}$, the optimal over-rotation is that with $λ=\tanα\sin2θ + \cos2θ$ not exceeding $λ_\text{max}$. It has average T-count $p\cdot$T-count, which is divided by $\sin2θ$ in the table to show a $θ$-independent quantity. The data is visible as a ``staircase'' in \Cref{fig:av_t_count_per_rot}. We list one possible gate sequence with lowest $T$-count. As we are only interested in the top-left entry $u$ of each over-rotation (after normalizing the determinant), multiple sequences will be equivalent. This table is an excerpt, with more data for lower $\tanα$ available in \Cref{app:matsumoto_amano}.}
    \label{tab:staircase}
\end{table}

The Matsumoto-Amano normal form allows every Clifford+T sequence to be written uniquely, and with the minimal T count. Following Ref.~\cite{Giles2019}, the normal form can be written using the notation of regular expressions,
\begin{equation}
    (T|\mathbb{1})(HT|SHT)^*\mathcal{C},
    \label{eq:matsumoto normal form}
\end{equation}
where $T$, $H$ and $S$ are T, Hadamard and S gates, respectively, and $\mathcal{C}$ represents one of the $24$ single-qubit Clifford gates, up to a phase. Using this normal form, it is straightforward to construct all Clifford+T unitaries up to $n_T$ T gates. Our C++ implementation makes use of a number theoretic representation \cite{kliuchnikov_fast_2013} to compute the gate sequences exactly, without floating-point error\footnote{Each entry of a $2\times2$ Clifford+T unitary can be written as $aω^3+bω^2+cω+d\in\mathbb{D}[ω],$ with $ω=e^{i\pi/4}$ and each $a,b,c,d\in\mathbb{D}:=\mathbb{Z}[\tfrac{1}{2}]$, for example $a = n/2^k$ with $n$ integer and the denominator exponent $k$ a non-negative integer. Computation in this ring can be performed exactly by storing and manipulating the underlying integers.}.  Only in the final step is $u$ converted to \texttt{double}: Each unitary's determinant is normalized to 1 to yield an element of $SU(2)$, as global phases are unphysical. Then $p/\sin2θ = 1/(2xy)$ \eqref{eq:probability} and $\tanα = (1-x^2)/(xy)$ \eqref{eq:tanalpha} are computed and compared with other results, keeping tally of the best over-rotations. We use the Cartesian form of the expressions to avoid expensive (and possibly inaccurate) trigonometric calculations.

While we have optimized our code with careful caching, ultimately the number of Clifford+T sequences at T count $n_T$ grows exponentially as $O(2^{n_T})$, limiting our exhaustive search. For this work we have generated and assessed all sequences up to T count $n_T=35$, which took 262 hours on a single core, see \Cref{app:matsumoto_amano} for the full results. Note that results towards the bottom of the table in the appendix might not be optimal: Conceivably, a higher T count over-rotation will have a lower $p$. An example of this is the 12 T-count unitary in \Cref{tab:staircase}; enumerating only up to T count 11 would have given suboptimal unitaries.

We will now explain in detail the first two rows of \Cref{tab:staircase} as examples.
\begin{itemize}
\item First row, over-rotation with T count 0:
\begin{equation}
    e^{i\pi/4} ISZ = \begin{pmatrix}e^{i\pi/4} & 0 \\ 0 & e^{-i\pi/4}\end{pmatrix}
\end{equation}
The global phase turns it into an $SU(2)$ unitary with $r=1, φ=\pi/4\approx0.785$. In order to be able to use the over-rotation, $\tanα\ge\tan\pi/4 = 1$ is required, which is equivalent to
\begin{equation}
    λ \ge \sin2θ + \cos2θ = 1 + 2θ + O(θ^2).
\end{equation}
For this range of target $λ$, a T count of zero can be used. It is shown in \cref{fig:av_t_count_per_rot} as a vertical line to $-\infty$ at that $λ$. Note that this is the same $λ$ as for the introductory example, \cref{eq:introductory example I S^dag Z channel}, where the rotation channel was also implemented without any T gates.
\item Second row, over-rotation with T count 1: 
\begin{equation}
    e^{i\pi/8}TSZ = \begin{pmatrix} e^{i\pi/8} & 0 \\ 0 & e^{-i\pi/8}\end{pmatrix},
\end{equation}
with $r=1, φ=\pi/8 \approx 0.393$. In order to be able to use the over-rotation, $\tanα\ge\tan\pi/8 = \sqrt{2}-1\approx 0.414$ is required, which is equivalent to
\begin{equation}
    λ \ge  (\sqrt{2}-1)\sin2θ + \cos2θ = 1 + 2(\sqrt{2}-1)θ + O(θ^2).
\end{equation}
The $p$ at that point $r=1,φ=\pi/8$ is $p=\frac{\sin 2θ}{\sin π/4}$, giving a T count of $\sqrt{2}\sin 2θ$. This is shown as a horizontal segment in the plot; once $λ\ge\sin2θ+\cos2θ > (\sqrt{2}-1)\sin2θ + \cos2θ$ it is preferable to use the solution with zero T count.
\end{itemize}

\subsection{Analytic asymptotic formula for \texorpdfstring{$λ-1 \ll θ \ll 1$}{λ-1 << θ << 1}}
\label{sec:asymptotic}

For resource estimation purposes, and to better understand the advantages of the small-angle method, a simple analytic formula estimating the average T count is important. In this section, we derive such a formula in the limit $δ :=λ-1 \ll θ \ll 1$.

\begin{figure}
    \includegraphics[width=0.5\textwidth]{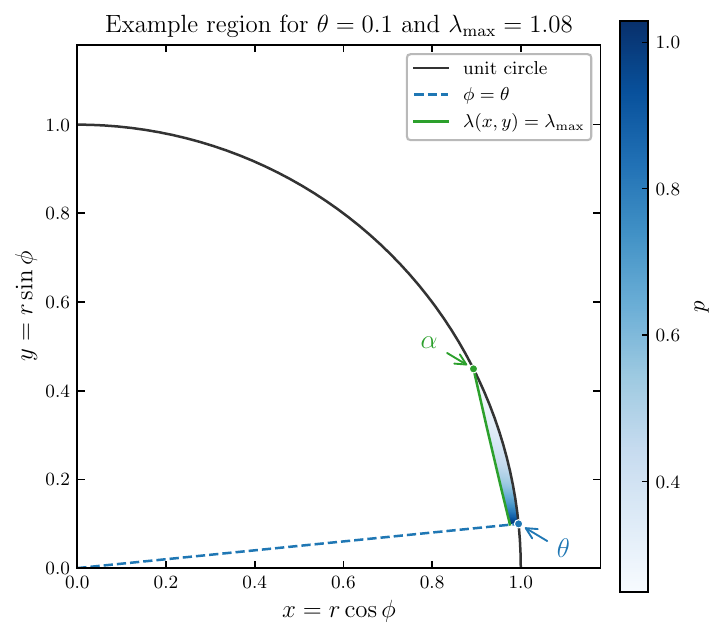}
    \caption{Example region, defined in \cref{eq:region}, for selecting the over-rotation with quasi-probabilities. Each unitary with upper left entry $u=x+iy$ inside the shaded region has a sample complexity $λ$,  \cref{eq:lambda}, within the desired target $λ_{\max}$. Here we take $\theta=0.1$ and $\lambda_{\max}=1.08$ as an illustrative example. The shading indicates the sampling probability $p$, \cref{eq:probability} (and thereby average T count, $p\cdot$T-count), which varies within the region. The lower end of the region is at the target angle $θ$ due to the assumption $φ>θ$, while the upper end of the region is at angle $\alpha$, \cref{eq:tanalpha}. As $λ_\text{max}$ increases, $α$ and the area size increase.}
    \label{fig:quasiregion}
\end{figure}

Given a target angle $θ$ and maximum target sample complexity $λ_\text{max}$, such a quasi-rotation can be achieved with any over-rotation fulfilling $λ\leλ_\text{max}$. The target region $A$ for $u$ is thus defined by \cref{eq:lambda} and its assumptions, giving the intersection
\begin{equation}
A = \{(x,y)\in \mathbb{C},\ λ(x,y)=\frac{1-x^2}{xy}\sin2θ + \cos2θ \le λ_\text{max}\} \cup \{re^{iφ}\in\mathbb{C},\ 0<r\le1, θ<φ<\pi/4\}.
\label{eq:region}
\end{equation}
This region is illustrated for example parameters in \Cref{fig:quasiregion}.

Our analysis is then based on the estimate that the lowest T count of a Clifford+T sequence lying in a region $A\subset\mathbb{C}$ is \cite{Kliunchnikov_shorter2023}
\begin{equation}
  \log_2(1/|A|),
  \label{eq:area t count}
\end{equation}
assuming $|A|$ is small. In turn, a small area $|A|$ is achieved in the regime $λ_\text{max}-1 \ll θ$. Contrarily, if the area is large (when $λ_\text{max}-1$ is not much smaller than $\theta$), the estimate \cref{eq:area t count} breaks down. Then the T count is a small integer, meaning the relationship between area size and T count becomes discrete. In the regime where the area is large, our asymptotic analysis is not valid, but the problem can be be solved exactly (\Cref{sec:matsumoto_amano}), resulting in the staircase in \Cref{fig:av_t_count_per_rot}.

We estimate average T count with the product
\begin{equation}
    T = p_\text{avg}(A) \log_2 \frac{1}{|A|}, \label{eq:estimate T count principle}
\end{equation}
of $p_\text{avg}(A)$, the average probability $p$, \cref{eq:probability}, inside the area $A$, and the $T$ count of a unitary in the region. Here, we have assumed that the over-rotation is equally distributed within the region and hence used $p_\text{avg}$. Rather than using the full permissible region defined in \cref{eq:region} for desired $θ,λ_\text{max}$, we reduce the region size by choosing a minimal angle $φ_0\in[θ,α]$, and requiring $\arctan(y/x) = φ \ge φ_0$. As can be seen in \Cref{fig:quasiregion}, $φ_0$ introduces a tradeoff between region size (T-count) and $p_\text{avg}(A)$, and $φ_0$ can be optimized.

We will compute \cref{eq:estimate T count principle} in the $λ-1\ll θ \ll1$ limit, neglecting subdominant orders. Let us compute the boundary curve of the region $A$ (see \cref{eq:region}, \Cref{fig:quasiregion}). For ease of notation, we will use $λ$ in place of $λ_\text{max}$ for the (maximum) target sample overhead. Let us begin with the point $α$, which follows from \cref{eq:lambda}:
\begin{align}
    \tanα = \frac{λ-\cos2θ}{\sin2θ} = \frac{λ-1}{\sin2θ} + \tanθ  \ \Rightarrow α \approx\frac{λ-1}{2θ} + θ
    \label{eq:tanalpha from lambda}
\end{align}
and therefore $φ\leα\ll1$ for all points $(r,φ)$ in the region. For simplicity and consistency we also truncate the quasi-probability \cref{eq:probability} up to first order:
\begin{equation}
    p = \frac{\sin2θ}{r^2\sin2φ} \approx \frac{θ}{r^2φ}.
\end{equation}
The boundary curve in polar coordinates can be computed by solving \cref{eq:tanalpha} for $r^2$:
\begin{align}
    2/r^2 &= \tanα\sin2φ +\cos2φ + 1 \approx 2 + 2φ(α-φ), \\
    \Rightarrow r^2(φ) &\approx 1 - φ(α-φ).
\end{align}
Now, area and average probability can be computed by integrating the area in polar coordinates:
\begin{align}
    |A| &= \int_{φ_0}^α dφ \int_{r(φ)}^1 r' dr'  = \int_{φ_0}^α dφ\, \frac{1-r^2(φ)}{2} \\
    & \approx \frac{1}{12}(α-φ_0)^2(α+2φ_0) \\
    p_\text{avg}(A) &= \frac{1}{|A|} \int_{φ_0}^α dφ \int_{r(φ)}^1 r' dr' \frac{θ}{r'^2φ} = \frac{1}{|A|} \int_{φ_0}^α dφ \frac{θ}{φ}(-1)\ln (r(φ)) \\
    &\approx\frac{1}{|A|} \int_{φ_0}^α dφ \frac{θ}{φ}(φ(α-φ))/2 = \frac{1}{|A|} θ\frac{(α-φ_0)^2}{4} \\
    &\approx 3θ\frac{1}{α+2φ_0}
\end{align}
where we have used $-\ln(\sqrt{1-x})\approx x/2$.
Altogether, we get an asymptotic $T$ count of
\begin{equation}
T = p_\text{avg}(A)\log_2\left(\frac{1}{|A|}\right) = \frac{3θ}{α+2φ_0}\log_2\frac{12}{(α-φ_0)^2(α+2φ_0)},
\label{eq:asymptotic}
\end{equation}
where $θ\leφ_0\leα$ must be chosen to minimize $T$. In \Cref{app:optimal phi0} we show that, at leading order, the minimizer is
\begin{equation}
    φ_0 \approx \max\left(α-\frac{α}{\ln(K/α)},θ\right).
\end{equation}

The minimizer is clipped to $φ_0=θ$ in the regime $δ\llθ^2$ (see \Cref{app:optimal phi0}). Then the probability $p_\text{avg}\to1$, while the area term
\begin{equation}
    \log_2\frac{1}{|A|} \approx \log_2\frac{16θ}{δ^2} \approx 2\log_2(1/δ) + \log_2(16θ)
\end{equation}
is dominant, such that the choice of $φ_0=θ$ maximizes the available area. It is this regime where fixing the under-rotation as the identity is not optimal any more, and lower average T counts can be achieved with other under-rotations: the angle-independent mixed approximation has a lower average T count of $T\approx 1.52\log_2(1/δ)-0.01$. \Cref{fig:schematic}(a) schematically shows the different regimes.

Conversely, in the regime $δ \gg θ^2$, the T count becomes
\begin{align}
2\frac{θ^2}{δ}\log_2\left(32\frac{θ^3}{δ^3}\left(\ln(2Kθ/δ)\right)^2\right).
\label{eq:asymptotic delta>>theta^2}
\end{align}
This is an improvement on the angle-independent formula \cref{eq:mixed_diagonal_theta_independent}. Up to the slowly varying logarithmic term, this is a power law $T\sim θ^2/δ$, and appears linear on the log-log plot, clearly visible in \Cref{fig:av_t_count_per_rot}. Thanks to this regime, Trotterization approaches a total constant T cost as the step size $t$ becomes smaller. The step size $t$ cancels in \cref{eq:asymptotic delta>>theta^2} with scalings $θ\sim t$, $N_\text{rotations}\sim1/ t$, and $δ\sim1/N_\text{rotations}$ to ensure limited overall sample complexity. See \Cref{sec:trotter} for more detailed discussion of Trotterization results.

\section{Fallback scheme}
\label{sec:mixed_fallback_scheme}

\begin{figure}
\centering


\includegraphics[width=0.42\linewidth]{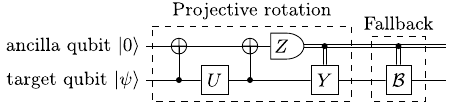}
\caption{Fallback protocol circuit, see Ref.~\cite[Fig.~1]{Kliunchnikov_shorter2023}. When the top-left entry of $U$ is $u=re^{iφ}$, the projective rotation results in the rotation $R_z(φ)$ with probability $r^2$. Otherwise, a fallback operation is applied to correct the erroneous operation.\label{fig:fallback circuit}}
\end{figure}

We next consider the use of fallback channels. In the fallback approach, we first perform a Clifford+T unitary together with a projective measurement, see \Cref{fig:fallback circuit} for the circuit diagram requiring one ancilla qubit. Let us again write the Clifford+T unitary as in \cref{eq:su2_def_2}, where the top-left element is $u = r e^{i\varphi}$ and the bottom-left element is $v$. In this case, the projective measurement succeeds with probability $r^2$, and fails with probability $1 - r^2$. If the projective measurement succeeds then the applied operation is
\begin{equation}
    | \psi \rangle \rightarrow e^{i \varphi Z} | \psi \rangle,
\end{equation}
while if the measurement fails then the applied operation is
\begin{equation}
    | \psi \rangle \rightarrow e^{i \arg(v) Z} | \psi \rangle. \label{eq:failure rotation}
\end{equation}
The latter case is typically substantial in error, and therefore one applies a fallback to correct this case. The fallback itself could be implemented in a range of different ways. We again follow the approach of Ref.~\cite{Kliunchnikov_shorter2023}, where the fallback is implemented by the mixed diagonal approximation.

The fallback channel is
\begin{equation}
    \mathcal{F}(\rho) = q  \mathcal{Z}_{\varphi}(\rho) + (1-q) \mathcal{B'}(\rho),
\end{equation}
where $q = r^2$ is the success probability. The channel $\mathcal{B'}$ is the composition of the failure rotation, \cref{eq:failure rotation}, and the applied fallback, $\mathcal{B}$. Since $\mathcal{F}$ is a well-defined quantum channel, it is possible to use this in a quasi-probability LCC, similarly to the ancilla-free approach in \Cref{sec:ancilla_free_scheme}.

\subsection{Quasi-probability LCC over fallback channels}
\label{sec:quasi_prob_fallback}

We define a quasi-probability LCC which contains under- and over-rotation fallback channels, together with the Pauli channels. We denote the under- and over-rotation unitaries for the projective step as $U_1$ and $U_2$. Following the notation of \cref{eq:under_and_over_rotation_def}, these have top-left elements $u_1 = r_1 e^{i \varphi_1}$ and $u_2 = r_2 e^{i \varphi_2}$. The under-rotation is defined as having $\varphi_1 < \theta$, while the over-rotation is defined as having $\varphi_2 > \theta$.

The LCC is defined
\begin{equation}
    \mathcal{Z}_{\theta}(\rho) = c_1 \mathcal{F}_1(\rho) + c_2 \mathcal{F}_2(\rho) + c_I \rho + c_X X \rho X + c_Y Y \rho Y +  c_Z Z \rho Z.
    \label{eq:fallback_lcc_general}
\end{equation}
The channels $\mathcal{F}$ can be written in the Pauli basis
\begin{equation}
    \mathcal{F}_k(\rho) = i a_{k, \times} (Z \rho I -  I \rho Z)  + \sum_{k} a_{k, j} P_j \rho P_j,
\end{equation}
for some coefficients $a_{k, j}$ and $P_j \in \{I, X, Y, Z \}$. That is, they consist of Pauli channels plus cross terms, $I(\cdot)Z - Z(\cdot) I$.

The coefficients $c_1$ and $c_2$ are then uniquely determined by the condition $c_1 + c_2 = 1$, together with the condition that the coefficients for the cross-terms $I(\cdot)Z - Z(\cdot) I$ match between the LHS and RHS of \cref{eq:fallback_lcc_general}. After specifying $c_1$ and $c_2$ in this way, the coefficients $\{ c_I, c_X, c_Y, c_Z \}$ for the Pauli channels then follow uniquely by matching the LHS and RHS of \cref{eq:fallback_lcc_general}. Due to the more complicated nature of the fallback channels, we do not expand out these expressions further, but the procedure described exactly matches that used in \Cref{sec:ancilla_free_scheme} for the ancilla-free scheme.

In the case where the projective under-rotation is the identity, $\mathcal{F}_1 = \mathcal{I}$, we can set $c_1 = 0$ since the identity term is accounted for by $c_I$.

The L1 norm of the LCC is
\begin{equation}
    \lambda = |c_1| + |c_2| + |c_I| + |c_X| + |c_Y| + |c_Z|,
\end{equation}
and average T-gate count is calculated as
\begin{equation}
    T = \frac{|c_1|}{λ}\cdot\text{T-count}(\mathcal{F}_1) + \frac{|c_2|}{λ}\cdot\text{T-count}(\mathcal{F}_2),
    \label{eq:av_t_count_fallback}
\end{equation}
where $\text{T-count}(\mathcal{F}_k) $ denotes the average T-gate count to implement the channel, $\mathcal{F}_k$; this consists of the T gate count for the initial projective step with the unitary U, plus the average T gate count for the applied fallback $\mathcal{B}$ with probability $1-q$,
\begin{equation}
    \text{T-count}(\mathcal{F}) = \text{T-count}(U) + (1-q) \cdot \text{T-count}(\mathcal{B}).
\end{equation}

\subsection{Numerical results}
\label{sec:numerical results fallback}

\begin{figure}
    \centering
    \includegraphics[width=1.0\linewidth]{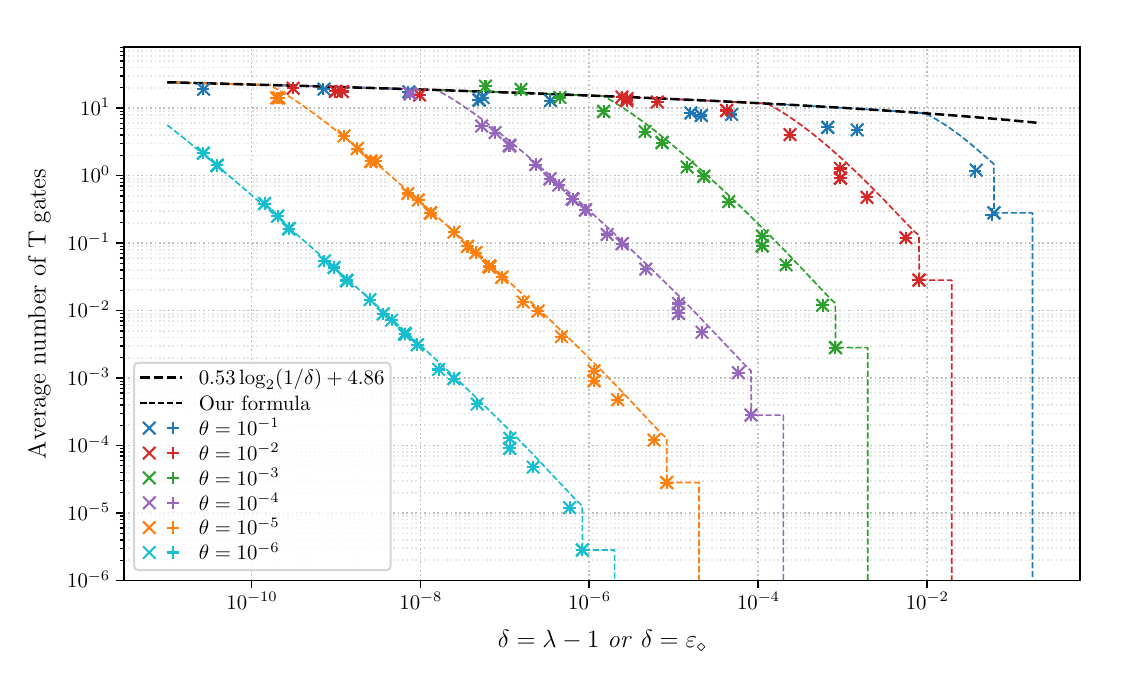}
    \caption{Average T gate counts to implement a rotation channel $\rzchannel$ in the fallback scheme and for two different formulations of our results, first in terms of quasi-probability mixtures with L1 norm $\lambda$ (results marked as ``$\times$''), and second in terms of strict probability mixtures with diamond norm error $\epsdiamond$ (results marked as ``$+$''). The dashed-dotted line indicates the previous angle-independent formula for the mixed fallback approximation from Kliuchnikov~\emph{et al}~\cite{Kliunchnikov_shorter2023}. Results for all rotation angles, $\theta$, return to the angle-independent result in the worst case. However, as for the ancilla-free scheme, we see significant improvements at small angles, dependent on $\delta$. In the small-angle regime, for a given $\delta$ the fallback scheme roughly halves the average T gate count compared to the ancilla-free scheme.}
    \label{fig:av_t_count_per_rot_fallback}
\end{figure}

Numerical results are presented in \Cref{fig:av_t_count_per_rot_fallback}, comparing the quasi-probability approach described in \Cref{sec:quasi_prob_fallback} (results marked ``$\times$'', assessed by $\lambda-1)$ to the probability approach of Ref.~\cite{Kliunchnikov_shorter2023} (results marked ``$+$'', assessed by $\epsdiamond$).
As for the ancilla-free scheme, we see that both quasi-probability and proper probability mixtures perform almost equally when identifying $λ-1$ with $\epsdiamond$. For this figure, we generated under- and over-rotations using a combination of enumeration for small T gate counts (up to 19 T gates) and solutions from gridsynth. The procedure for generating the under- and over-rotations involved some ad-hoc choices which are not necessarily optimal. In particular:
\begin{enumerate}
    \item When enumerating solutions, we defined the optimal under- and over-rotations as those which maximize $r^2/\sin^2(\gamma)$, where the top-left element of the unitary is $u = r e^{i\varphi}$ and $\gamma = \varphi - \theta$. This is based on the heuristic that we should choose Clifford+T sequences which minimize the angle error, $\gamma$, and maximize the success probability for the projective measurement, $q = r^2$.
    
    \item After choosing the unitaries $U_1$ and $U_2$ for the projective step, we must decide on the value of $\epsdiamond$ used to generate the mixed diagonal fallbacks, $\mathcal{B}_1$ and $\mathcal{B}_2$; see \Cref{fig:mixed_diagonal_u_region}. Ref~\cite[Eq.~(36)]{Kliunchnikov_shorter2023} defines $\varepsilon_1$, $\varepsilon_2$ and $\varepsilon_3$ for the projective step and two mixed diagonal fallback steps, respectively. Particularly for large values of $\varepsilon_1$, we found it beneficial to set $\varepsilon_2$ and $\varepsilon_3$ smaller, hence we took $\varepsilon_2 = \varepsilon_3$ = $\varepsilon_1 / 10$.
\end{enumerate}
While these choices are ad-hoc and based on experimentation, they are sufficient to allow us to establish both the angle-independent and angle-dependent regions in our data. Because the fallback case has additional parameters compared to the ancilla-free case, there is more potential for optimization. We leave such investigation for future work.

As for the ancilla-free scheme, we see that for $\delta \gtrsim \theta^2$, the average T gate count can be significantly below the existing angle-independent formula. This corresponds to the regime where the under-rotation is the identity, $\mathcal{F}_1 = \mathcal{I}$. For $\delta \lesssim \theta^2$ the optimal under-rotation is a non-identity fallback channel, and results return to the angle-independent formula.

We next study the $\mathcal{F}_1 = \mathcal{I}$ case in more detail to derive a formula for resource estimation.

\subsection{Fallback scheme for small angles}
\label{sec:fallback_small_angle}

We will show that the small-angle result in the fallback scheme is closely-related to the small-angle result in the ancilla-free scheme, derived in \Cref{sec:small angles}. We begin from the LCC defined in \cref{eq:fallback_lcc_general}, and expanding $\mathcal{F}_1 = q_1 \mathcal{Z}_{\varphi_1} + (1-q_1) \mathcal{B}_1'$ and $\mathcal{F}_2 = q_2 \mathcal{Z}_{\varphi_2} + (1-q_2) \mathcal{B}_2'$,
\begin{equation}
    \mathcal{Z}_{\theta}(\rho) = c_1 (r_1^2 \mathcal{Z}_{\varphi_1} + (1-r_1^2) \mathcal{B}_1') + c_2 (r_2^2 \mathcal{Z}_{\varphi_2} + (1-r_2^2) \mathcal{B}_2') + c_I \rho + c_X X \rho X + c_Y Y \rho Y +  c_Z Z \rho Z,
\end{equation}
where we have used $q_1 = r_1^2$ and $q_2 = r_2^2$. Since $c_i(1 - r_i^2)$ is small ($r_i \approx 1$), we neglect the contribution from the fallback steps $\mathcal{B}_1'$ and $\mathcal{B}_2'$,
\begin{equation}
    \mathcal{Z}_{\theta}(\rho) \approx c_1 r_1^2 \mathcal{Z}_{\varphi_1} + c_2 r_2^2 \mathcal{Z}_{\varphi_2} + c_I \rho + c_X X \rho X + c_Y Y \rho Y +  c_Z Z \rho Z.
\end{equation}
Now, notice that $r_1^2 \mathcal{Z}_{\varphi_1}$ is exactly the twirl $\mathcal{T}_{U_1}$ defined in \cref{eq:twirl}, except that the $X(\cdot)X$ and $Y(\cdot)Y$ terms are $0$, and similarly for the over-rotation. From this we see that, provided we neglect the fallback steps $\mathcal{B}_1'$ and $\mathcal{B}_2'$, the LCC for the fallback case is the same as for the ancilla-free case, except that we set $c_X = c_Y = 0$. Therefore, we may write
\begin{align}
    \lambda_{\mathrm{FB}} &= \lambda - 2|c_X|, \\
    &= \lambda - \sum_i c_i (1 - r_i^2),
\end{align}
where we have used $\lambda_{\mathrm{FB}}$ to denote the L1 norm of the fallback LCC, and $\lambda$ to denote the LCC for the equivalent ancilla-free case studied in \Cref{sec:ancilla_free_scheme}, see \cref{eq:lambda channel coefficients}.

Next, we specialize to the small-angle case where $\mathcal{F}_1 = \mathcal{I}$. In this case we may set $c_1 = 0$ (since the identity contribution is accounted for in $c_I$) and relabel $c_2$ as $p$, with $p$ taking the same value as in \cref{eq:probability}. We also drop the ``2'' subscript for quantities referring to the over-rotation. Recalling the derivation of \cref{sec:small angles}, we again define $x := r \cos \theta$ and $y := r\sin \theta$.

Let us consider the asymptotic $\theta \ll 1$, in which case $y \ll 1$,
\begin{align}
    \lambda-1 &\approx 2\theta \tan \alpha \approx 2 \theta \frac{1-x^2}{xy}, \\
    \lambda_{\mathrm{FB}}-1 &\approx \lambda-1 - p_i(1-r^2) \\
    &\approx 2 \theta \frac{1-x^2}{xy} - \frac{2\theta}{2xy}(1-x^2-y^2), \\
    &\approx 2\theta \frac{1-x^2}{2xy}, \\
    &= \frac{\lambda-1}{2}.
\end{align}
Therefore, in the small-angle limit and neglecting the fallback steps, $\lambda-1$ and $\lambda_{\mathrm{FB}}-1$ differ by a factor of $1/2$. To achieve a given average T gate count, and in the small-angle regime, $\delta = \lambda - 1$ can be set twice as large when using fallback channels compared to the ancilla-free scheme.

Compared to the ancilla-free scheme, the effect of the factor of $1/2$ is
\begin{equation}
    f_{\textrm{small-angle}}(\theta, \delta) \rightarrow f_{\textrm{small-angle}}(\theta, 2\delta).
\end{equation}
While this is simple, in the regime $\delta \ll \theta \ll 1$ and $\delta \gtrsim \theta^2$ it matches the data very well. For the fallback scheme we do not derive a staircase of solutions, except for the first two solutions corresponding to $S^{\dagger}$ and $T^{\dagger}$; these can be viewed as special cases of fallback channels where the success probability is exactly $1$. A formula for both angle-independent and angle-dependent regimes is given in \cref{sec:summary_fallback}, and including the $S^{\dagger}$ and $T^{\dagger}$ solutions, is plotted as ``Our formula'' in \Cref{fig:av_t_count_per_rot_fallback}.

\section{Application to Trotterized quantum dynamics}
\label{sec:trotter}

In this section we use our derived formulas to perform an analysis of the T gate cost of Trotterization for example chemistry problems. Trotterization has been widely studied in fault-tolerant resource estimation papers \cite{reiher2017, Kivlichan2020, Campbell2022, blunt_2022, Bay-Smidt2025}. Trotter circuits are particularly simple; in their most basic form, they consist of multi-qubit Pauli rotations, which can be compiled to single-qubit $R_z$ rotations plus Clifford gates. As such, the non-Clifford cost of Trotterization is usually dominated by the cost of compiling rotation gates to the Clifford+T gate set. We note that alternative schemes such as Hamming weight phasing (HWP) have been proposed to reduce this cost for circuits with a particular structure, for example in the Hubbard model \cite{Kivlichan2020, Campbell2022, Bay-Smidt2025}.

Prior resource estimation studies have used well-known angle-independent formulas, for example the  mixed-diagonal and mixed-fallback results discussed in this paper (although some alternative methods have been used, such as TE-PAI \cite{Kiumi2025}; see the discussion in \Cref{app:comparison trotter results}). However, particularly for \emph{ab initio} chemistry Hamiltonians, most of the rotation gates in the Trotter circuit have a very small rotation angle. This suggests that significant savings can be made using the results in this paper. We investigate these savings using our derived formulas in this section.

Consider a Hamiltonian in a qubit representation,
\begin{equation}
    H = \sum_i^L a_i P_i,
\end{equation}
where $a_i$ are real numbers and $P_i$ are multi-qubit Pauli operators. In this paper we use the Jordan-Wigner transformation to map fermionic Hamiltonians to the qubit representation.

We consider the simplest first-order product formula
\begin{equation}
    S_1(t) = \prod_i^L e^{-i a_i P_i t },
\end{equation}
to approximate time evolution $U(t) = e^{-i H t}$. For any Pauli $P_i$, we can find a Clifford $C_i$ such that
\begin{equation}
    P_i = C_i  \, (Z \otimes \mathbb{1} \otimes \ldots \otimes \mathbb{1}) \, C_i^{\dagger},
\end{equation}
i.e., conjugation with $C_i$ maps $P_i$ to a single-qubit $Z$ Pauli. Therefore,
\begin{equation}
    S_1(t) = \prod_i^L C_i e^{-i a_i t Z \otimes \mathbb{1} \otimes \ldots \otimes \mathbb{1}} C_i^{\dagger}.
\end{equation}

Evolution to total time $t_{\mathrm{total}}$ can be perform in $r$ Trotter steps of size $t = t_{\mathrm{total}} / r$,
\begin{align}
    U(t_{\mathrm{total}}) \approx S_1(t)^r &= \Big( \prod_i^L C_i e^{-i a_i t Z \otimes \mathbb{1} \otimes \ldots \otimes \mathbb{1}} C_i^{\dagger} \Big)^r, \\
    &= \Big( \prod_i^L C_i R_z(\theta_i) C_i^{\dagger} ,\Big)^r,
    \label{eq:trotter decomp}
\end{align}
with $\theta_i = -a_i t_{\mathrm{total}} / r = -a_i t$. Each Trotter step consists of $L$ single-qubit rotations with angles $ \{ \theta_i \}$, and in the full circuit each rotation is repeated $r$ times. Therefore, after constructing the qubit representation of $H$, it is simple to calculate the average number of T gates to implement the full circuit with both the angle-dependent or angle-independent formulas. Higher-order product formulas have a similar form - they consist of a product of single-qubit rotations conjugated by Cliffords - and so results would be similar, but we only consider the first-order formula in the following.

In this paper we focus on T count. However, the total Clifford+T gate count is at most a constant factor higher: first of all, for each unitary sampled in place of a rotation, the Matsumoto-Amano normal form \cref{eq:matsumoto normal form} shows that its Clifford count is at most a constant factor higher than its T count, with the exception of 0 T count unitaries. In addition to the sampled unitary, each factor in the Trotter step \cref{eq:trotter decomp} contains conjugating Cliffords $C_i$ (and $C_i^\dagger$). We can absorb these into the final Clifford \eqref{eq:matsumoto normal form} of the preceding (and current) unitary. For 0 T count unitaries, such as the identity when it is sampled as the under-rotation, the entire factor in the Trotter step becomes Clifford and can be absorbed into the preceding unitary. Note that rotations in the small-angle regime will be mostly sampled as the identity, where the corresponding $C_i$ and $C_i^\dagger$ Cliffords cancel and can be removed.

\subsection{Performing the costing by assigning \texorpdfstring{$δ$}{δ} for each rotation}
\label{sec:assigning_delta}

In order use the formula $T(\theta_i,\delta_i)$ to cost each rotation of angle $θ_i$ in the circuit, we must choose a value of $δ_i$ for each rotation. All $N_\text{rot} = rL$ rotations combined give a total sampling overhead (in the case of using quasi-probabilities) of
\begin{equation}
    λ_\text{total} = \prod_{i=1}^{N_\text{rot}}(1+δ_i),
    \label{eq:trotter lambda combined}
\end{equation}
or a total diamond norm error (in the case of using proper probabilities)
\begin{equation}
    ε_\text{total} = \sum_{i=1}^{N_\text{rot}} δ_i.
    \label{eq:trotter eps combined}
\end{equation}
In either case, in order to allocate $\{δ_i\}$, we will fix
\begin{equation}
    δ_{\text{total}} := \sum_{i=1}^{N_\text{rot}}δ_i.
    \label{eq:delta total}
\end{equation}
When costing using the angle-independent formula \cref{eq:mixed_diagonal_theta_independent}, the optimal choice of $\{δ_i\}$ is all equal,
\begin{equation}
    δ_i = \frac{δ_\text{total}}{N_\text{rot}}.
\end{equation}
When costing using our formula \cref{eq:min_of_small_and_worst} that includes improvements for small angles, we use the following heuristic to choose $\{δ_i\}$. We sketch the shape of $T(θ,δ)$ very roughly with a cutoff for angles above $θ_{\max}$, with an ensuing polynomial decay below this cutoff:
\begin{equation}
    \tilde T(θ,δ) \sim 2\frac{\min(θ^2,θ_{\text{max}}^2)}{δ}.
    \label{eq:heuristic tilde T}
\end{equation}
The two regions represent where the angle-independent formula gives an advantage in \cref{eq:min_of_small_and_worst}, and  the lead behavior in the $θ^2\llδ\llθ$ regime, see \cref{eq:asymptotic delta>>theta^2}. This rough shape can also be seen in \Cref{fig:av_t_count_per_rot}. The parameter $θ_\text{max}$ can be optimized.  We minimize the objective function $\sum_{i=0}^{N_\text{rot}} \tilde T(θ_i,δ_i)$ subject to the constraint \cref{eq:delta total} using a Lagrange multiplier. The Lagrangian
\begin{equation}
    \mathcal{L}(\{θ_i\},\{δ_i\}) = \sum_{i=1}^{N_\text{rot}}\min(θ_i^2,θ_{\max}^2)/δ_j - γ\left(δ_\text{total}-\sum_{i=1}^{N_\text{rot}}δ_i\right)
\end{equation}
leads to the first order condition for each $δ_j$
\begin{equation}
    \frac{∂\mathcal{L}}{∂δ_i} = 0\  \Rightarrow\ δ_j = \frac{\min(|θ_j|, θ_\text{max})}{\sqrt{γ}}.
\end{equation}
Therefore, to respect the sum constraint, the optimal allocation is
\begin{equation}
    δ_i = δ_\text{total}\frac{\min(|θ_i|,θ_\text{max})}{\sum_{k=1}^{N_\text{rot}}\min(|θ_k|,θ_\text{max})}.
    \label{eq:delta allocation}
\end{equation}
We emphasize that we only use the heuristic $\tilde T(θ,δ)$ from \cref{eq:heuristic tilde T} to assign $\{δ_i\}$ as in the above formula, but use our full formula for costing. Further optimization of $\{δ_i\}$ and reduction in T count is possible, which would further improve results from the angle-dependent formula compared to the angle-independent setting.

For our costings, we calculate the total average T gate count as
\begin{equation}
    T_{\textrm{full circuit}} = r \sum_{i=1}^{L} T(θ_i, δ_i),
\end{equation}
and only taking the sum over all angles within one Trotter step, as each Trotter step is the same.
For probability mixtures, we choose a typical total error of
\begin{equation}
    ε_{\text{total}} = rε_\text{trotter} = 10^{-3}
\end{equation}
and use $δ_\text{total} = ε_\text{total}$ as suggested by Eqs.~\eqref{eq:trotter eps combined} and \eqref{eq:delta total}.
For quasi-probability mixtures, we set $δ_\text{total}=1$. This results in the total sampling overhead
\begin{align}
    λ_\text{total} = \prod_{i=1}^{L}(1+δ_i)^r \to \prod_{i=1}^L e^{rδ_i} = e^{δ_\text{total}} = e
    \label{eq:e_sampling_overhead}
\end{align}
in the limit of a large number of Trotter steps, $r\to\infty$. For a finite number of Trotter steps, $λ_\text{total}$ is bounded by $e$. To take the limit factor by factor, note that $rδ_i$ is a small number depending on the precise angles and on $L$, but not on $r$.  This means that our quasi-probability results have a sample complexity overhead of only $e^2 \approx 7.39$, corresponding to less than 10 times more samples required. As we will see in the section discussing the results (\Cref{sec:numerical_results_ancilla_free}), quasi-probabilities enable a T count up to $1000$ times lower than proper probabilities, redeeming the higher sample complexity. \Cref{app:prob_vs_quasi_prob} also further elucidates more generally how larger $λ_i - 1$ than $ε_i$ can be used, leading to favorable quasi-probability T count.

It is interesting to note that in the small angle regime, with $|θ_i| = |a_i| t < θ_{\max}$, our allocation \cref{eq:delta allocation} becomes $δ_i = δ_\text{total}|θ_i|/\sum_k |θ_k|$, such that the rough total T count is
\begin{align}
    \sum_i\tilde T(\{θ_i\},\{δ_i\}) &= r\sum_{i=1}^{L} 2(|a_i|t)^2\frac{r\sum_{k=1}^{L} |a_k|t}{δ_\text{total}|a_i|t}
    = 2(rt)^2\left(\sum_{i=1}^L |a_i|\right)^2 = 2t_\text{total}^2 \left(\sum_{i=1}^L |a_i|\right)^2.
    \label{eq:qdrifty}
\end{align}
The total T count is constant, regardless of size of the steps, which we will also see in our T cost results beyond this simple heuristic $\tilde T$ for small step sizes. This result and scaling is reminiscent of the scaling of qDRIFT and other randomized methods \cite{Campbell2019}, see also \Cref{app:comparison trotter results}.

\subsection{Definition of chemistry Hamiltonians}
\label{sec:chemistry_hamiltonians}

For our results we consider two example molecules: (i) a pentacene monomer in a (22e,22o) active space consisting of valence $\pi$ orbitals; and Fe(II)-Porphyrin (Fe(P)) in a (32e,29o) active space, consisting of 20 C $2p_z$, 4 N $2p_z$ and 5 Fe $3d$ orbitals. We primarily use localized orbitals, but in the case of pentacene we present results with both localized and also canonical (restricted Hartree--Fock) orbitals, which are delocalized. The geometries, active spaces and orbitals correspond to those from Ref.~\cite{Blunt2021}. For Fe(P), orbitals were optimized by performing CASSCF before localizing the orbitals, following the approach of Smith \emph{et al.} \cite{smith2017, smith_github}; the active space for Fe(P) was first suggested in Ref.~\cite{LiManni2016}. The PySCF package was used to set up chemical Hamiltonians \cite{pyscf, pyscf_2}.

After generating the chemical Hamiltonian in the appropriate orbital basis, we map it to a qubit representation using the Jordan-Wigner transformation as implemented in OpenFermion \cite{openfermion}. To ensure a fairer comparison between angle-dependent and angle-independent results, we then truncate any Hamiltonian coefficients, $a_i$, which are smaller in magnitude than $10^{-7}$ Ha. This ensures a fairer comparison since very small terms will typically be removed in a Trotter resource analysis; below a threshold one expects the loss in accuracy by discarding such terms to be negligible. We do not attempt to obtain an optimal choice of the cutoff, which is in general a computationally intractable problem, but we choose $10^{-7}$ Ha as a typical value used in resource estimation studies.

\subsection{Results}
\label{sec:trotter_results}

\begin{figure}
    \centering
    \includegraphics[width=1.0\linewidth]{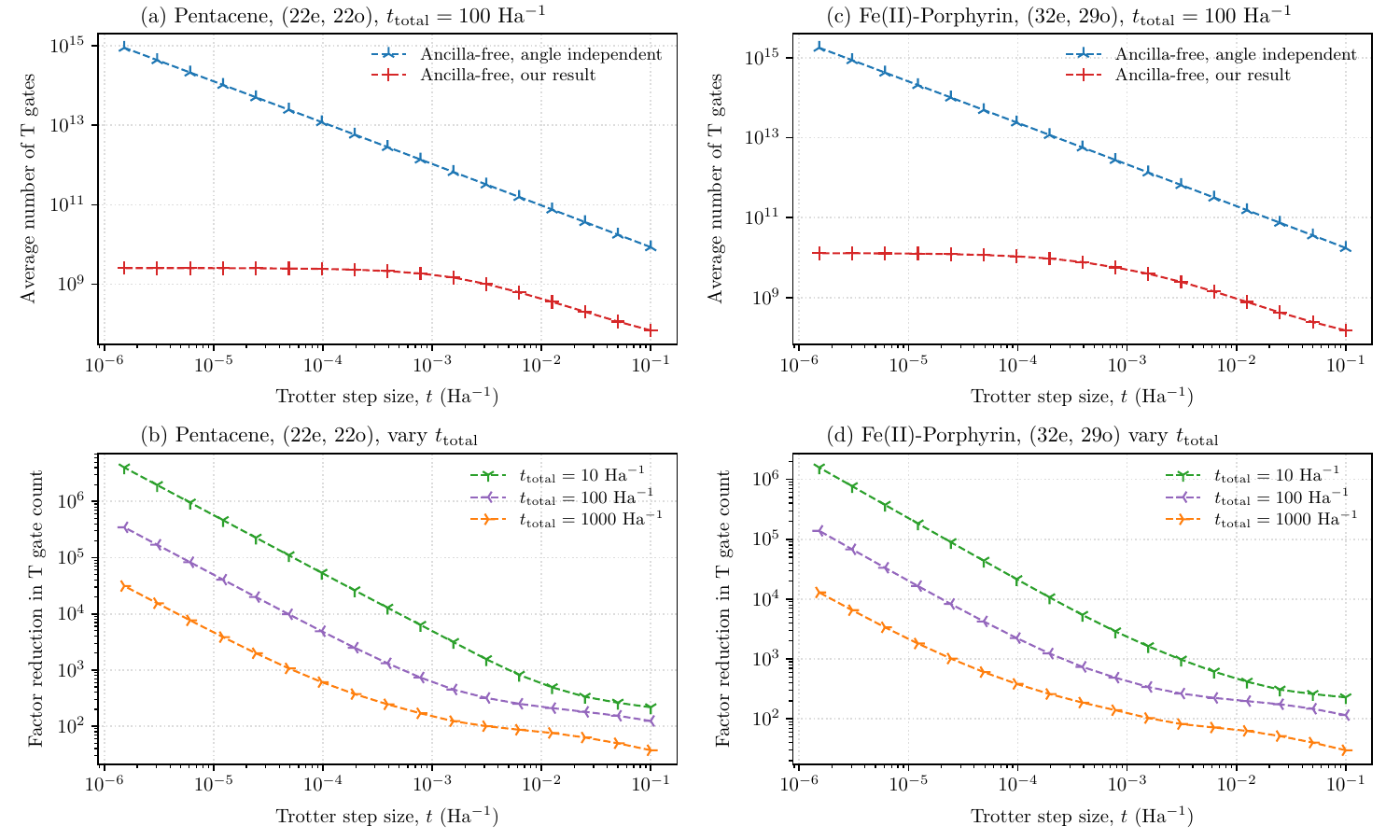}
    \caption{Results for the two systems studied: pentacene in (a) and (b); and Fe(II)-Porphyrin in (c) and (d). Localized orbitals were used in each case. Here, we perform costing with $\delta_{\mathrm{total}} = 1$, which is appropriate when using quasi-probabilities and leads to a sampling overhead of $\lambda_{\text{total}} \approx e^2$; see the discussion in Section~\ref{sec:assigning_delta}. Results in this plot use the ancilla-free scheme. Results in the top row fix the total evolution time to $t_{\mathrm{total}} = 100$ Ha$^{-1}$ and show the average number of T gates, while results in the bottom row vary the total evolution time and show the factor reduction in average T gate count from the angle-dependent formula compared to the angle-independent formula. In the limit of small Trotter step sizes, the T gate count using the angle-dependent formula converges to a constant.}
    \label{fig:quasi_prob_comparison}
\end{figure}

We first investigate results in the quasi-probability case, see \Cref{fig:quasi_prob_comparison}. Here, we take $\delta_{\mathrm{total}} = 1$; as shown in \cref{eq:e_sampling_overhead}, using quasi-probabilities, this leads to a sampling overhead for the full Trotter circuit of $\lambda_{\text{total}}^2 \approx e^2$, which is less than a factor of ten times. Based on approximate optimization, we took $\theta_{\text{max}} = 10^{-4}$ in \cref{eq:delta allocation}. For subplots in the top row we fix the total evolution time to $t_{\mathrm{total}} = 100$ Ha$^{-1}$ and vary the Trotter step size, $t$. We then plot the average T gate count using both the previous (angle-independent) formula \eqref{eq:mixed_diagonal_theta_independent} and newly-derived angle-dependent formula as implemented in \cref{app:costing script}. We use the ancilla-free formula, rather than the fallback result. For each of the three systems, we see that the average T gate cost is significantly reduced by using the angle-dependent formula even at large $t$; meanwhile in the limit of small $t$ the cost tends to a constant. At large $t$, only some rotation angles are sufficiently small to enjoy the small-angle savings, while at small $t$, all of them are in the small-angle regime $θ_i^2 \ll δ_i$. In contrast, the cost using the angle-independent varies as $\tilde{\mathcal{O}}(t^{-1})$. At $t=0.1$ Ha$^{-1}$ (which is a large step size) the average T gate cost is reduced by roughly two orders of magnitude.

In the bottom row we plot the factor reduction in T gate count for three different total evolution times, $t_{\mathrm{total}} = 10$, $100$ and $1000$ Ha$^{-1}$. This reduction factor is defined as T gate count using the angle-independent formula divided by the T gate count using the angle-dependent formula. In the small $t$ limit, the reduction factor goes as $\tilde{\mathcal{O}}(t_{\mathrm{total}}^{-1})$. This is because, in the small $t$ limit, the average T gate cost using the angle-dependent formula scales as $\tilde{\mathcal{O}}(t_{\mathrm{total}}^2)$, while the same quantity using the angle-independent formula scales as $\tilde{\mathcal{O}}(t_{\mathrm{total}})$; the former is similar behavior to that seen in the qDRIFT method, for example. However, we emphasize that the T gate cost using the angle-dependent formula can never exceed that obtained by the previous angle-independent formula. Overall, the results in \Cref{fig:quasi_prob_comparison} show a transformative reduction in the T gate cost to implement Trotter circuits; this reduction can be multiple orders of magnitude, depending on $t$ and $t_{\mathrm{total}}$. Another benefit is that the dependence of the T gate cost on the step size, $t$, is significantly reduced. This is valuable because it lessens the importance of performing a tight Trotter error analysis. Such a tight Trotter error analysis is notoriously difficult, and has been the subject of significant research \cite{childs2021theory, Su2021, Zhao2022, Yi2022, Mizuta2025, Blunt2025, baysmidt2026}.

\begin{figure}
    \centering
    \includegraphics[width=1.0\linewidth]{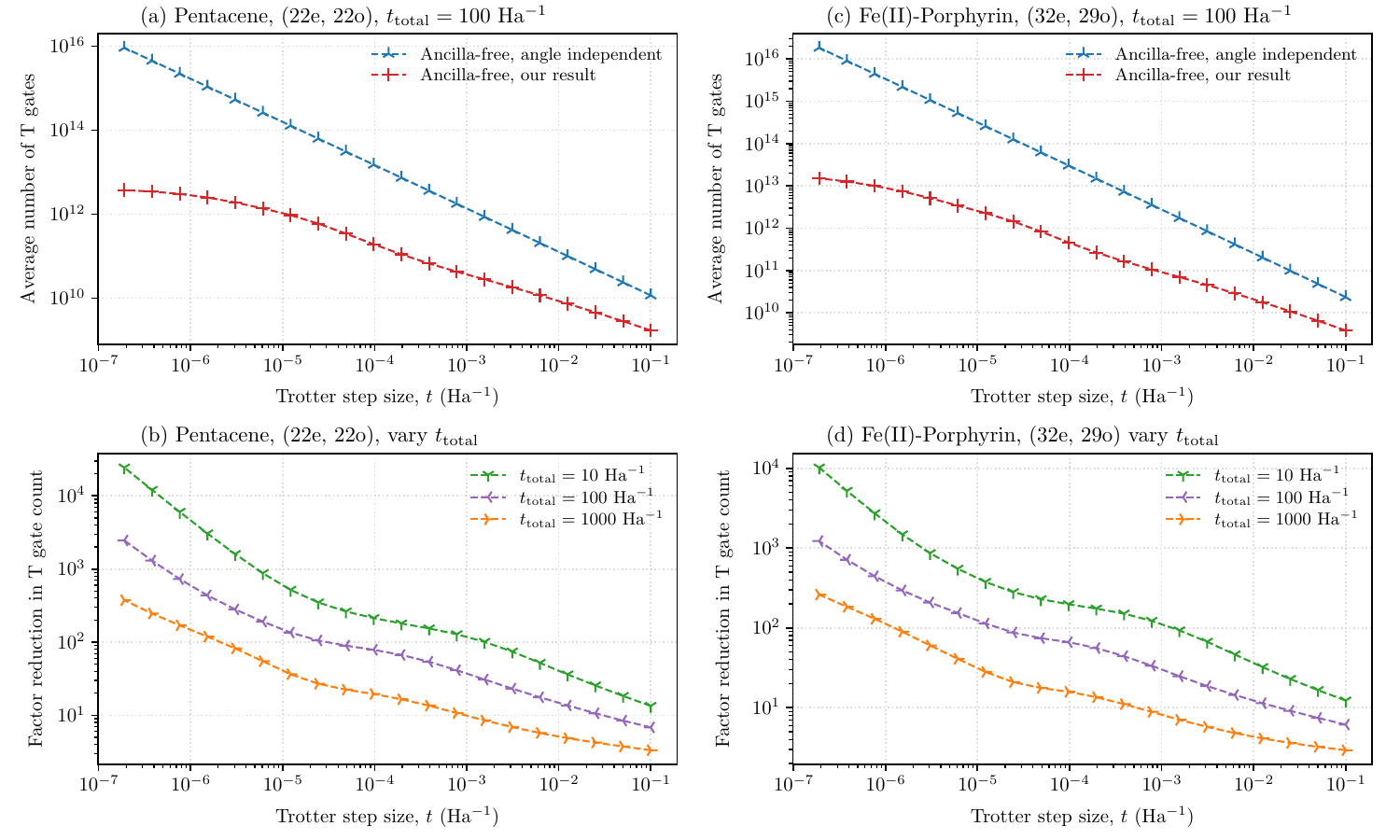}
    \caption{Results for the two systems studied: pentacene in (a) and (b); and Fe(II)-Porphyrin in (c) and (d). Localized orbitals were used in each case. Here, we perform costing with $\delta_{\mathrm{total}} = 10^{-3}$, which is appropriate when using mixtures over strict probabilities, as in the mixed diagonal approximation \cite{Kliunchnikov_shorter2023}, to achieve $\varepsilon_{\text{total}} = 10^{-3}$; see the discussion in Section~\ref{sec:assigning_delta}. Results in this plot use the ancilla-free scheme. Results in the top row fix the total evolution time to $t_{\mathrm{total}} = 100$ Ha$^{-1}$ and show the average number of T gates, while results in the bottom row vary the total evolution time and show the factor reduction in average T gate count from the angle-dependent formula. In the limit of small Trotter step sizes, the T gate count of the angle-dependent formula converges to a constant, although convergence is not fully reached for the range of step sizes here.}
    \label{fig:prob_comparison}
\end{figure}

Next, we consider the same results but using $\delta_{\text{total}} = 10^{-3}$. This would be appropriate when using probability mixtures via the mixed diagonal approximation, where we want the total diamond norm error across the full circuit to be $\epstotal = 10^{-3}$, taking a typical small value. Based on approximate optimization, we took $\theta_{\text{max}} = 10^{-6}$ in \cref{eq:delta allocation}. Results are shown in \Cref{fig:prob_comparison}. Note that we plot step sizes down to $t = 10^{-7}$ Ha$^{-1}$, which is lower than in the \Cref{fig:quasi_prob_comparison}. We see that results from the angle-independent formula are roughly the same in both Figures~\ref{fig:quasi_prob_comparison} and \ref{fig:prob_comparison}. This is expected since this formula has only logarithmic dependence on $\delta_{\text{total}}$. In contrast, results using the angle-dependent formula are quite different between the two. In particular, the advantage of the angle-dependent formula over the angle-independent formula is lessened. This is also expected because each rotation must be implemented with a much smaller value of $\delta_i$. As such, many more rotations are in the ``angle-independent'' regime with higher cost, i.e. the regime where $δ_i \lesssim \theta_i^2$. Still, we see significant reductions in the cost of Trotterization using the angle-dependent formula. This is highly-dependent on the values of $t$ and $t_{\mathrm{total}}$ (and the Hamiltonian), but reductions by orders of magnitudes are common. Moreover, the T gate cost of Trotterization becomes independent of $t$ in the limit of small $t$, even in the case of probability mixtures.

Let us further compare the use of quasi-probabilities with $\delta_{\text{total}} = 1$ (corresponding to unbiased estimators but with a sample overhead $\lambda_{\text{total}}^2 \approx e^2$), and proper probabilities with $\epstotal = 10^{-3}$ (no sampling overhead but with a systematic bias quantified by the diamond norm error). Comparing \cref{fig:quasi_prob_comparison}(a) and \cref{fig:prob_comparison}(a) in the large step size case of $t = 0.1$ Ha$^{-1}$, the reduction in T gate count is $\approx 16$ times. In the small-$t$ limit, the reduction in the T gate count by using quasi-probabilities is over $1000$ times. These improvements are despite only having $\lambda_{\text{total}}^2 \approx 7.38$. Therefore, we see that, even accounting for increased sampling overhead, quasi-probabilities can significantly reduce the cost of fault-tolerant Trotterization, in addition to providing unbiased estimators.

Results above demonstrate that there can be significant savings by reducing the total evolution time, $t_{\mathrm{total}}$, and by using the quasi-probability approach. These results suggest that there may be benefits in performing Trotterized phase estimation by statistical phase estimation techniques \cite{Lin2022, Wan2022, Dutkiewicz2022, Wang2023, Blunt2023, Ding2023}. These can significantly reduce the total evolution time required to reach a given precision \cite{Wang2023, Blunt2023, Ding2023}, and also work by estimating expectation values, and are therefore readily applicable to approaches based on quasi-probability distributions.

\begin{figure}
    \centering
    \includegraphics[width=0.6\linewidth]{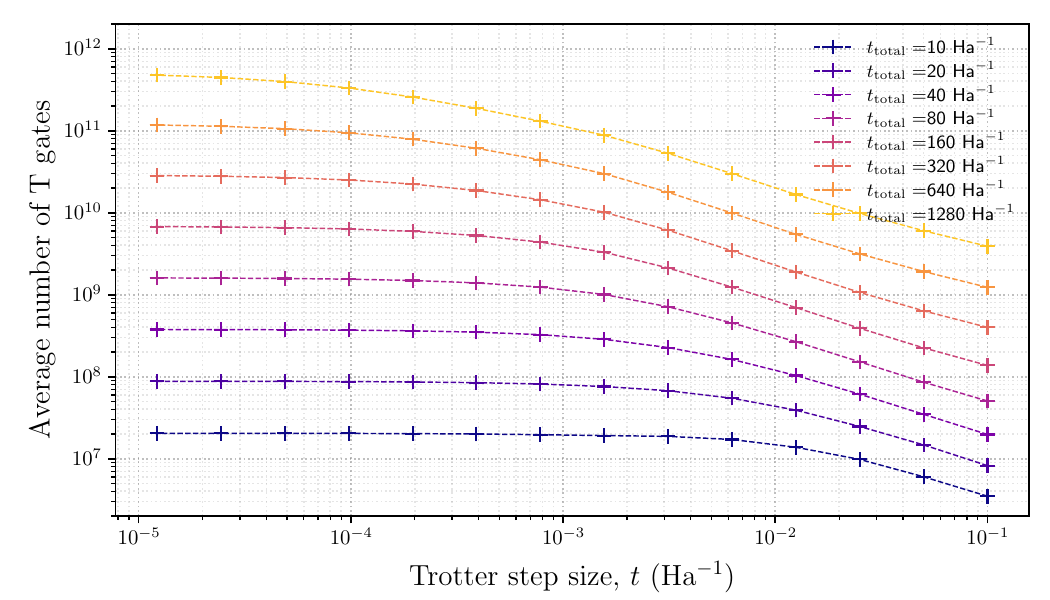}
    \caption{Average T gate count to perform Trotterization as a function of step size, $t$, using the quasi-probability method, for pentacene in a localized orbital basis set. The total evolution time, $t_{\mathrm{total}} = rt$, is varied. These results use the angle-dependent formula for the ancilla-free scheme, as implemented in \Cref{app:costing script}.}
    \label{fig:pentacene_vary_T}
\end{figure}

For the remaining results we return to the using $\delta_{\text{total}}=1$, as appropriate for the quasi-probability approach, and consider the pentacene (22e,22o) example. In \Cref{fig:pentacene_vary_T} we show the average T gate count against Trotter step size $t$, using the angle-dependent formula and for several different values of the total evolution time, $t_{\mathrm{total}}$. In the limit of small $t$, we see that varying the total evolution time $t_{\mathrm{total}}$ by roughly two orders of magnitude varies the the average T gate count by roughly four orders of magnitude. This confirms our earlier claim that in the small $t$ limit the average T gate cost goes as $\mathcal{O}(t_{\mathrm{total}}^2)$, which is similar to qDRIFT. We also note that as $t_{\mathrm{total}}$ increases, smaller step sizes $t$ are needed to reach this regime where the T gate count is independent of $t$. Still, in practice we would expect to be working with larger values of $t$, and so these small-$t$ regimes might be less relevant.

\begin{figure}
    \centering
    \includegraphics[width=1.0\linewidth]{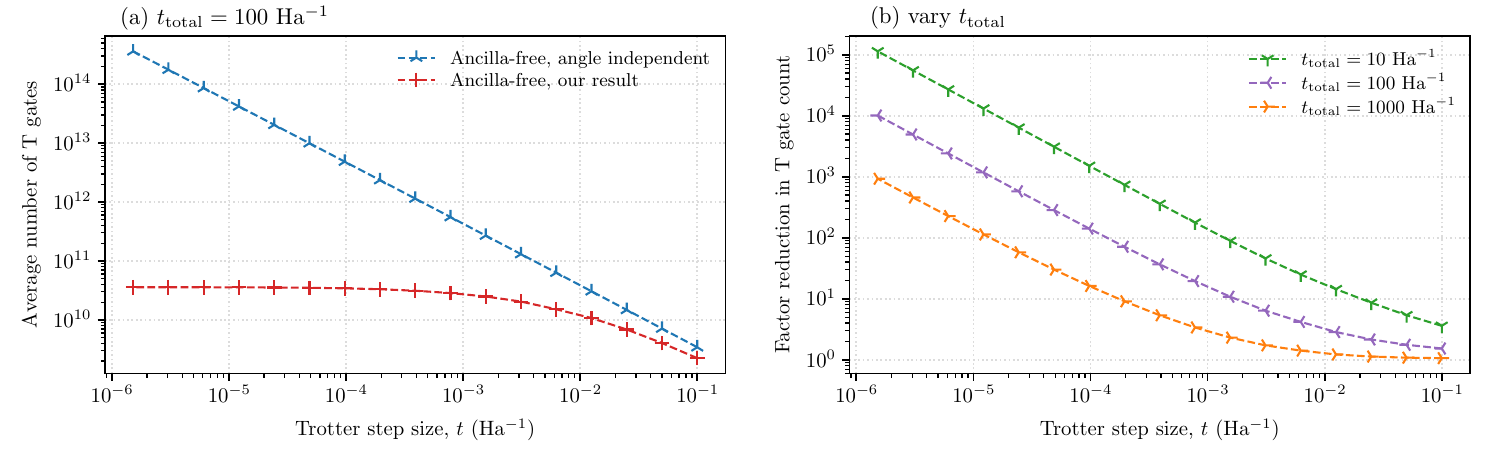}
    \caption{Results using the quasi-probability method for pentacene in a canonical (delocalized) orbital basis set, and varying the maximum evolution time. This can be compared to \Cref{fig:quasi_prob_comparison} (a) and (b), which shows equivalent results for a localized orbital basis for the same system. (a) Average T gate cost as a function of step size, $t$, and with maximum evolution time $t_{\mathrm{total}} = 100$ Ha$^{-1}$, comparing the previous angle-independent formula (blue) to our angle-dependent formula (red). (b) Factor reduction in average T gate count for $t_{\mathrm{total}} = 10$, $100$ and $1000$ Ha$^{-1}$.}
    \label{fig:pentacene_canonical}
\end{figure}

In \Cref{fig:pentacene_canonical} we show results for pentacene but using a canonical orbital basis set, obtained by performing a restricted Hartree--Fock calculation. These orbitals are delocalized across the molecule. The results are otherwise equivalent to \Cref{fig:quasi_prob_comparison} (a) and (b), which considered a localized orbital basis set, i.e. where each orbital is localized on a single atom. When working with localized orbitals, the integrals defining Hamiltonian elements will decay with distance (exponentially quickly in some cases), such that there will be many small Hamiltonian elements and hence small-angle rotations to implement. This contrasts to using canonical orbitals, where we expect more Hamiltonian elements to be large. We might then expect the angle-dependent formula to give a smaller improvement when using canonical orbitals compared to localized orbitals. Comparing \Cref{fig:quasi_prob_comparison}(a) and (b) to \Cref{fig:pentacene_canonical}(a) and (b), this is indeed what is seen. The factor reduction in T gate count from the angle-dependent formula is smaller when using canonical orbitals. Moreover, by comparing the subplots (a), we see that using localized orbitals gives the lowest T gate counts overall. We therefore believe that using localized orbitals is sensible to reduce the cost of Trotterization, although details may very depending on the precise system under study.

\section{Conclusion}
\label{sec:conclusion}

In this paper we developed a quasi-probability approach to Clifford+T rotation synthesis and performed a study of Clifford+T compilation of rotation gates in the small-angle regime. It is widely claimed that the cost of Clifford+T rotation synthesis is largely independent of the rotation angle; our results show that this claim is incorrect. While existing angle-independent formulas are accurate for $\delta \lesssim \theta^2$, they can substantially over-estimate the average T gate count in the small-angle regime, where $\delta \gtrsim \theta^2$. We obtained numerical data to demonstrate this, and derived formulas which accurately match this data. These formulas can be used for resource estimation of fault-tolerant quantum algorithms, and as an example we studied Trotterized dynamics of chemistry Hamiltonians. Our results show that properly capturing the small-angle cost of Clifford+T synthesis can have a transformative effect on the cost of fault-tolerant quantum simulation by Trotter product formulas.

In addition to studying the small-angle regime of existing mixed schemes for Clifford+T rotation synthesis, we developed a quasi-probability approach to solve the same problem. In the probabilistic approach, Clifford+T compilation is performed approximately with diamond norm error $\epsdiamond$. In the quasi-probability approach that we presented in this paper, the linear combination of channels is exact but has a sampling overhead $\lambda^2$ to estimate expectation values of observables. We compared these two approaches, demonstrating numerically and analytically the correspondence in T cost when $\epsdiamond \approx \lambda-1$. Despite this correspondence, we emphasize that both approaches have important advantages: the quasi-probability approach can significantly reduce resource requirements (even when accounting for additional sampling overhead) to achieve a given total accuracy, however it is primarily applicable to scenarios where we seek to calculate the expectation value of an observable. In contrast, the scheme based on strict probabilities can be more expensive but can be applied to any quantum circuit. Separately, note that in this paper we considered quasi-probability methods for correcting compilation errors, however such methods can also be used to correct gate and decoherence errors \cite{Temme2017} and in combination with fault-tolerant compilation \cite{Camilo2026}.

We showed that the gate cost of Trotter circuits compiled to the Clifford+T gate set becomes independent of the Trotter step size, $t$, in the limit of small $t$, simply by virtue of mixed Clifford+T compilation as in our paper. Moreover, compared to existing angle-independent formulas, we often see significant reductions in the T gate cost of Trotterization even for large Trotter step sizes. This reduction is significant enough that we believe the cost of fault-tolerant Trotterization should be reassessed for a variety of applications. We note that alternative quantum simulation approaches have been proposed that take advantage of savings for small Hamiltonian terms, for example qDRIFT \cite{Campbell2019}, partially-randomized methods \cite{Gunther2025} and TE-PAI \cite{Kiumi2025}, as discussed in \Cref{app:comparison trotter results}. The approach presented in this paper has the advantage that it obtains the small-angle saving, while also returning to state-of-the-art Clifford+T results, specifically the mixed diagonal and mixed fallback approximations \cite{Kliunchnikov_shorter2023}, in the worst case. We emphasize that these savings are obtained automatically by simply performing the Clifford+T rotation synthesis of deterministic product formulas by a mixed approximation. Moreover, while we focused on the application to Trotter circuits in this paper, the presented Clifford+T compilation results can be applied to fault-tolerant quantum algorithms generally. We expect that the presented results can also simplify practical compilation of circuits to the Clifford+T gate set. In particular, in the small-angle regime, the under-rotation is chosen as the identity while the over-rotation can be selected from a lookup table; we presented such a table up to $35$ T gates in \Cref{app:matsumoto_amano}. Solutions can also be obtained generally from existing approaches based on the Ross-Selinger algorithm \cite{ross2016optimal}.

It will be particularly interesting to apply these results to perform resource estimation for Trotterized quantum phase estimation, which has been considered expensive compared to modern phase estimation strategies \cite{reiher2017, lee2021, blunt_2022}. We showed that the improvement over existing resource estimation schemes is amplified for shorter total evolution times. Statistical versions of phase estimation \cite{Lin2022, Wan2022, Dutkiewicz2022, Wang2023, Blunt2023, Ding2023} can allow much shorter evolution times to achieve a given accuracy, and can also be readily combined with the quasi-probability version of our results. Combined with improvements in Trotter error analysis \cite{childs2021theory, Zhao2022, Yi2022, Mizuta2025, Blunt2025, baysmidt2026}, we expect this to lead to significant improvements for this task.

Lastly, we anticipate that these results will be valuable for early error-corrected experiments. The quasi-probability approach in particular allows the possibility to reduce the Clifford+T depth in exchange for increased sampling overhead, and the reduction in T gate counts generally will reduce the burden on methods like magic state cultivation \cite{gidney2024magic}. The reduction in average T gate counts also extends the applications reachable by magic state cultivation without further distillation, which may have important consequences for fault-tolerant architectures. Taken together, we hope these insights lead to interesting new directions for quantum simulation on fault-tolerant quantum computers.

\section{Data Availability}

The formula used for resource estimation, as plotted in \cref{fig:av_t_count_per_rot} and used in Trotterization results, is included as a Python script in \cref{app:costing script} and can be downloaded at \href{https://arxiv.org/src/2605.31544/anc/small_angle_costing.py}{the following link}. Data for over-rotations from enumeration up to 35 T gates are provided in Tables~\ref{tab:staircase app} and \ref{tab:staircase gate sequences app} in \Cref{app:matsumoto_amano}.

\acknowledgments

M.B. is a Sustaining Innovation Postdoctoral Research Associate at Astex Pharmaceuticals and thanks Astex Pharmaceuticals for funding, as well as his Astex colleague Patrick Schoepf for his support.

\appendix
\crefalias{section}{appendix}

\section{Script for resource estimation}
\label{app:costing script}
In this appendix, we provide a simple Python script to run our costing formula for the ancilla-free synthesis method. The script is also downloadable on this paper's arXiv page under ``Ancillary files,'' \href{https://arxiv.org/src/2605.31544/anc/small_angle_costing.py}{access script with this direct link}.

\textbf{Inputs:} target rotation angle $θ$, target $δ$ (where $δ=λ-1$ or $δ=\epsdiamond$)

\textbf{Outputs:} average T gate count, actual $δ$ (might be lower than the target value)

Our analysis in \Cref{sec:small angles} is valid for $θ\in(0,\pi/4)$.
First, the Python script normalizes $θ$ to the range $θ\in[0,\pi/8)$ by using pure Clifford operations. Multiplication by $S$ allows to rotate by angle $-π/4$, while conjugating with $X$ allows sign reversal of the target angle. We map to the range $θ\in[0,\pi/8)$ rather than $θ\in[0,\pi/4)$, since the Clifford mapping $θ=\pi/4-η\toθ=-η \toθ=η$ will result in a smaller angle with lower cost (for small $η>0$).
The script then attempts to find a suitable over-rotation in the staircase \Cref{tab:staircase app}, which has been precomputed exactly for $θ\lesssim 500δ$, see \Cref{app:matsumoto_amano}. For this, it computes $\tanα$ using the second formula in \cref{eq:tanalpha from lambda} for improved numerical stability. Beyond the staircase, it uses our asymptotic formula \cref{eq:asymptotic}, valid for $δ \ll θ \ll 1$, see \Cref{sec:asymptotic}.
Finally, the script compares with the result from the angle-independent formula and returns the lower T count.

Beyond this costing script, further improvements could be considered. If $θ\in[\pi/16,π/8]$, especially when $θ$ is close to $π/8$, we believe that it is more efficient to use a $T$ gate as a fixed over-rotation, and find a suitable under-rotation (rather than fixing the under-rotation to the identity and finding a suitable over-rotation). Our analysis can be adapted to that setting, with the average T count being $|p|\cdot\text{T-count}+|p_T|$ to account for the $T$ under-rotation occurring with quasi-probability $p_T$. Alternatively, $θ\in[π/16,π/8]$ could be rotated into $θ\in[0,π/16]$ by rotation by a T gate. Our analysis and asymptotic formula can then be used directly, where the one extra T gate must be added to total cost. The cost is similar to the method choosing the over-rotation as $T$ since typically $|p_T|$ is close to 1.

\lstdefinestyle{appendixpython}{
    language=Python,
    basicstyle=\ttfamily\scriptsize,
    keywordstyle=\bfseries,
    commentstyle=\itshape\color{gray},
    stringstyle=\color{black},
    numbers=none,
    numberstyle=\tiny\color{gray},
    numbersep=6pt,
    breaklines=true,
    showstringspaces=false,
    columns=fullflexible,
    keepspaces=true,
    xleftmargin=2em,
    frame=leftline,
    framerule=0.5pt,
    rulecolor=\color{gray}
}

\begin{lstlisting}[style=appendixpython,language=Python]
import bisect
from typing import Tuple

import numpy as np

staircase_tanalpha = [
    1.000000000000000000, 0.414213562373095090, 0.350834874673672581, 0.312876580442609076, 0.212012989773690319, 
    0.199053696635809713, 0.137553374901583453, 0.114743436552311937, 0.110968924780318418, 0.106108155871760396, 
    0.095599761041589515, 0.093857467246492854, 0.089772198956584240, 0.081007642852468559, 0.080059368825502769, 
    0.071823156445232433, 0.068074811618681924, 0.063557891353411389, 0.056565932310567113, 0.054687766545393701, 
    0.043853474353372793, 0.043184230632607387, 0.041757558494795996, 0.035525578867165862, 0.026620221579097662, 
    0.023740068332829375, 0.023219765631210493, 0.023041225970146049, 0.021820124509931472, 0.021584661590176659, 
    0.021095252610628751, 0.020170619047074331, 0.016419959844448571, 0.013250840260523361, 0.013201874372092005, 
    0.011678064337090286, 0.011610273857837769, 0.010123357001352220, 0.009178937379670129, 0.008939807749050285, 
    0.008911641286055169, 0.008891831473781506, 0.008223547022012874, 0.008134840314959944, 0.005676564243201448, 
    0.005490790602226593, 0.005438659165343685, 0.005011421293695884, 0.004445849949711543, 0.003802274290667123, 
    0.003414481239748043, 0.003400465772733619, 0.003362918351896612, 0.002623446891891916, 0.002421456525235684, 
    0.001942671784383428, ]

staircase_avg_t_count = [
    0.000000000000000000, 1.414213562373094923, 8.358523998839725522, 14.90836057126281666, 18.66666666666666430, 
    35.41820150062702055, 36.69143914061946532, 45.81060988897981900, 59.62423066692823426, 79.37357444754127300, 
    82.26631789232057201, 91.53596103867951683, 100.1969478876198707, 103.3309855545793710, 114.8865692342005786, 
    129.9467535443398845, 130.1425954431041134, 137.7083130201563108, 162.2185949989647042, 177.3704252435253466, 
    190.7728291979915696, 234.8268288886290804, 253.6524018919697312, 274.1441976350207597, 277.6535538936209377, 
    337.6541634661525109, 418.3216273097789895, 546.1909215474321400, 569.3392538941667453, 610.1284182110173333, 
    621.5720973965859457, 648.8713606032779353, 670.4052913138149279, 869.0702760915087310, 1061.190329568459902, 
    1126.943755563780542, 1206.586675638444831, 1334.124470449861747, 1395.777866999348362, 1472.699658090156618, 
    1580.987320181555106, 1690.149558553591305, 1831.306963333395743, 1910.882476918264274, 2018.310491944398791, 
    2324.117154310660226, 2486.215289670361472, 3293.182410105669078, 3605.545686445599586, 4083.705682496006830, 
    4431.716899082452983, 4707.761035937885026, 5205.974979638611330, 6671.197660262269892, 7123.787980357737979, 
    8537.336639973320416, ]

staircase_phi = [
    0.785398163397448168, 0.392699081698724195, 0.255495373648521762, 0.284924126622062401, 0.192835807949161497, 
    0.173552440720655482, 0.124107795648728134, 0.110082423147095237, 0.109906358777430102, 0.101528462773912573, 
    0.091710893591349804, 0.093403544968404958, 0.070187845172285837, 0.077752676679104127, 0.078669855732249924, 
    0.069493297706496718, 0.057791644573391976, 0.061888637249731267, 0.055597998797776847, 0.053665989739915342, 
    0.041987261143899557, 0.042637172618535335, 0.041443167083348088, 0.034682096952168145, 0.019816715253174161, 
    0.022220137471257971, 0.022717871336363209, 0.022893844150929561, 0.021083637492522535, 0.021313565565117158, 
    0.020116119052836046, 0.020040204171539192, 0.016410933443297120, 0.013234079689553997, 0.013194264811155256, 
    0.011536666869366122, 0.011604021540317999, 0.010119687083072504, 0.008597823205676720, 0.008827793348440431, 
    0.008855693342139597, 0.008875421608560755, 0.008191240044674157, 0.008111792864880870, 0.005202477715701859, 
    0.005378493332280404, 0.005430047050166927, 0.005010434945759594, 0.004437666869261012, 0.003795608808731670, 
    0.003384718928535352, 0.003398669385652424, 0.003361547449303978, 0.002623229168641762, 0.002386380112804906, 
    0.001932691904085085, ]

staircase_tanalpha = staircase_tanalpha[::-1]
staircase_avg_t_count = staircase_avg_t_count[::-1]
staircase_phi = staircase_phi[::-1]


def get_t_count_for_rot(theta: float, delta: float) -> Tuple[float, float]:
    """
    Parameters
    ----------
    theta : float
        The angle of the rotation.
    delta : float
        The desired value of delta (either the diamond norm error or lambda-1).

    Returns
    -------
    av_t_count : float
        The calculated average T gate count.
    delta_true : float
        The actual value of delta used. This might be less than the requested
        value of delta if the solution is from the staircase.
    """
    # Map theta to range [0, pi/8] by applying Clifford gates as needed.
    if np.abs(theta) > np.pi/8:
        theta = (theta + np.pi/8) % (np.pi/4) - np.pi/8
    theta = np.abs(theta)

    if theta == 0:
        return 0.0, 0.0

    tanalpha = delta / np.sin(2 * theta) + np.tan(theta)
    if np.sin(2 * theta) == 0 or tanalpha == 0:
        print("Error: accuracy too poor, sin(2 * theta) = 0 or tanalpha = 0")

    # Find index of largest tanalpha value that is smaller than or equal the given tanalpha.
    index = bisect.bisect_right(staircase_tanalpha, tanalpha) - 1
    if (
        index >= 0 and np.tan(theta) <= staircase_phi[index]
    ):
        # We have found a solution in the staircase that we can use.
        av_t_count = staircase_avg_t_count[index] * np.sin(2 * theta)
        delta_true = (staircase_tanalpha[index] - np.tan(theta)) * np.sin(2 * theta)
    else:
        # We did not find a solution in the staircase. Use the asymptotic formula instead.
        alpha = delta / (2 * theta) + theta
        K = (2 * np.sqrt(2 * np.e**3) / 3) ** (2 / 3)
        phi_0 = max(alpha - alpha / np.log(K / alpha), theta)
        av_t_count = (3 * theta / (alpha + 2 * phi_0)) * np.log2(
            12 / (((alpha - phi_0) ** 2) * (alpha + 2 * phi_0))
        )
        delta_true = delta

    # Finally, take the minimum of the above result and the angle-independent result.
    worst_case_t_count = 1.52 * np.log2(1 / delta) - 0.01
    if av_t_count > worst_case_t_count:
        av_t_count = worst_case_t_count
        delta_true = delta

    return av_t_count, delta_true
\end{lstlisting}

\section{Numerical optimization of quasi-probability distributions}
\label{app:cvx}

In this appendix we present a numerical approach to optimize a linear combination of channels (LCC) using CVXPY \cite{cvxpy_paper_1, cvxpy_paper_2}. Although our final results use LCCs whose coefficients are derived analytically, this numerical approach has been a key tool throughout our investigation. A similar numerical approach has recently been used in Ref.~\cite{Luthra2025}.

We want to construct an LCC for $\rzchannel$, and suppose that we have selected a set of channels $\{ \mathcal{U}_i \}$ that we are able to implement. In general, such a set  will be over-complete. In this case, we choose the optimal coefficients $\{ c_i \}$ as the ones which minimize the L1 norm, $\lambda = \sum_i |c_i|$. The task of finding $\{ c_i \}$ to minimize $\lambda$ given the constraint
\begin{equation}
    \rzchannel = \sum_i c_i \mathcal{U}_i
    \label{eq:rzchannel_lcc_appendix}
\end{equation}
is a convex optimization problem which can be solved efficiently with existing numerical libraries such as CVXPY, using the following approach.

First, we write each $\mathcal{U}_i$ in the Pauli basis, i.e.
\begin{equation}
    \mathcal{U}_i(\rho) = \sum_{jk} A_{jk,i} P_j \rho P_k.
\end{equation}
Similarly, we write the target $\rzchannel$ in the Pauli basis.
\begin{equation}
    \rzchannel(\rho) = \sum_{jk} b_{jk} P_j \rho P_k.
\end{equation}
Therefore, to satisfy \cref{eq:rzchannel_lcc_appendix},
\begin{equation}
    b_{jk} = \sum_i A_{jk,i} x_i, \;\;\;\ \forall \; j, k.
\end{equation}
Merging the indices $j$ and $k$ into a single index, we obtain the linear system
\begin{equation}
    A x = b.
\end{equation}
Therefore, the optimal LCC is obtained by solving
\begin{equation}
    \textrm{minimize } \sum_i |c_i| \textrm{ subject to } Ax = b.
    \label{eq:optimal_lcc_problem}
\end{equation}
Given $A$ and $b$, this is a convex optimization problem which can be solved efficiently using existing libraries such as CVXPY \cite{cvxpy_paper_1, cvxpy_paper_2}. Taking $\{ \mathcal{U}_i \}$ as a set containing Clifford+T and fallback channels allows one to numerically investigate the quasi-probability approach to rotation synthesis. We used this approach to explore solutions in the early stages of this work, eventually settling on the analytic solutions given in \Cref{sec:analytic coefficients}. One downside of the numerical approach is that the problem becomes ill-conditioned when optimizing over Clifford+T channels with high accuracy, which is resolved using the solution in \Cref{sec:analytic coefficients}.

\section{Exact solutions for \texorpdfstring{$λ-1\gtrsimθ/500$}{λ-1 >~ θ/500}}
\label{app:matsumoto_amano}

In this work we have extensively studied the quasi-probability method using an over-rotation, and with the under-rotation fixed as the identity. In the case of large $λ-1\gtrsimθ/500$, we have used exhaustive search to find optimal over-rotation unitaries; see \Cref{sec:matsumoto_amano} for details, which presents solutions in \Cref{tab:staircase}.

Tables~\ref{tab:staircase app} and \ref{tab:staircase gate sequences app} present a longer version of \Cref{tab:staircase}, with T gate count up to $n_T=35$. The exhaustive search required 262 hours on a single core. All results of matrix multiplications $(HT|SHT)^*$ of the middle term of the Matsumoto-Amato normal form \cref{eq:matsumoto normal form} were cached up to T count 12 to avoid recalculating matrix products at every step. Longer normal forms can be efficiently computed by multiplying matrices from the cache.

In order to use an over-rotation from the table, one must check that its angle $φ$ exceeds the target angle $θ$. However, as we will argue, this requirement is not a real restriction of the applicability of the table. Consider a case where the over-rotation angle $φ$ from the table is in fact $φ<θ$. Luckily, as can be seen by inspecting the table, $\tanα\approx\phi$ (except for the first row), such that we also have $\tanα\lesssim θ$. Therefore the target $λ$ for the desired rotation is
\begin{equation}
    λ-1 = \tanα\sin(2θ)+\cos(2θ) - 1\approx 2θ\tanα \lesssim 2θ^2.
\end{equation}
Note that the regime $λ-1\lesssimθ^2$ is the one in which it is optimal to choose an under-rotation different from the identity, see \Cref{fig:schematic}(a). The staircase however is based on choosing the under-rotation as the identity. Altogether that means, when $φ<θ$ for the over-rotation from the staircase, one would not want to use the staircase in the first place.

Note that results towards the bottom of the table might not be optimal: conceivably, a higher T count over-rotation will have a lower $p$. An example of this is the 12 T-count unitary; enumerating only up to T count 11 would have given suboptimal unitaries. 

Thanks to \Cref{app:comparison kliuchnikov}, the tables can also (approximately) be used for probability mixtures by identifying $\epsdiamond = \lambda-1$. This is also clear from the numerical results, see \Cref{fig:av_t_count_per_rot}.

\begin{table}[p!]
{\small
\begin{tabular}{cc|ccc}
 &  & \multicolumn{3}{c}{Over-rotation unitary with $u=re^{iφ}$} \\
{$\tan \alpha$} & {$(p/\sin2θ)\cdot\text{T-count}$} & {T-count} & {$1-r$} & {$\varphi$} \\
\hline
\num{1.00e+00} & \num{0.00e+00} & 0 & \num{0.00e+00} & \num{7.85e-01} \\
\num{4.14e-01} & \num{1.41e+00} & 1 & \num{0.00e+00} & \num{3.93e-01} \\
\num{3.51e-01} & \num{8.36e+00} & 4 & \num{1.08e-02} & \num{2.55e-01} \\
\num{3.13e-01} & \num{1.49e+01} & 8 & \num{2.68e-03} & \num{2.85e-01} \\
\num{2.12e-01} & \num{1.87e+01} & 7 & \num{1.57e-03} & \num{1.93e-01} \\
\num{1.99e-01} & \num{3.54e+01} & 12 & \num{2.01e-03} & \num{1.74e-01} \\
\num{1.38e-01} & \num{3.67e+01} & 9 & \num{7.86e-04} & \num{1.24e-01} \\
\num{1.15e-01} & \num{4.58e+01} & 10 & \num{2.30e-04} & \num{1.10e-01} \\
\num{1.11e-01} & \num{5.96e+01} & 13 & \num{3.37e-05} & \num{1.10e-01} \\
\num{1.06e-01} & \num{7.94e+01} & 16 & \num{2.13e-04} & \num{1.02e-01} \\
\num{9.56e-02} & \num{8.23e+01} & 15 & \num{1.66e-04} & \num{9.17e-02} \\
\num{9.39e-02} & \num{9.15e+01} & 17 & \num{8.42e-06} & \num{9.34e-02} \\
\num{8.98e-02} & \num{1.00e+02} & 14 & \num{6.80e-04} & \num{7.02e-02} \\
\num{8.10e-02} & \num{1.03e+02} & 16 & \num{1.20e-04} & \num{7.78e-02} \\
\num{8.01e-02} & \num{1.15e+02} & 18 & \num{4.81e-05} & \num{7.87e-02} \\
\num{7.18e-02} & \num{1.30e+02} & 18 & \num{7.68e-05} & \num{6.95e-02} \\
\num{6.81e-02} & \num{1.30e+02} & 15 & \num{2.94e-04} & \num{5.78e-02} \\
\num{6.36e-02} & \num{1.38e+02} & 17 & \num{4.91e-05} & \num{6.19e-02} \\
\num{5.66e-02} & \num{1.62e+02} & 18 & \num{2.53e-05} & \num{5.56e-02} \\
\num{5.47e-02} & \num{1.77e+02} & 19 & \num{2.60e-05} & \num{5.37e-02} \\
\num{4.39e-02} & \num{1.91e+02} & 16 & \num{3.86e-05} & \num{4.20e-02} \\
\num{4.32e-02} & \num{2.35e+02} & 20 & \num{1.11e-05} & \num{4.26e-02} \\
\num{4.18e-02} & \num{2.54e+02} & 21 & \num{6.02e-06} & \num{4.14e-02} \\
\num{3.55e-02} & \num{2.74e+02} & 19 & \num{1.44e-05} & \num{3.47e-02} \\
\num{2.66e-02} & \num{2.78e+02} & 11 & \num{6.74e-05} & \num{1.98e-02} \\
\num{2.37e-02} & \num{3.38e+02} & 15 & \num{1.68e-05} & \num{2.22e-02} \\
\num{2.32e-02} & \num{4.18e+02} & 19 & \num{5.65e-06} & \num{2.27e-02} \\
\num{2.30e-02} & \num{5.46e+02} & 25 & \num{1.64e-06} & \num{2.29e-02} \\
\num{2.18e-02} & \num{5.69e+02} & 24 & \num{7.73e-06} & \num{2.11e-02} \\
\num{2.16e-02} & \num{6.10e+02} & 26 & \num{2.85e-06} & \num{2.13e-02} \\
\num{2.11e-02} & \num{6.22e+02} & 25 & \num{9.82e-06} & \num{2.01e-02} \\
\num{2.02e-02} & \num{6.49e+02} & 26 & \num{1.28e-06} & \num{2.00e-02} \\
\num{1.64e-02} & \num{6.70e+02} & 22 & \num{6.20e-08} & \num{1.64e-02} \\
\num{1.33e-02} & \num{8.69e+02} & 23 & \num{1.06e-07} & \num{1.32e-02} \\
\num{1.32e-02} & \num{1.06e+03} & 28 & \num{4.51e-08} & \num{1.32e-02} \\
\num{1.17e-02} & \num{1.13e+03} & 26 & \num{8.13e-07} & \num{1.15e-02} \\
\num{1.16e-02} & \num{1.21e+03} & 28 & \num{3.33e-08} & \num{1.16e-02} \\
\num{1.01e-02} & \num{1.33e+03} & 27 & \num{1.68e-08} & \num{1.01e-02} \\
\num{9.18e-03} & \num{1.40e+03} & 24 & \num{2.50e-06} & \num{8.60e-03} \\
\num{8.94e-03} & \num{1.47e+03} & 26 & \num{4.93e-07} & \num{8.83e-03} \\
\num{8.91e-03} & \num{1.58e+03} & 28 & \num{2.47e-07} & \num{8.86e-03} \\
\num{8.89e-03} & \num{1.69e+03} & 30 & \num{7.18e-08} & \num{8.88e-03} \\
\num{8.22e-03} & \num{1.83e+03} & 30 & \num{1.32e-07} & \num{8.19e-03} \\
\num{8.13e-03} & \num{1.91e+03} & 31 & \num{9.28e-08} & \num{8.11e-03} \\
\num{5.68e-03} & \num{2.02e+03} & 21 & \num{1.23e-06} & \num{5.20e-03} \\
\num{5.49e-03} & \num{2.32e+03} & 25 & \num{3.02e-07} & \num{5.38e-03} \\
\num{5.44e-03} & \num{2.49e+03} & 27 & \num{2.32e-08} & \num{5.43e-03} \\
\num{5.01e-03} & \num{3.29e+03} & 33 & \num{2.37e-09} & \num{5.01e-03} \\
\num{4.45e-03} & \num{3.61e+03} & 32 & \num{1.81e-08} & \num{4.44e-03} \\
\num{3.80e-03} & \num{4.08e+03} & 31 & \num{1.26e-08} & \num{3.80e-03} \\
\num{3.41e-03} & \num{4.43e+03} & 30 & \num{5.03e-08} & \num{3.38e-03} \\
\num{3.40e-03} & \num{4.71e+03} & 32 & \num{3.03e-09} & \num{3.40e-03} \\
\num{3.36e-03} & \num{5.21e+03} & 35 & \num{2.28e-09} & \num{3.36e-03} \\
\num{2.62e-03} & \num{6.67e+03} & 35 & \num{2.78e-10} & \num{2.62e-03} \\
\num{2.42e-03} & \num{7.12e+03} & 34 & \num{4.18e-08} & \num{2.39e-03} \\
\num{1.94e-03} & \num{8.54e+03} & 33 & \num{9.64e-09} & \num{1.93e-03} \\
\end{tabular}}
\caption{Optimal over-rotations up to T count 35, found by exhaustive search (\Cref{sec:matsumoto_amano}) see \Cref{app:matsumoto_amano}. For given target rotation $θ$ and target sample complexity $λ_\text{max}$, the optimal over-rotation is that with $λ=\tanα\sin2θ + \cos2θ$ not exceeding $λ_\text{max}$. It has average T-count $p\cdot$T-count, which is divided by $\sin2θ$ in the table to show a $θ$-independent quantity. Gate sequences are listed in \Cref{tab:staircase gate sequences app}. Note that extending the search to higher T counts could yield improvements over this table.}
\label{tab:staircase app}
\end{table}

\begin{table}
{\scriptsize
\begin{tabular}{c|l}
{$\tan \alpha$} & {Over-rotation gate sequence}\\
\hline
\num{1.00e+00} & ISZ. \\
\num{4.14e-01} & TSZ. \\
\num{3.51e-01} & ISHTHTSHTSHTHZ. \\
\num{3.13e-01} & ISHTSHTSHTHTHTHTHTSHTSHSI. \\
\num{2.12e-01} & IHTSHTHTHTSHTSHTSHTSHX. \\
\num{1.99e-01} & ISHTSHTHTHTHTHTSHTSHTSHTSHTSHTSHTHX. \\
\num{1.38e-01} & IHTHTSHTSHTHTSHTHTSHTSHTHSI. \\
\num{1.15e-01} & IHTHTHTSHTSHTHTHTSHTSHTHTHSI. \\
\num{1.11e-01} & ISHTSHTSHTSHTHTHTSHTHTSHTHTHTSHTSHTSHZ. \\
\num{1.06e-01} & IHTHTSHTHTHTSHTSHTSHTSHTHTSHTHTHTSHTSHTSHTSI. \\
\num{9.56e-02} & IHTHTHTSHTHTSHTHTSHTHTHTSHTHTSHTSHTSHTHSY. \\
\num{9.39e-02} & IHTHTSHTHTHTSHTSHTSHTSHTHTSHTSHTSHTSHTHTHTSHTHI. \\
\num{8.98e-02} & IHTSHTHTHTSHTSHTHTHTHTSHTSHTSHTHTSHTSHSZ. \\
\num{8.10e-02} & ISHTSHTHTSHTHTHTHTSHTSHTHTHTHTSHTSHTSHTSHTSHSX. \\
\num{8.01e-02} & IHTSHTSHTSHTHTSHTSHTSHTHTHTHTHTSHTSHTSHTHTSHTSHTSHZ. \\
\num{7.18e-02} & ISHTHTSHTSHTSHTHTHTSHTHTSHTHTHTHTSHTSHTSHTHTHTSHY. \\
\num{6.81e-02} & IHTHTSHTHTHTSHTSHTSHTHTHTSHTSHTSHTHTHTHSZ. \\
\num{6.36e-02} & ISHTHTSHTSHTSHTSHTHTHTSHTSHTSHTHTHTSHTSHTSHTSHTHSZ. \\
\num{5.66e-02} & TSHTSHTSHTHTSHTHTSHTSHTSHTHTHTSHTSHTHTHTHTHTSHI. \\
\num{5.47e-02} & ISHTSHTHTSHTSHTHTHTSHTSHTHTHTSHTSHTSHTHTHTHTHTSHTHSY. \\
\num{4.39e-02} & IHTHTSHTSHTSHTSHTHTHTSHTSHTSHTSHTSHTSHTHTHTHSZ. \\
\num{4.32e-02} & IHTHTHTHTSHTSHTHTHTSHTSHTSHTSHTSHTSHTHTHTSHTSHTHTHTHZ. \\
\num{4.18e-02} & ISHTHTHTSHTHTSHTHTHTSHTHTSHTSHTHTSHTHTHTHTSHTHTSHTHTSHI. \\
\num{3.55e-02} & IHTSHTHTSHTSHTSHTHTHTSHTHTSHTSHTSHTSHTSHTHTHTHTHTIY. \\
\num{2.66e-02} & ISHTSHTHTSHTHTHTSHTHTHTSHTHTSHZ. \\
\num{2.37e-02} & ISHTHTHTHTSHTSHTSHTSHTSHTSHTSHTSHTSHTHTHTHSI. \\
\num{2.32e-02} & ISHTSHTHTHTHTHTHTHTSHTHTHTSHTHTHTSHTSHTHTHTSHTHSX. \\
\num{2.30e-02} & IHTHTHTHTHTHTHTHTHTSHTSHTSHTSHTSHTSHTSHTSHTSHTHTHTHTSHTSHTHTSHTSHX. \\
\num{2.18e-02} & TSHTHTHTSHTHTHTSHTSHTSHTHTSHTSHTSHTSHTSHTSHTSHTSHTSHTHTHTHTHTHI. \\
\num{2.16e-02} & IHTSHTSHTHTSHTHTHTSHTSHTHTSHTHTSHTHTHTSHTHTHTSHTHTSHTSHTHTHTHTHTHZ. \\
\num{2.11e-02} & ISHTSHTSHTHTHTSHTHTSHTSHTHTHTSHTSHTSHTHTSHTSHTHTSHTHTSHTHTHTHTHTSHX. \\
\num{2.02e-02} & IHTHTHTHTHTHTHTSHTSHTHTSHTSHTHTSHTSHTSHTHTHTSHTSHTHTHTHTSHTHTHTSX. \\
\num{1.64e-02} & THTHTHTSHTHTHTSHTHTHTSHTHTHTHTSHTHTHTSHTHTHTSHTHTHSZ. \\
\num{1.33e-02} & TSHTHTSHTHTHTHTHTSHTSHTHTSHTSHTSHTSHTHTSHTSHTHTHTHTHTSHTHSZ. \\
\num{1.17e-02} & THTHTSHTSHTSHTSHTSHTHTHTSHTHTSHTSHTHTHTHTHTSHTHTHTHTSHTHTHTHTHSY. \\
\num{1.16e-02} & ISHTHTHTSHTHTSHTHTHTSHTHTHTSHTHTSHTSHTHTSHTSHTHTHTSHTHTHTSHTHTHTHTSHTHSX. \\
\num{1.01e-02} & ISHTHTSHTHTSHTHTSHTHTHTSHTHTHTSHTSHTHTHTSHTSHTHTSHTHTHTSHTHTHTHTHTSHX. \\
\num{9.18e-03} & TSHTHTHTSHTSHTHTHTSHTSHTSHTSHTSHTSHTSHTSHTSHTHTSHTHTHTHTSHTHTHI. \\
\num{8.94e-03} & ISHTHTSHTSHTSHTSHTHTHTSHTSHTHTSHTSHTSHTHTHTHTSHTSHTSHTSHTHTSHTSHTHTSHTHSZ. \\
\num{8.91e-03} & IHTSHTSHTSHTHTSHTSHTSHTHTHTHTSHTSHTHTSHTSHTSHTSHTHTHTHTSHTSHTSHTHTSHTHTHTHI. \\
\num{8.89e-03} & IHTHTHTHTHTSHTSHTSHTSHTSHTSHTHTHTSHTSHTHTHTSHTHTHTSHTHTHTHTSHTHTSHTHTHTHTSHI. \\
\num{8.22e-03} & THTHTHTSHTHTSHTHTSHTHTHTHTSHTSHTSHTSHTHTSHTSHTSHTSHTHTHTHTSHTHTSHTHTSHTHTHI. \\
\num{8.13e-03} & TSHTSHTSHTHTHTSHTHTHTSHTHTHTSHTHTHTHTSHTHTHTSHTHTSHTHTSHTSHTSHTSHTSHTSHTSHTSHTHSI. \\
\num{5.68e-03} & THTHTHTHTHTHTHTSHTHTHTHTHTHTHTSHTHTHTHTHTHTHSI. \\
\num{5.49e-03} & ISHTHTHTSHTSHTSHTSHTSHTSHTHTSHTSHTHTHTSHTHTSHTSHTHTSHTHTHTSHTHTSHTHSY. \\
\num{5.44e-03} & IHTSHTSHTSHTSHTSHTSHTHTSHTHTHTSHTSHTHTSHTSHTSHTSHTSHTHTSHTSHTHTHTHTHTHTHX. \\
\num{5.01e-03} & THTSHTHTSHTSHTHTSHTHTSHTHTSHTHTSHTHTHTSHTHTHTSHTSHTHTHTSHTHTSHTSHTHTSHTHTHTHTHTSHZ. \\
\num{4.45e-03} & THTSHTHTHTSHTSHTSHTSHTSHTSHTHTHTSHTSHTSHTSHTSHTHTSHTHTHTHTSHTHTHTSHTSHTSHTSHTHTHTSHSX. \\
\num{3.80e-03} & IHTSHTSHTSHTHTHTSHTHTHTSHTHTHTHTSHTSHTHTSHTHTHTSHTHTSHTSHTHTSHTHTHTSHTHTHTHTHSZ. \\
\num{3.41e-03} & THTSHTSHTSHTHTSHTHTHTSHTHTSHTHTSHTHTSHTHTSHTHTHTHTHTSHTHTHTSHTSHTSHTHTHTHSI. \\
\num{3.40e-03} & IHTHTHTSHTHTSHTSHTSHTSHTSHTSHTSHTSHTSHTHTSHTHTSHTHTSHTSHTHTHTSHTHTHTSHTHTHTHTHTHTHX. \\
\num{3.36e-03} & IHTSHTSHTHTSHTSHTHTHTSHTHTSHTHTHTHTHTHTHTHTHTSHTSHTHTSHTSHTHTSHTHTHTHTSHTHTHTHTHTHTHSI. \\
\num{2.62e-03} & IHTHTHTHTHTHTSHTHTSHTSHTSHTSHTHTHTHTHTSHTSHTHTSHTSHTHTHTHTHTSHTSHTSHTSHTHTSHTHTHTHTHTHSZ. \\
\num{2.42e-03} & IHTHTSHTHTHTSHTHTSHTSHTSHTHTSHTSHTSHTSHTSHTSHTSHTSHTHTSHTHTHTSHTSHTHTHTHTHTHTSHTHTHTSHTSHSZ. \\
\num{1.94e-03} & THTHTSHTSHTSHTHTSHTHTHTSHTHTHTSHTSHTHTHTHTHTHTHTHTHTHTHTSHTHTSHTHTHTHTHTHTHSX. \\
\end{tabular}}
\caption{Gate sequences for the over-rotations in \Cref{tab:staircase app} in Matsumoto-Amano normal form, \cref{eq:matsumoto normal form}. We list one possible gate sequence with lowest $T$-count. As we are only interested in the top-left entry $u$ of each over-rotation (after normalizing the determinant), multiple sequences will be equivalent.}
\label{tab:staircase gate sequences app}
\end{table}

\section{Adapted gridsynth implementation}
\label{app:gridsynth}

In this appendix, we describe an adapted implementation of gridsynth, that can find over-rotations for our ancilla-free protocol when the under-rotation is fixed as the identity.

Note that the numerical results shown throughout the paper, particularly \Cref{fig:av_t_count_per_rot}, use a different approach. This was described in Sections~\ref{sec:numerical_results_ancilla_free} and \ref{sec:numerical results fallback}. These involve searching for both over- and under-rotation (which can be, but is not forced to be, the identity). That approach therefore naturally encompasses both parameter regions (\Cref{fig:schematic}(a)).

In contrast, here we present a streamlined adaptation of gridsynth for the specific setting of \Cref{sec:small angles}, i.e.~with the under-rotation fixed as the identity. Here, the region can be chosen slightly differently, i.e.~compare \Cref{fig:mixed_diagonal_u_region} and \Cref{fig:quasiregion}. In the regime $δ \ll θ^2$, where this choice of under-rotation is favorable, we ultimately find that it gives essentially the same results as the more general approach.

Gridsynth is an algorithm developed by Ross and Selinger \cite{ross2016optimal} for Clifford+T synthesis of rotations. At its core, it allows finding Clifford+T sequences with the top left entry $u$ in a certain given region of the complex plane. We repurposed the algorithm and extended the Python pygridsynth implementation \cite{pygridsynth_github, pygridsynth_paper} to our setting of quasi-probabilities, with the under-rotation fixed as the identity. The appropriate region is given in \cref{eq:region}.

Standard gridsynth terminates once it has found one solution inside the region. However, we enumerate all\footnote{Using a factoring oracle, all Clifford+T sequences in the region can be enumerated. In the absence of such an oracle, some might be skipped. At least in the context of \cite{ross2016optimal}, this has negligible impact.} solutions inside the region up to a certain T count, and compute $p\cdot$T-count from \cref{eq:probability} for each of them. Note that we must explore higher T count than that of the first unitary found, as higher T count sequences might lead to lower $p$ and overall lower average T count. We find our modified pygridsynth to be a very practical approach to explicitly finding good Clifford+T over-rotations. Moreover, note that if the best over-rotation for one $(θ,λ)$ pair has been found, the same over-rotation is the best over-rotation for other $(θ,λ)$ pairs: As for the staircase in \Cref{sec:matsumoto_amano} and \Cref{app:matsumoto_amano}, the $θ$ dependence of $λ$ and $p$ can be updated to a different $θ$, subject to the requirement $φ>θ$. Still, this approach fixes the under-rotation to the identity, and so gives sub-optimal solutions for $\delta \lesssim \theta^2$.

In this appendix we will first discuss how we specify the region for the over-rotation unitary, and then turn to discussing how overall phase of the unitary impacts the algorithm.

\subsection{Region for over-rotation}

In \Cref{sec:asymptotic}, \cref{eq:region} we have seen that we are searching for over-rotations whose top-left entry $u=x+iy=re^{iφ}$ lies in the region
\begin{equation}
\{(x,y)\in \mathbb{C},\ λ(x,y)=\frac{1-x^2}{xy}\sin2θ + \cos2θ \le λ_\text{max}\} \cup \{re^{iφ}\in\mathbb{C},\ 0<r\le1, θ<φ<\pi/4\}.
\label{eq:region app}
\end{equation}

The region is implemented by a new child class in pygridsynth inheriting \texttt{ConvexSet}.
It must implement 3 interface functions: An inside region test, calculation of the intersection of the region's boundary with a given line, and a bounding ellipse. In all cases we must respect a scale $s$ which is the scale for $r^2$, i.e. we must use $r^2/s$ instead of $r^2$, and $x/\sqrt{s},y/\sqrt{s}$ instead of $x,y$.

\subsubsection{Inside function}

For given $θ$ and $δ=λ_\text{max}-1$, we need to test whether a given point $x=r\cosφ, y=r\sinφ$ is in the target region. 
We have the following 4 conditions. To prevent prevent loss of accuracy when performing trigonometric functions and to improve numerical stability overall, we transform the conditions directly to cartesian coordinates and try to avoid numerically unstable operations.
\begin{itemize}
    \item In the upper right quadrant $x \ge 0, y \ge 0$.
    \item Inside the $λ$-hyperbole
    \begin{align}
    λ_\text{max} - \cos2θ &\ge \sin2θ\frac{s-x^2}{xy}, \\
    \Leftrightarrow\underbrace{(δ + 2\sin^2θ)}_A xy &\ge (s-x^2)\underbrace{\sin2θ}_B.
\end{align}

\item Inside unit disk
\begin{equation}
    r^2/s \le 1\ \Leftrightarrow\ x^2 + y^2 \le s.
\end{equation}

\item Angle $φ\geφ_0\geθ$. For performance we allow an optional $φ_0\geqθ$, as in the asymptotics,
\begin{align}
    \sin(φ-φ_0) &\ge 0 \\
    \Leftrightarrow
    y\cosφ_0 &\ge x\sinφ_0.
\end{align}
    
\end{itemize}

\subsubsection{Intersect function}
We must find the two intersection points $t$ of the line
\begin{equation}
    g(t) = \vec u + t\vec v
\end{equation}
with the boundary of the region, or return \texttt{None} if there is no intersection.

Note that, if we wanted, we could make the region described by the \texttt{intersect} function slightly bigger than it is by ignoring the $φ\geθ$ constraint. This would reduce efficiency of the algorithm as more solutions will filtered out with \texttt{inside}. However, here we can use the exact region.

Intersection points of the line $g(t)$ with the boundary curves from above lead to (up to) quadratic equations in $t$:
\begin{itemize}
    \item Inserting into the $λ$-hyperbole equation:
    \begin{align}
        A(u_x+tv_x)(u_y + tv_y) &= (s-(u_x+tv_x)^2)B \\
        t^2Av_xv_y + tA(v_xu_y + v_yu_x) + Au_xu_y &= -t^2Bv_x^2 - t2B(u_xv_x) +sB-u_x^2B .
    \end{align}
    \item Inserting into the circle:
    \begin{align}
        (u_x+tv_x)^2 + (u_y+tv_y)^2 &= s \\
        t^2(v_x^2 + v_y^2) + t(2 \vec u\cdot\vec v) + (u_x^2+u_y^2-s) &= 0.
    \end{align}
    \item Inserting into the $φ=θ$ line:
    \begin{align}
        (u_y+tv_y)\cosφ_0 &= (u_x + tv_x) \sinφ_0 \\
        t(v_y\cosφ_0 -v_x\sinφ_0) &= u_x\sinφ_0 -u_y\cosφ_0 \\
        t =\frac{u_x\sinφ_0 -u_y\cosφ_0}{v_y\cosφ_0 -v_x\sinφ_0}.
    \end{align}
\end{itemize}
Provided that any solutions are real, they are intersection points if they are inside the region. For numerical reasons we do not check the inequality whose equality has been solved for $t$, but only the others.

\subsubsection{Bounding ellipse}
Finally, we must find an ellipse
\begin{equation}
    (\vec{x}-\vec{p})^T D (\vec x - \vec p) \le 1
\end{equation}
with centre-point $\vec{p}$ and $D>0$ that bounds the region.  The tighter the bounding ellipse, the more efficient the result of the uprighting procedure of gridsynth will be, a core contributor of its efficacy \cite{ross2016optimal}. If the ellipse is too small (i.e., not actually a bounding ellipse), gridsynth will still work and find all solutions as the region is correctly delineated by the intersect and inside functions, but the efficiency improvement from the uprighting procedure will be reduced.

The following points should be contained in the ellipse:
\begin{itemize}
\item The point $(r=1,α)$ (intersection of $λ$-hyperbole and unit circle) in Cartesian coordinates is
\begin{equation}
    \vec\alpha = \frac{(B,A)}{\sqrt{sA^2+sB^2}},
\end{equation}
as can be seen by plugging $s-x^2=y^2$ into the hyperbole. We have $α = \arctan(A/B)$.

\item 
The point $(r=1, φ_0)$ (intersection of the unit circle and the $φ_0$ ray) in Cartesian coordinates is
\begin{equation}
    (\cosφ_0, \sinφ_0).
\end{equation}
\item 
The intersection point between the $φ_0$ ray and the $λ$ hyperbole can be found by using the intersect function with the ray. 

\item 
The point $(r=1, (φ_0+α)/2)$ on the unit circle, halfway between the other two points on the unit circle.

\item
The intersection point between the origin and the previous point, i.e.~the halfway point on the $λ$-hyperbole.
\end{itemize}

While these points do not form a convex enclosing region, it is reasonable to expect that a minimum area ellipse containing them will will give a good approximation of the location, size, and orientation of the region.

We use an implementation of a Khachiyan-style MVEE loop to find a bounding ellipse. This is an algorithm that allows one to find the minimum area ellipse covering a set of points.
While even $0$ iterations give a good result, we use $10$ iterations to give a smaller region but keeping good performance. Further, we inflate the resulting approximate ellipse to ensure all the above points are contained within it.

\subsection{Impact of phase}

All Clifford+T unitaries may be generated by
\begin{equation}
     X, \; Y, \; Z, \; H =\frac{1}{\sqrt{2}}\begin{pmatrix}1&1\\1&-1\end{pmatrix}, \; S = \begin{pmatrix}1&0\\0&i\end{pmatrix}, \;
    T =\begin{pmatrix}1&0\\0&ω\end{pmatrix}, \; \text{where}\ ω = e^{i\pi/4}.
\end{equation}
Every such unitary has the form
\begin{equation}
    U = \begin{pmatrix}u & -t^\dag ω^\ell \\ t & u^\dagω^\ell\end{pmatrix}, \; u,t\in \mathbb{Z}[1/\sqrt{2}, i], \; \ell\in\mathbb{Z},\ u^\dag u +v^\dag v = 1, \quad\Rightarrow\det U = ω^\ell,
\end{equation}
(see \cite[Eq.~11]{ross2016optimal}).

Our approach, and that of \cite{Kliunchnikov_shorter2023}, requires $\det U = 1$ in order to apply the twirling proposition (see \Cref{sec:analytic coefficients}). However, as the overall phase in a quantum algorithm does not matter, we can first normalize the determinant, and apply our machinery to $u/\sqrt{ω^\ell}$.
We only need to consider $\ell=0$ ($\det U=1$) and $\ell=-1$ ($\det U = ω^{-1}$), as the $\ell$ can be increased by 2 by multiplying with $XSX$, without any T gates.

For $\ell=0$, we can directly use the main gridsynth approach. In pygridsynth this is called \texttt{hasphase=false}. The T count is related to the denominator exponent $k$ by: \cite[Lemma 7.3]{ross2016optimal}:
\begin{itemize}
\item $k=0\rightarrow T = 0$
\item $k>0\rightarrow T = 2k-2$ or $2k$. If the latter and we are only interested in the top-left entry, then $T^\dag UT$ has T count $2k-2$.
\end{itemize}

For $\ell=-1$, the determinant is $ω^{-1}=e^{-i\pi/4}$. This corresponds to ``Approximation up to a phase" in \cite{ross2016optimal}, and \texttt{hasphase=true} in pygridsynth. The unitary $U$ can be brought into $SU(2)$ form by multiplying with $e^{i\pi/8}=(1+ω)/|1+ω|$, using $(1+ω)/ω=(1+ω)^\dag$:
\begin{align}
    U &= \begin{pmatrix}u & -t^\dag ω^{-1}\\ t & u^\dagω^{-1}\end{pmatrix} = e^{-i\pi/8} U',\ U'\in SU(2), \\
    U' &= e^{i\pi/8}U = \frac{1}{|1+ω|}\begin{pmatrix}(1+ω)u & - (1+ω)^\dag t^\dag \\
    (1+ω)t & (1+ω)^\dag u^\dag\end{pmatrix} =: \frac{1}{|1+ω|}\begin{pmatrix}u' & -t'^\dag \\
    t' & u'^\dag\end{pmatrix}.
\end{align}
Physically, there is no difference between $U$ and $U'$. For our analysis however, we have determined the required region for
\begin{equation}
    u'/|1+ω| = (1+ω)u/|1+ω|,
\end{equation}
which is the top-left entry of the det=1 unitary.
In fact, thanks to the scaling chosen above, $|1+ω|U'$ has elements in $\mathbb{Z}[1/\sqrt{2}, 1]$, such that the usual gridsynth algorithm can be run for $u'$ with the scaling modifications of the area $A$ and the unit disc:
\begin{gather}
    u' \in |1+ω|\cdot A \\
    u'^\dag u' = |1+ω| - v'^\dag v' \le |1+ω|\ \Rightarrow\ u'\in |1+ω|\cdot \text{UnitDisc}
\end{gather}
The T count for \texttt{hasphase=true} is now \cite[Lemma 9.7]{ross2016optimal}:
\begin{itemize}
\item $k=0\rightarrow T = 1$
\item $k>0\rightarrow T = 2k-1$ or $2k+1$. If the latter and we are only interested in the top-left entry, then $T^\dag UT$ has T count $2k-1$.
\end{itemize}

\section{Minimizer \texorpdfstring{$φ_0$}{φ\_0} for asymptotic T count}
\label{app:optimal phi0}

In this appendix, we derive the value of $\varphi_0$ which minimizes the average T gate count in the derivation of \Cref{sec:asymptotic}. Here, the under-rotation is fixed as the identity and we are concerned with quasi-probabilities.
We have derived an asymptotic formula in \Cref{sec:asymptotic} for the average T count in the regime $λ-1 \ll θ \ll 1$. It depends on the chosen lower boundary $φ_0$ of the region, with $θ \le φ_0 \le α$. We will find the approximate minimizer $φ_0$, with which our asymptotic formula gives highly accurate average T count estimates, as seen in \Cref{fig:av_t_count_per_rot}.

Setting the derivative $\partial /\partial φ_0$ of \cref{eq:asymptotic} to zero gives the condition
\begin{align}
    \ln\frac{12}{(α-φ_0)^2(α+2φ_0)} = \frac{3φ_0}{α-φ_0}.
    \label{eq:minimisation condition}
\end{align}
We set $φ_0 = (1-t)α,\ t\in(0,1)$, giving
\begin{align}
    &\ln\frac{12}{t^2(3-2t)α^3} = \frac{3}{t}-3.
\end{align}
Thanks to $α\ll1$, the logarithm is large, such that the solution $t$ on the RHS will be small, $t\ll1$.  Therefore, we approximate $(3-2t) \approx 3$, and the equation becomes
\begin{align}
    &\Leftrightarrow \frac{12e^3}{t^23α^3} = e^{3/t} \\
    &\Leftrightarrow \frac{4e^3}{α^3} = t^2e^{3/t} \\
    &\Leftrightarrow \sqrt{\frac{4e^3}{α^3}} = te^{3/(2t)} \\
    &\Leftrightarrow \sqrt{\frac{4e^3}{α^3}} = \frac{3}{-2t'} e^{-t'},\ t':=-\frac{3}{2t}\\
    &\Leftrightarrow -\frac{3}{4}\sqrt{\frac{α^3}{e^3}} = t'e^{t'}.
\end{align}
The solution for $t$ is given by the Lambert-W function. We require the $W_{-1}$ branch for $t' \in (-\infty,-1) \Leftrightarrow t \in (0, 3/2)$ as we are interested in $t\to0^+ \Leftrightarrow t'\to-\infty$.
Therefore, we have the solution
\begin{align}
&t'= W_{-1}\left(-\frac{3}{4}\sqrt\frac{α^3}{e^3}\right) \\
&\Leftrightarrow t = \frac{3}{2}\frac{1}{-W_{-1}\left(-\frac{3}{4}\sqrt\frac{α^3}{e^3}\right)} .
\end{align} Due to $α\ll1$ we can further use the the first order of the expansion of the Lambert-W function $-W_{-1}(-x) = \ln\frac{1}{x} + \ln\ln\frac{1}{x} + \cdots$ as $x\to0^+$, giving the following value for $φ_0$:
\begin{equation}
φ_0 \approx \left(1-\frac{1}{\ln(K/α)}\right)α,\ \text{with}\ K:=\left(\frac{4\sqrt{e^3}}{3}\right)^{2/3}\approx3.29,
\end{equation}
subject to the requirement $θ<φ_0<α$. The upper bound is automatically met as $α \ll 1 < K/e$. Using $α\approx\frac{δ}{2θ}+θ$ (c.f.~\cref{eq:tanalpha}), the lower bound will be violated if
\begin{align}
    δ < \frac{2θ^2}{\ln(K/α)-1} \approx \frac{2θ^2}{\ln(1/θ)},
\end{align}
where in the second step we have approximated $(K/e)α\approxθ$ (which follows from the first step). In that case, $φ_0$ must be set to $φ_0=θ$.

\section{Comparison to Kliuchnikov \emph{et al}.}
\label{app:comparison kliuchnikov}

In this appendix, we compare the scheme from Kliuchnikov \emph{et al}.~\cite{Kliunchnikov_shorter2023} using proper probabilities with the quasi-probability scheme introduced in this paper.
We show that using the same over- and under-rotation unitaries, these methods have near equal average T counts with diamond norm error $ε_\diamond$ (for probabilities) and sampling overhead $λ$ (for quasi-probabilities) identified as $λ - 1 = ε_\diamond$.
This is also visible in the main plot,~\Cref{fig:av_t_count_per_rot}, where the points are almost indistinguishable. The two methods are also discussed and compared in \Cref{sec:analytic coefficients}.

Despite this equivalence, there can be significant advantage to using quasi-probabilities: For the same desired total error, $λ-1$ can be chosen orders of magnitude larger than the equivalently required $\epsdiamond$, see \Cref{app:prob_vs_quasi_prob}.

\begin{proposition}[{Comparison to Kliuchnikov \emph{et al}., Under-rotation${}=\mathbb{1}$}]
\label{prop:identity}
Let $θ$ be a target angle and $u=re^{i\phi}$ a chosen over-rotation, with the under-rotation fixed as the identity.
According to our quasi-probability method (\Cref{sec:small angles}), let $p$ be the quasi-probability of the twirled over-rotation, and $λ$ the sampling overhead. According to the mixed approximation in Appendix C of~\cite{Kliunchnikov_shorter2023}, let $p'$ be the probability of the over-rotation, and $ε_\diamond$ the diamond norm error. Then we have:
\begin{align}
    \frac{p}{p'} = 1 -O(θ^2) +O(p)
\end{align}
and
\begin{align}
   λ-1 = ε_\diamond - O(θ^2).
\end{align}
That is, in the regime $λ-1 \gg θ^2$, where using the identity as the under-rotation gives an advantage, $p\approx p'$ and $λ-1\approxε_\diamond$.
\end{proposition}
Proof:
The notation used in \cite[Appendix C]{Kliunchnikov_shorter2023} describes the identity as $δ_1=-θ$ and the over-rotation as $δ_2 = φ-θ$. Let us begin by gathering $p$ from \cref{eq:probability}, and $p'$ from \cite[Appendix C]{Kliunchnikov_shorter2023}:
\begin{align}
    p &= \frac{\sin2θ}{r^2\sin2φ} \\
    p' &= \frac{\sin2θ}{\sin2θ +r^2\sin(2(φ-θ))}
\end{align}
These are almost equal which we show by calculating the ratio:
\begin{align}
    \frac{p}{p'} &= p + pr^2\frac{\sin(2(φ-θ))}{\sin2θ} \\
    &= p + pr^2 \frac{\sin2φ\cos2θ - \cos2φ\sin2θ}{\sin2θ} \\
    &=p + \cos2θ - pr^2 \cos2φ \\
    &= \cos2θ + p(1-r^2\cos2φ) \\
    &= 1 -O(θ^2) + O(p)
\end{align}
In the regime $λ-1\ggθ^2$, it is $p\approx θ/(r^2φ) < θ/φ < θ/α < 2θ^2/(λ-1) \ll1$, such that $p\approx p'$.
Therefore, the average T count is approximately the same in our approach and the Kliuchnikov Appendix C approach with under-rotation fixed as the identity and for the same over-rotation.

Next, let us gather $λ$ from \cref{eq:lambda}
\begin{align}
λ &= p + \cos^2θ - pr^2\cos^2φ + p(1-r^2)-\sin^2θ + pr^2\sin^2φ \\
λ-1 & = 2\left(\cos^2θ - 1 + p(1-r^2\cos^2φ)\right) \\
\end{align}
$ε_\diamond$ from \cite[Appendix C]{Kliunchnikov_shorter2023}, where we may replace $p'\approx p$:
\begin{align}
ε_\diamond &= 2(p'(1-r^2\cos^2(φ-θ)) + (1-p')(1-\cos^2θ)) \\
&= 2\left(1-\cos^2θ + p(-r^2\cos^2(φ-θ) +\cos^2θ)\right).
\end{align}
Then the difference is
\begin{align}
    ε_\diamond - (λ-1) &= (1-\cos^2θ)(4-2p) + 2pr^2(\cos^2φ -\cos^2(φ-θ))
\end{align}
The first term is $O(θ^2)$, and the second term is also $O(θ^2)$:
\begin{align}
    & pr^2 (\cos^2φ - \cos^2(φ-θ) \\
    &= \frac{\sin2θ}{\sin2φ}(-\sin(2φ-θ)\sinθ) \\
    &= -\sin2θ\sinθ \frac{\sin(2φ-θ)}{\sin2φ} \\
    &= O(θ^2)
\end{align}
as the fraction is $O(1)$ as $θ\leφ<1$.

\begin{proposition}[{Comparison to Kliuchnikov \emph{et al}., Under-rotation${}\neq\mathbb{1}$}]
\label{prop:notidentity}
Let $θ$ be a target angle, and $u_1=r_1e^{iφ_1},u_2=r_2e^{iφ_2}$ chosen under- and over-rotations. According to the quasi-probability method, let $p_1,p_2$ be the quasi-probabilities of the rotations \eqref{eq:quasiprobabilities 2 rotations} ($p_1+p_2=1$, but potentially negative) and $λ$ the sample complexity. According to Kliuchnikov's mixed approximation \cite{Kliunchnikov_shorter2023}, let $p'_1,p'_2$ ($p_1'+p_2'=1$) be the probabilities of the rotations and $ε_\diamond$ the diamond norm error. We assume that the over- and under-rotations are randomly distributed in their corresponding regions in \cite[Section~3.4]{Kliunchnikov_shorter2023}, which is a valid assumption for $ε_\diamond\llθ^2$. Then we have:
\begin{equation}
    |p_1-p'_1|,|p_2-p'_2|\le O(\sqrt{ε_\diamond}\tanθ)
\end{equation}
and
\begin{equation}
    \frac{|(λ-1)-ε_\diamond|}{ε_\diamond} \le O(\sqrt{ε_\diamond}\tanθ).
\end{equation}
That means, $λ-1\approxε_\diamond$, and as the probabilities are of the order $p'_1\sim p'_2 \sim 0.5$ (under the assumption), also $p_1\approx p_1',p_2\approx p'_2$.
\end{proposition}

Proof:
The notation used in \cite{Kliunchnikov_shorter2023} describes over- and over-rotation by $δ_i = φ_i-θ$, where $δ_1<0<δ_2$.
Kliuchnikov \emph{et al}.~use the same calculation and equation as \cref{eq:channel coeffients}, except that the whole expression has been rotated by $θ$ inside the diamond norm, such that  $θ\mapsto0$ and $φ_i \mapsto φ_i-θ=δ_i$, and using the probabilities $p'_i$ instead of $p_i$. Let these so-called ``rotated'' coefficients be $q_I', q_X',q_Y',q_Z',q_\times'$. For example, both rotated and unrotated cross-terms
\begin{align}
    q_\times &= \sum_ip_ir_i^2\sin2φ_i  -\sin2θ = 0\\
    q_\times' &=
    \sum_ip'_ir_i^2\sin2δ_i = 0
\end{align}
are zero, as that is the condition for choosing $\{p_i\},\{p_i'\}$. The condition $q_\times'=0$ is explicitly written in \cite[Eq.~(109)]{Kliunchnikov_shorter2023}.
Similarly to the sampling overhead, \cref{eq:lambda}, the diamond error $ε_\diamond$ can be calculated from these coefficients \cite[Eq.~(115)]{Kliunchnikov_shorter2023}:
\begin{align}
    λ &= |p_1| +|p_2| + |q_1|+2|q_X| +|q_Z|\\
    \epsdiamond &= |q'_1| + 2|q'_X| +|q'_Z|.
\end{align}

Let us define $\Delta p_i = p_i-p'_i$, and $\Delta p =\Delta p_1 = - \Delta p_2$ (as both sets of (quasi)probabilities sum to 1). With this, $q_\times$ becomes:
\begin{align}
    \sin2θ &= \sum p_ir_i^2\sin2φ_i \\
    &= \cos2θ \sum (p'_i+\Delta p_i) r_i^2\sin2δ_i + \sin2θ\sum(p'_i+\Delta p_i) r_i^2\cos2δ_i \\
    &= \cos2θ\underbrace{\sum p_i'r_i^2\sin2δ_i}_{=q_\times'=0} + \cos2θ \Delta p(r_1^2\sin2δ_1 - r_2^2\sin2δ_2) \\
    &\ + \sin2θ \underbrace{\sum p'_ir_i^2(\cos^2δ_i - \sin^2δ_i)}_{=q_1'+1 - q'_Z} + \sin2θ \Delta p(r_1^2\cos2δ_1 - r_2^2\cos2δ_2)
\end{align}
Solving for $\Delta p$ gives
\begin{align}
    \Delta p
    &=\frac{\sin2θ(1 -q'_1 - 1 + q'_Z)}{\cos2θ(r_1^2\sin2δ_1 - r_2^2\sin2δ_2) + \sin2θ(r_1^2\cos2δ_1 - r_2^2\cos2δ_2)} \\
    &= \tanθ \frac{-q'_1 + q'_Z}{(r_1^2\sin2δ_1 - r_2^2\sin2δ_2) + \tanθ(r_1^2\cos2δ_1 - r_2^2\cos2δ_2)}
\end{align}
We wish to bound $|\Delta p|$ from above. The numerator is bounded by $ε_\diamond$. For the denominator, we assume that $\sin 2δ_i \sim 2δ_i \sim \pm \sqrt{ε_\diamond/2}$ (and accordingly $\cos2δ_i \sim 1$) to highest order in $ε_\diamond$. This comes from the assumption that the over- and under-rotations are uniformly randomly distributed in the area described by Kliuchnikov et al. The area is delineated by $|δ_i|<\arcsin\sqrt{ε_\diamond/2}$. For randomly distributed sequences, they will each sit somewhere closer to the middle rather than at an asymptotically smaller $δ_i$. This is a similar assumption as in our asymptotic derivation (\Cref{sec:asymptotic}), where we use the average probability $p_\text{avg}$ inside the region, i.e.~assumed that the over-rotation approximately sits in the middle. The assumption is valid in the regime where the optimal under-rotation is not the identity ($ε_\diamond\llθ^2$), otherwise the under-rotation wouldn't be randomly distributed. It also matches Kliuchnikov's approximate costing assumptions, where they distribute error equally between over- and under-rotation, meaning they are both similar distance from 0 \cite[Eq.~(22)]{Kliunchnikov_shorter2023}. For the same reason, we can assume $p'_1\sim p_2'\sim 0.5$ in highest order of $ε_\diamond$.  We also have $r_i\ge\sqrt{1-ε_\diamond/2}$. Therefore the denominator can be bounded as
\begin{align}
|(r_1^22δ_1-r_2^22δ_2) + \tanθ(r_1^2-r_2^2)| &\ge 2|r_1^2δ_1-r_2^2δ_2| - |r_1^2-r_2^2|\tanθ \\
&\gtrsim 2\sqrt{ε_\diamond} - ε_\diamond\tanθ \approx \sqrt{ε_\diamond},
\end{align}
where higher order terms in $ε_\diamond$ have been suppressed.
The $ε_\diamond\tanθ$ term is subdominant because $\tanθ < 1/\sqrt{ε_\diamond}$ is always achieved for $θ<\pi/4, ε_\diamond\le1$.
Overall, we have the claimed bound
\begin{align}
    |\Delta p| &\lesssim \sqrt{ε_\diamond}\tanθ.
    \label{eq:delta p}
\end{align}

Now, let us bound the difference between $ε_\diamond$ and $λ-1$.
For this we consider all coefficient terms in turn:
\begin{align}
    q_1 &= -\cos^2θ + \sum_ip_ir_i^2\cos^2(θ+δ_i) \\
    &= -\cos^2θ + \sum_ip_ir_i^2(\cos^2θ\cos^2δ_i + \sin^2θ\sin^2δ_i -\frac{1}{2}\sin2θ\sin2δ_i) \\
    &= -\cos^2θ + \cos^2θ(q'_1+1)+\cos^2θ\sum_i\Delta p_i r_i^2\cos^2δ_i \label{eq:q1rotated correction1}\\
    &\qquad + \sin^2θq'_Z + \sin^2θ\sum_i\Delta p_ir_i^2\sin^2δ_i \label{eq:q1rotated correction2}\\
    &\qquad - \frac{1}{2}\sin2θ \underbrace{q'_\times}_{=0} - \frac{1}{2}\sin2θ\sum_i\Delta p_i r_i^2\sin2δ_i \label{eq:q1rotated correction3}\\
    &= \cos^2θq_1'  +  \sin^2θq_Z'  +O(ε_\diamond\Delta p)
\end{align}
The $q_1$ coefficient is a rotated version of the $q'_1,q'_Z$ coefficients, with correction terms. The positive or negative correction terms are at most $O(ε_\diamond\Delta p)$ in highest order of $ε_\diamond$ by use of the same assumption on $δ_i$,$r_i$ as above. For the first correction term \eqref{eq:q1rotated correction1} we have used
\begin{equation}\cos^2θ\sum\Delta p_ir_i^2\cos^2δ_i \approx \cos^2θ\Delta p (r_1^2-r_2^2) = O(\Delta p\, ε_\diamond).\end{equation}
The second \eqref{eq:q1rotated correction2} and third  \eqref{eq:q1rotated correction3} correction terms are even smaller as $\sin δ_i\approx\sqrt{ε_\diamond}$. On this occasion the $θ$ dependence has been simply bounded as $|\cosθ|,|\sinθ|\le1$.

A similar calculation gives
\begin{align}
    q_Z &= \sin^2θq'_1 + \cos^2θq'_Z + O(ε_\diamond\Delta p) \\
    q_X &= q'_X + O(ε_\diamond\Delta p)
\end{align}
Altogether, we therefore have the maximal difference
\begin{align}
    |(λ-1) - ε_\diamond| &= \left| \underbrace{|p_1| + |p_2|}_{1} + |q_1| + 2|q_X| + |q_Z| -1  - |q'_1| - 2|q'_X| - |q'_Z| \right|\\
    &\le O(ε_\diamond\Delta p) = O(ε_\diamond^{3/2}\tanθ)
\end{align}
as claimed. Note that the probabilities $p_1,p_2$ are positive for sufficiently low $\sqrt{ε_\delta}\tanθ$, since they are close to $p_1', p_2'\sim 0.5$, see \cref{eq:delta p}, therefore we could take $|p_1| + |p_2| = 1$.

\section{Probabilities vs quasi-probabilities for estimating expectation values}
\label{app:prob_vs_quasi_prob}

\begin{figure}
    \centering
    \includegraphics[width=0.7\linewidth]{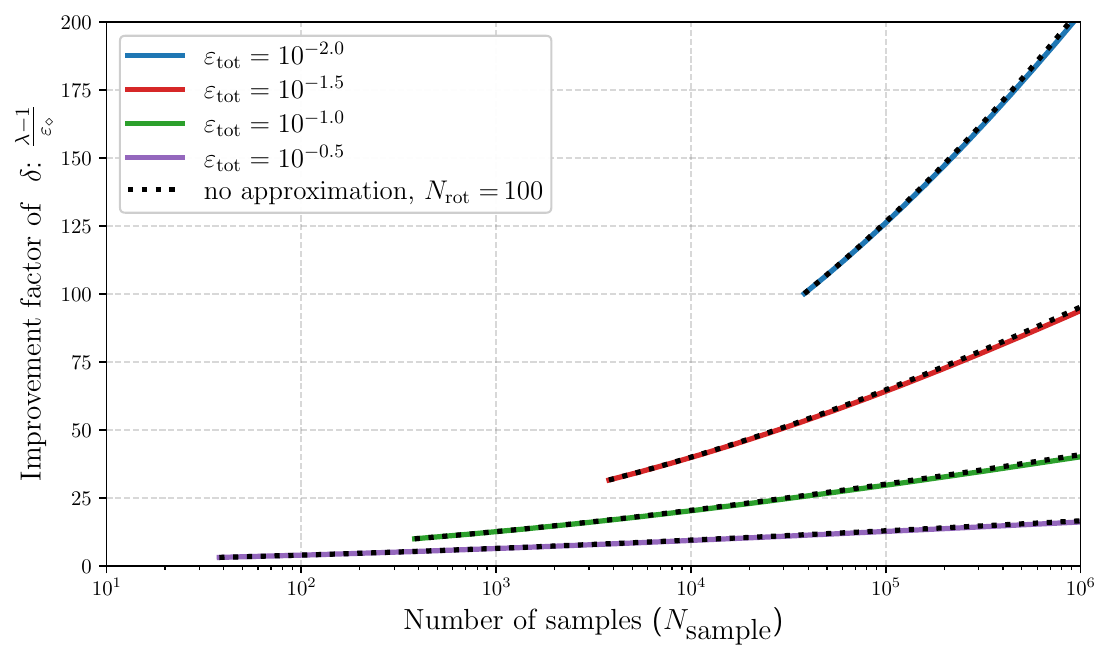}
    \caption{The improvement factor $\frac{λ-1}{\epsdiamond}$ from \cref{eq:app eps lambda-1 advantage} for using quasi-probabilities compared to using probabilities. A larger $λ-1$ than $ε_\diamond$ for same total error $ε_\text{total}$ means lower average T count. The improvement factor varies with the number of samples taken. The curves only begin at $ε_\text{total} = \sqrt{\frac{2\ln(2/\eta)}{N_\text{sample}}}$ as this is the usual sample requirement to achieve such a sample error. Here we have chosen $\eta=0.3$.
    The derivation of \cref{eq:app eps lambda-1 advantage} included the assumption $N_\text{rot}\gg1$ for \eqref{eq:lambda prop 1/N_rot}. For each $ε_\text{total}$, we show the exact ratio $\frac{λ-1}{\epsdiamond}$ with $N_\text{rot}=100$ with black dotted curves. The good correspondence shows that the approximation is well justified.}
    \label{fig:improvement factor eps lambda-1}
\end{figure}

While our numerical results and Propositions 1 and 2 in \Cref{app:comparison kliuchnikov} show a correspondence between T counts with quasi-probability LCCs and probabilistic mixtures when $λ-1$ and $ε_\diamond$ are identified, quasi-probabilities can have a significant advantage. In particular, when we are interested in sampling the expectation value of an observable, such that quasi-probabilities are applicable, they can significantly outperform probabilistic mixtures, as much lower $\lambda - 1$ than $\epsdiamond$ can be sufficient for total target error $ε_\text{total}$. We explain this point in this appendix.

Consider a setting where we wish to learn the expectation value of some observable $O$, with bounded value $||O||\leq1$, taken over some final state $\rho$, i.e. $\Tr(O \rho)$. The present analysis will include both sampling error as well as approximation error (diamond error), treating both (quasi)-probabilistic mixtures and diamond error together in one generalized setting. Therefore, assume we have access to $N_\text{sample}$ shots of an approximate final state $\tilde{ρ}$, where $\tilde{ρ}$ need not be positive semidefinite nor obey $\Tr\tilde{ρ}=1$, due to the quasi-probabilistic nature of the output state. The measurement protocol means that we take $N_\text{sample}$ samples of the random variable $X_i := \text{Meas}_i[O \tilde\rho]$, over which we take the average, $S:= \frac{1}{N_\text{sample} }\sum_i X_i$. The difference between the true observable and our sample average is 
\begin{align}
     \varepsilon_\text{total}&=| \Tr(O\rho) - S| \leq |\Tr(O\rho) - \Tr(O\tilde{\rho})| + |  \Tr(O\tilde{\rho}) -S| \\
     &\le ||\rho -\tilde \rho||_\diamond + \sqrt{2\frac{λ_\text{total}^2}{N_\text{sample}}\ln(2/η)} \\
     &\le N_\text{rot}ε_\diamond + ε_\text{sample}λ^{N_\text{rot}},\ \text{with}\ ε_\text{sample}=\sqrt{\frac{2\ln(2/η)}{N_\text{sample}}},\ λ_\text{total}=|\tilde{ρ}|_*\label{eq:eps_tot}
\end{align}
Here, the diamond norm accuracy of a single rotation is $ε_\diamond$, as used previously in the paper. The second term is the sampling error given by Hoeffding's inequality,
where $1-\eta$ is the probability that the sampling error is within the stated quantity, and $λ_\text{total}$ is the normalisation of the total probability distribution,
i.e.~the nuclear norm of $\tilde{ρ}$. 

In our approach using proper probabilities, $ρ$ is positive definite with normalisation $λ=1$. In our approach using quasi-probabilities, we have $ε_\diamond=0$ as the LCC exactly gives the desired rotation, but have a sampling overhead $λ>1$ per rotation. The required $ε_\diamond$ or $λ-1$ for both methods can be compared from \cref{eq:eps_tot} in the large $N_\text{rot}$ limit, where $x^{1/N_\text{rot}}-1 \approx \ln(x)/N_\text{rot}$:
\begin{align}
    ε_\diamond &= (ε_\text{total}-ε_\text{sample})\frac{1}{N_\text{rot}}, \\
    λ-1 &= \left(\frac{ε_\text{total}}{ε_\text{sample}}\right)^\frac{1}{N_\text{rot}} -1 \approx \left(\lnε_\text{total} - \lnε_\text{sample}\right)\frac{1}{N_\text{rot}}. \label{eq:lambda prop 1/N_rot}
\end{align}
The approximation for large $N_\text{rot}$ is valid whenever $λ-1\ll1$; this is exactly the regime we are interested in in this paper as it prevents an exponential blow-up of sampling overhead.
The scaling behavior of both $ε_\diamond$ and $λ-1$ is $\propto 1 / N_\text{rot}$. The prefactor in $λ-1$ is larger, as the logarithm $\lnε$ is very steep for small $ε>0$, amplifying the difference $ε_\text{total}-ε_\text{sample}$. The exact advantage of $λ-1$ depends on the chosen $ε_\text{total}$ and parameters $N_\text{sample}, \eta$ for $ε_\text{sample}$, but typically is substantial. For a quantitative comparison we define the ratio $x:= ε_\text{sample} / ε_\text{total},\ 0<x<1$ and compute the improvement factor
\begin{equation}
    \frac{λ-1}{ε_\diamond} \approx \frac{\lnε_\text{total}-\lnε_\text{sample}}{ε_\text{total}-ε_\text{sample}} = \frac{1}{ε_\text{total}}\frac{1 - ε_\text{sample}/ε_\text{total}}{\ln(ε_\text{total}/ε_\text{sample})}{1 - ε_\text{sample}/ε_\text{total}} = \frac{1}{ε_\text{total}}\frac{\ln(1/x)}{1-x}\ge\frac{1}{ε_\text{total}}\frac{2}{x+1}\ge\frac{1}{ε_\text{total}}. \label{eq:app eps lambda-1 advantage}
\end{equation}
We plot the improvement factor \cref{eq:app eps lambda-1 advantage} in \Cref{fig:improvement factor eps lambda-1}. Even the lower bound above gives a significant advantage to quasi-probabilities as $λ-1$ can be bigger than $ε_\diamond$. Further, the ratio $x\propto 1/\sqrt{N_\text{sample}}$ can be reduced by increasing the sample complexity, as long as $\ln(N_\text{sample})/N_\text{rot}\ll1$ to ensure the assumption $λ-1 \ll 1$ stays valid \eqref{eq:lambda prop 1/N_rot}.

As seen in \Cref{fig:improvement factor eps lambda-1}, with the same number of samples, required $δ=ε_\diamond$ can be 2 orders of magnitude smaller than required $δ=λ-1$. Since $T$-count dependence is nearly identical on $δ$ in both methods (see Propositions~\ref{prop:identity} and \ref{prop:notidentity}), this gives an advantage to quasi-probabilities. Especially in the small angle case $δ\gtrsimθ^2$, the $T$-count depends on $δ$ as $1/δ$ (see derivation leading to \cref{eq:asymptotic delta>>theta^2}), such that quasi-probabilities give an average T count \emph{2 orders of magnitude lower} than proper probabilities. A similarly dramatic reduction can also be seen in our Trotter results, comparing the T counts of \Cref{fig:quasi_prob_comparison} with \Cref{fig:prob_comparison}.

\section{Comparison of Trotterization results to alternative approaches}
\label{app:comparison trotter results}

In \Cref{sec:trotter_results} we have seen significant reductions in the cost of implementing Trotterized dynamics by using an angle-dependent formula, which much more accurately captures the cost of mixed Clifford+T synthesis in the small-angle regime, specifically when $\delta \gtrsim \theta^2$. There are other quantum simulation approaches in the literature which also aim to reduce the cost of Hamiltonian simulation by taking advantage of savings for small-angle rotations. Here, we briefly compare our results to existing approaches.

We noted in \Cref{sec:trotter} that we obtain results reminiscent of the qDRIFT algorithm.  Indeed, the result of \cref{eq:qdrifty} matches the expression for the gate complexity of qDRIFT.   However, there are several subtleties that show our results are actually a dramatic improvement.  Firstly, our \cref{eq:qdrifty} counts the average number of T gates used, whereas the standard qDRIFT formula counts the number of Pauli rotations and incurs an additional synthesis overhead to convert this into a T gate count.  Since synthesis overheads are typically 1--2 orders of magnitude, this gives an estimate of the improvement with respect to using qDRIFT. In the worst case, the approach described here will return to the cost of Trotterization using existing Clifford+T formulas, while qDRIFT can be significantly more expensive.

G\"{u}nther \emph{et al.} studied a partially randomized time evolution procedure, where small Hamiltonian terms are treated with qDRIFT, while large terms are treated with deterministic product formulas \cite{Gunther2025}. This approach is closer to what is achieved by the Clifford+T synthesis results in this work. In particular, from \Cref{fig:av_t_count_per_rot} we see clear small-angle and large-angle regimes of behavior. The small-angle behavior is reminiscent of qDRIFT, as discussed above, while the large-angle regime matches existing angle-independent Clifford+T formulas. Compared to the partially randomized approach, our results do not require introducing any separation between small and large Hamiltonian terms, nor do they require introducing quantum simulation methods beyond deterministic product formulas. Rather, we show that the small-angle and large-angle behavior emerges automatically by compiling a deterministic product formula to a Clifford+T gate set using a mixed approximation. The approach of Ref.~\cite{Gunther2025} has a benefit that it is formulated such that Hamming weight phasing can be applied.

We additionally introduce a quasi-probabilistic method that may trade increased sample complexity for shorter circuits, which has parallels with the results of Wan \emph{et. al.} \cite{Wan2022}, though again here we benefit from having compiled down T gates without any additional overhead.

Our method showed that the average T count converges to a constant in the small Trotter step size limit. Similar observations have been made in other settings: Wan \emph{et al.} \cite{Wan2022} show that Trotterization error can be reduced to zero for a randomized LCU approach to statistical phase estimation; Granet \textit{et al.} noted vanishing Trotter error in their method, calculating expectation values of time-evolved observables by sampling circuits \cite{Granet2024}; and Luthra \textit{et al.} \cite{Luthra2025} demonstrated vanishing Trotter error using a quasi-probabilistic decomposition into gates from the Clifford hierarchy. A similar result was shown for TE-PAI (see below) \cite{Kiumi2025}. As such, this appears to be a generic feature of randomized approaches. Though, our work establishes that this holds when directly compiling into Clifford+T gates, which can be regarded as the most established scalable approach to quantum computing, and is demonstrated for both probabilities and quasi-probabilities.

Recently, Kiumi and Koczor proposed TE-PAI \cite{Kiumi2025}, which builds on the Probabilistic Angle Interpolation (PAI) technique \cite{Koczor2024}. In this method, rotation gates are performed as a quasi-probability mixture over rotation angles $\{0, \pm \Delta, \pi/2 \}$, where typically $\Delta$ is chosen such that $|\theta| \le \Delta \le \pi/2$. Ref.~\cite{Kiumi2025} primarily focuses on the case where $\Delta$ is chosen in the Clifford hierarchy, i.e. $\Delta = \pi/2^l$ for $l \in \mathbb{N}$. Given such an angle, they consider implementing $R_z(\Delta)$ by injecting the appropriate resource state using a catalyst tower and then using a repeat-until-success (RUS) procedure. In Section~4.1 of Ref.~\cite{Kiumi2025} the authors also discuss the possibility of approximately implementing $R_z(\Delta)$ using Clifford+T synthesis. TE-PAI achieves savings in the small-angle limit because the identity will be sampled most of the time. Compared to our results, the approach based on catalyst towers is very different in terms of the underlying circuits, requiring significantly more ancilla space. Implementing $R_z(\Delta)$ as a Clifford+T sequence results in a more similar but nonetheless different approach for the following reasons. Firstly, we present results both using quasi-probability and strict probability mixtures. This is valuable because there are many quantum circuits where the use of quasi-probability mixtures is not possible, while probability mixtures can be used in the compilation of any quantum circuit. Secondly, rather than introducing an intermediate gate $R_z(\Delta)$ and then performing subsequent synthesis (to some acceptable precision, $\varepsilon$, which must be analyzed) we directly perform compilation to a Clifford+T gate set, and analyze this case. Thirdly, we do not enforce that the under-rotation is the identity. This is important because more accurate results can be obtained for large-angle (or small $\delta$) rotations by also allowing the under-rotation to be a Clifford+T unitary, and we show in this case that we return to mixed diagonal and mixed fallback results; that is, our results achieve both the small-angle benefit and also achieve state-of-the-art results for large angles. The PAI approach has the benefit that $\Delta$ can be chosen arbitrarily; this can be useful if one wishes to choose rotations in the Clifford hierarchy, allowing alternative methods to perform non-Clifford logic, as discussed above.


\begin{thebibliography}{62}%
\makeatletter
\providecommand \@ifxundefined [1]{%
 \@ifx{#1\undefined}
}%
\providecommand \@ifnum [1]{%
 \ifnum #1\expandafter \@firstoftwo
 \else \expandafter \@secondoftwo
 \fi
}%
\providecommand \@ifx [1]{%
 \ifx #1\expandafter \@firstoftwo
 \else \expandafter \@secondoftwo
 \fi
}%
\providecommand \natexlab [1]{#1}%
\providecommand \enquote  [1]{``#1''}%
\providecommand \bibnamefont  [1]{#1}%
\providecommand \bibfnamefont [1]{#1}%
\providecommand \citenamefont [1]{#1}%
\providecommand \href@noop [0]{\@secondoftwo}%
\providecommand \href [0]{\begingroup \@sanitize@url \@href}%
\providecommand \@href[1]{\@@startlink{#1}\@@href}%
\providecommand \@@href[1]{\endgroup#1\@@endlink}%
\providecommand \@sanitize@url [0]{\catcode `\\12\catcode `\$12\catcode `\&12\catcode `\#12\catcode `\^12\catcode `\_12\catcode `\%12\relax}%
\providecommand \@@startlink[1]{}%
\providecommand \@@endlink[0]{}%
\providecommand \url  [0]{\begingroup\@sanitize@url \@url }%
\providecommand \@url [1]{\endgroup\@href {#1}{\urlprefix }}%
\providecommand \urlprefix  [0]{URL }%
\providecommand \Eprint [0]{\href }%
\providecommand \doibase [0]{https://doi.org/}%
\providecommand \selectlanguage [0]{\@gobble}%
\providecommand \bibinfo  [0]{\@secondoftwo}%
\providecommand \bibfield  [0]{\@secondoftwo}%
\providecommand \translation [1]{[#1]}%
\providecommand \BibitemOpen [0]{}%
\providecommand \bibitemStop [0]{}%
\providecommand \bibitemNoStop [0]{.\EOS\space}%
\providecommand \EOS [0]{\spacefactor3000\relax}%
\providecommand \BibitemShut  [1]{\csname bibitem#1\endcsname}%
\let\auto@bib@innerbib\@empty
\bibitem [{\citenamefont {Horsman}\ \emph {et~al.}(2012)\citenamefont {Horsman}, \citenamefont {Fowler}, \citenamefont {Devitt},\ and\ \citenamefont {Meter}}]{horsman2012surface}%
  \BibitemOpen
  \bibfield  {author} {\bibinfo {author} {\bibfnamefont {D.}~\bibnamefont {Horsman}}, \bibinfo {author} {\bibfnamefont {A.~G.}\ \bibnamefont {Fowler}}, \bibinfo {author} {\bibfnamefont {S.}~\bibnamefont {Devitt}},\ and\ \bibinfo {author} {\bibfnamefont {R.~V.}\ \bibnamefont {Meter}},\ }\bibfield  {title} {\bibinfo {title} {Surface code quantum computing by lattice surgery},\ }\href@noop {} {\bibfield  {journal} {\bibinfo  {journal} {New Journal of Physics}\ }\textbf {\bibinfo {volume} {14}},\ \bibinfo {pages} {123011} (\bibinfo {year} {2012})}\BibitemShut {NoStop}%
\bibitem [{\citenamefont {Fowler}\ and\ \citenamefont {Gidney}(2018)}]{fowler2018low}%
  \BibitemOpen
  \bibfield  {author} {\bibinfo {author} {\bibfnamefont {A.~G.}\ \bibnamefont {Fowler}}\ and\ \bibinfo {author} {\bibfnamefont {C.}~\bibnamefont {Gidney}},\ }\bibfield  {title} {\bibinfo {title} {Low overhead quantum computation using lattice surgery},\ }\href@noop {} {\bibfield  {journal} {\bibinfo  {journal} {arXiv preprint arXiv:1808.06709}\ } (\bibinfo {year} {2018})}\BibitemShut {NoStop}%
\bibitem [{\citenamefont {Litinski}(2019)}]{litinski2019game}%
  \BibitemOpen
  \bibfield  {author} {\bibinfo {author} {\bibfnamefont {D.}~\bibnamefont {Litinski}},\ }\bibfield  {title} {\bibinfo {title} {A game of surface codes: Large-scale quantum computing with lattice surgery},\ }\href@noop {} {\bibfield  {journal} {\bibinfo  {journal} {Quantum}\ }\textbf {\bibinfo {volume} {3}},\ \bibinfo {pages} {128} (\bibinfo {year} {2019})}\BibitemShut {NoStop}%
\bibitem [{\citenamefont {Chamberland}\ and\ \citenamefont {Campbell}(2022)}]{chamberland2022circuit}%
  \BibitemOpen
  \bibfield  {author} {\bibinfo {author} {\bibfnamefont {C.}~\bibnamefont {Chamberland}}\ and\ \bibinfo {author} {\bibfnamefont {E.~T.}\ \bibnamefont {Campbell}},\ }\bibfield  {title} {\bibinfo {title} {Circuit-level protocol and analysis for twist-based lattice surgery},\ }\href@noop {} {\bibfield  {journal} {\bibinfo  {journal} {Physical Review Research}\ }\textbf {\bibinfo {volume} {4}},\ \bibinfo {pages} {023090} (\bibinfo {year} {2022})}\BibitemShut {NoStop}%
\bibitem [{\citenamefont {Gottesman}\ and\ \citenamefont {Chuang}(1999)}]{gottesman1999demonstrating}%
  \BibitemOpen
  \bibfield  {author} {\bibinfo {author} {\bibfnamefont {D.}~\bibnamefont {Gottesman}}\ and\ \bibinfo {author} {\bibfnamefont {I.~L.}\ \bibnamefont {Chuang}},\ }\bibfield  {title} {\bibinfo {title} {Demonstrating the viability of universal quantum computation using teleportation and single-qubit operations},\ }\href@noop {} {\bibfield  {journal} {\bibinfo  {journal} {Nature}\ }\textbf {\bibinfo {volume} {402}},\ \bibinfo {pages} {390} (\bibinfo {year} {1999})}\BibitemShut {NoStop}%
\bibitem [{\citenamefont {Bravyi}\ and\ \citenamefont {Kitaev}(2005)}]{bravyi2005universal}%
  \BibitemOpen
  \bibfield  {author} {\bibinfo {author} {\bibfnamefont {S.}~\bibnamefont {Bravyi}}\ and\ \bibinfo {author} {\bibfnamefont {A.}~\bibnamefont {Kitaev}},\ }\bibfield  {title} {\bibinfo {title} {Universal quantum computation with ideal clifford gates and noisy ancillas},\ }\href@noop {} {\bibfield  {journal} {\bibinfo  {journal} {Physical Review A—Atomic, Molecular, and Optical Physics}\ }\textbf {\bibinfo {volume} {71}},\ \bibinfo {pages} {022316} (\bibinfo {year} {2005})}\BibitemShut {NoStop}%
\bibitem [{\citenamefont {Bravyi}\ and\ \citenamefont {Haah}(2012)}]{bravyi2012magic}%
  \BibitemOpen
  \bibfield  {author} {\bibinfo {author} {\bibfnamefont {S.}~\bibnamefont {Bravyi}}\ and\ \bibinfo {author} {\bibfnamefont {J.}~\bibnamefont {Haah}},\ }\bibfield  {title} {\bibinfo {title} {Magic-state distillation with low overhead},\ }\href@noop {} {\bibfield  {journal} {\bibinfo  {journal} {Physical Review A—Atomic, Molecular, and Optical Physics}\ }\textbf {\bibinfo {volume} {86}},\ \bibinfo {pages} {052329} (\bibinfo {year} {2012})}\BibitemShut {NoStop}%
\bibitem [{\citenamefont {Chamberland}\ and\ \citenamefont {Noh}(2020)}]{chamberland2020very}%
  \BibitemOpen
  \bibfield  {author} {\bibinfo {author} {\bibfnamefont {C.}~\bibnamefont {Chamberland}}\ and\ \bibinfo {author} {\bibfnamefont {K.}~\bibnamefont {Noh}},\ }\bibfield  {title} {\bibinfo {title} {Very low overhead fault-tolerant magic state preparation using redundant ancilla encoding and flag qubits},\ }\href@noop {} {\bibfield  {journal} {\bibinfo  {journal} {npj Quantum Information}\ }\textbf {\bibinfo {volume} {6}},\ \bibinfo {pages} {91} (\bibinfo {year} {2020})}\BibitemShut {NoStop}%
\bibitem [{\citenamefont {Gidney}\ \emph {et~al.}(2024)\citenamefont {Gidney}, \citenamefont {Shutty},\ and\ \citenamefont {Jones}}]{gidney2024magic}%
  \BibitemOpen
  \bibfield  {author} {\bibinfo {author} {\bibfnamefont {C.}~\bibnamefont {Gidney}}, \bibinfo {author} {\bibfnamefont {N.}~\bibnamefont {Shutty}},\ and\ \bibinfo {author} {\bibfnamefont {C.}~\bibnamefont {Jones}},\ }\bibfield  {title} {\bibinfo {title} {Magic state cultivation: growing t states as cheap as cnot gates},\ }\href@noop {} {\bibfield  {journal} {\bibinfo  {journal} {arXiv preprint arXiv:2409.17595}\ } (\bibinfo {year} {2024})}\BibitemShut {NoStop}%
\bibitem [{\citenamefont {Harrow}\ \emph {et~al.}(2002)\citenamefont {Harrow}, \citenamefont {Recht},\ and\ \citenamefont {Chuang}}]{harrow2002efficient}%
  \BibitemOpen
  \bibfield  {author} {\bibinfo {author} {\bibfnamefont {A.~W.}\ \bibnamefont {Harrow}}, \bibinfo {author} {\bibfnamefont {B.}~\bibnamefont {Recht}},\ and\ \bibinfo {author} {\bibfnamefont {I.~L.}\ \bibnamefont {Chuang}},\ }\bibfield  {title} {\bibinfo {title} {Efficient discrete approximations of quantum gates},\ }\href@noop {} {\bibfield  {journal} {\bibinfo  {journal} {Journal of Mathematical Physics}\ }\textbf {\bibinfo {volume} {43}},\ \bibinfo {pages} {4445} (\bibinfo {year} {2002})}\BibitemShut {NoStop}%
\bibitem [{\citenamefont {Ross}\ and\ \citenamefont {Selinger}(2016)}]{ross2016optimal}%
  \BibitemOpen
  \bibfield  {author} {\bibinfo {author} {\bibfnamefont {N.~J.}\ \bibnamefont {Ross}}\ and\ \bibinfo {author} {\bibfnamefont {P.}~\bibnamefont {Selinger}},\ }\bibfield  {title} {\bibinfo {title} {Optimal ancilla-free clifford+ t approximation of z-rotations.},\ }\href@noop {} {\bibfield  {journal} {\bibinfo  {journal} {Quantum Inf. Comput.}\ }\textbf {\bibinfo {volume} {16}},\ \bibinfo {pages} {901} (\bibinfo {year} {2016})}\BibitemShut {NoStop}%
\bibitem [{\citenamefont {Campbell}(2017)}]{Campbell2016}%
  \BibitemOpen
  \bibfield  {author} {\bibinfo {author} {\bibfnamefont {E.}~\bibnamefont {Campbell}},\ }\bibfield  {title} {\bibinfo {title} {Shorter gate sequences for quantum computing by mixing unitaries},\ }\href {https://doi.org/10.1103/PhysRevA.95.042306} {\bibfield  {journal} {\bibinfo  {journal} {Phys. Rev. A}\ }\textbf {\bibinfo {volume} {95}},\ \bibinfo {pages} {042306} (\bibinfo {year} {2017})}\BibitemShut {NoStop}%
\bibitem [{\citenamefont {Hastings}(2017)}]{Hastings2017}%
  \BibitemOpen
  \bibfield  {author} {\bibinfo {author} {\bibfnamefont {M.~B.}\ \bibnamefont {Hastings}},\ }\bibfield  {title} {\bibinfo {title} {Turning gate synthesis errors into incoherent errors},\ }\href@noop {} {\bibfield  {journal} {\bibinfo  {journal} {Quantum Info. Comput.}\ }\textbf {\bibinfo {volume} {17}},\ \bibinfo {pages} {488–494} (\bibinfo {year} {2017})}\BibitemShut {NoStop}%
\bibitem [{\citenamefont {Lloyd}(1996)}]{lloyd1996universal}%
  \BibitemOpen
  \bibfield  {author} {\bibinfo {author} {\bibfnamefont {S.}~\bibnamefont {Lloyd}},\ }\bibfield  {title} {\bibinfo {title} {Universal quantum simulators},\ }\href@noop {} {\bibfield  {journal} {\bibinfo  {journal} {Science}\ }\textbf {\bibinfo {volume} {273}},\ \bibinfo {pages} {1073} (\bibinfo {year} {1996})}\BibitemShut {NoStop}%
\bibitem [{\citenamefont {Childs}\ \emph {et~al.}(2021)\citenamefont {Childs}, \citenamefont {Su}, \citenamefont {Tran}, \citenamefont {Wiebe},\ and\ \citenamefont {Zhu}}]{childs2021theory}%
  \BibitemOpen
  \bibfield  {author} {\bibinfo {author} {\bibfnamefont {A.~M.}\ \bibnamefont {Childs}}, \bibinfo {author} {\bibfnamefont {Y.}~\bibnamefont {Su}}, \bibinfo {author} {\bibfnamefont {M.~C.}\ \bibnamefont {Tran}}, \bibinfo {author} {\bibfnamefont {N.}~\bibnamefont {Wiebe}},\ and\ \bibinfo {author} {\bibfnamefont {S.}~\bibnamefont {Zhu}},\ }\bibfield  {title} {\bibinfo {title} {Theory of trotter error with commutator scaling},\ }\href@noop {} {\bibfield  {journal} {\bibinfo  {journal} {Physical Review X}\ }\textbf {\bibinfo {volume} {11}},\ \bibinfo {pages} {011020} (\bibinfo {year} {2021})}\BibitemShut {NoStop}%
\bibitem [{\citenamefont {Reiher}\ \emph {et~al.}(2017)\citenamefont {Reiher}, \citenamefont {Wiebe}, \citenamefont {Svore}, \citenamefont {Wecker},\ and\ \citenamefont {Troyer}}]{reiher2017}%
  \BibitemOpen
  \bibfield  {author} {\bibinfo {author} {\bibfnamefont {M.}~\bibnamefont {Reiher}}, \bibinfo {author} {\bibfnamefont {N.}~\bibnamefont {Wiebe}}, \bibinfo {author} {\bibfnamefont {K.~M.}\ \bibnamefont {Svore}}, \bibinfo {author} {\bibfnamefont {D.}~\bibnamefont {Wecker}},\ and\ \bibinfo {author} {\bibfnamefont {M.}~\bibnamefont {Troyer}},\ }\bibfield  {title} {\bibinfo {title} {Elucidating reaction mechanisms on quantum computers},\ }\href {https://doi.org/10.1073/pnas.1619152114} {\bibfield  {journal} {\bibinfo  {journal} {Proceedings of the National Academy of Sciences}\ }\textbf {\bibinfo {volume} {114}},\ \bibinfo {pages} {7555} (\bibinfo {year} {2017})},\ \Eprint {https://arxiv.org/abs/https://www.pnas.org/doi/pdf/10.1073/pnas.1619152114} {https://www.pnas.org/doi/pdf/10.1073/pnas.1619152114} \BibitemShut {NoStop}%
\bibitem [{\citenamefont {Lee}\ \emph {et~al.}(2021)\citenamefont {Lee}, \citenamefont {Berry}, \citenamefont {Gidney}, \citenamefont {Huggins}, \citenamefont {McClean}, \citenamefont {Wiebe},\ and\ \citenamefont {Babbush}}]{lee2021}%
  \BibitemOpen
  \bibfield  {author} {\bibinfo {author} {\bibfnamefont {J.}~\bibnamefont {Lee}}, \bibinfo {author} {\bibfnamefont {D.~W.}\ \bibnamefont {Berry}}, \bibinfo {author} {\bibfnamefont {C.}~\bibnamefont {Gidney}}, \bibinfo {author} {\bibfnamefont {W.~J.}\ \bibnamefont {Huggins}}, \bibinfo {author} {\bibfnamefont {J.~R.}\ \bibnamefont {McClean}}, \bibinfo {author} {\bibfnamefont {N.}~\bibnamefont {Wiebe}},\ and\ \bibinfo {author} {\bibfnamefont {R.}~\bibnamefont {Babbush}},\ }\bibfield  {title} {\bibinfo {title} {Even more efficient quantum computations of chemistry through tensor hypercontraction},\ }\href {https://doi.org/10.1103/PRXQuantum.2.030305} {\bibfield  {journal} {\bibinfo  {journal} {PRX Quantum}\ }\textbf {\bibinfo {volume} {2}},\ \bibinfo {pages} {030305} (\bibinfo {year} {2021})}\BibitemShut {NoStop}%
\bibitem [{\citenamefont {Blunt}\ \emph {et~al.}(2022)\citenamefont {Blunt}, \citenamefont {Camps}, \citenamefont {Crawford}, \citenamefont {Izsák}, \citenamefont {Leontica}, \citenamefont {Mirani}, \citenamefont {Moylett}, \citenamefont {Scivier}, \citenamefont {S{\"u}nderhauf}, \citenamefont {Schopf}, \citenamefont {Taylor},\ and\ \citenamefont {Holzmann}}]{blunt_2022}%
  \BibitemOpen
  \bibfield  {author} {\bibinfo {author} {\bibfnamefont {N.~S.}\ \bibnamefont {Blunt}}, \bibinfo {author} {\bibfnamefont {J.}~\bibnamefont {Camps}}, \bibinfo {author} {\bibfnamefont {O.}~\bibnamefont {Crawford}}, \bibinfo {author} {\bibfnamefont {R.}~\bibnamefont {Izsák}}, \bibinfo {author} {\bibfnamefont {S.}~\bibnamefont {Leontica}}, \bibinfo {author} {\bibfnamefont {A.}~\bibnamefont {Mirani}}, \bibinfo {author} {\bibfnamefont {A.~E.}\ \bibnamefont {Moylett}}, \bibinfo {author} {\bibfnamefont {S.~A.}\ \bibnamefont {Scivier}}, \bibinfo {author} {\bibfnamefont {C.}~\bibnamefont {S{\"u}nderhauf}}, \bibinfo {author} {\bibfnamefont {P.}~\bibnamefont {Schopf}}, \bibinfo {author} {\bibfnamefont {J.~M.}\ \bibnamefont {Taylor}},\ and\ \bibinfo {author} {\bibfnamefont {N.}~\bibnamefont {Holzmann}},\ }\bibfield  {title} {\bibinfo {title} {Perspective on the current state-of-the-art of quantum computing for drug discovery applications},\ }\href {https://doi.org/10.1021/acs.jctc.2c00574} {\bibfield  {journal} {\bibinfo
   {journal} {Journal of Chemical Theory and Computation}\ }\textbf {\bibinfo {volume} {18}},\ \bibinfo {pages} {7001} (\bibinfo {year} {2022})},\ \bibinfo {note} {pMID: 36355616},\ \Eprint {https://arxiv.org/abs/https://doi.org/10.1021/acs.jctc.2c00574} {https://doi.org/10.1021/acs.jctc.2c00574} \BibitemShut {NoStop}%
\bibitem [{\citenamefont {Kliuchnikov}\ \emph {et~al.}(2023)\citenamefont {Kliuchnikov}, \citenamefont {Lauter}, \citenamefont {Minko}, \citenamefont {Paetznick},\ and\ \citenamefont {Petit}}]{Kliunchnikov_shorter2023}%
  \BibitemOpen
  \bibfield  {author} {\bibinfo {author} {\bibfnamefont {V.}~\bibnamefont {Kliuchnikov}}, \bibinfo {author} {\bibfnamefont {K.}~\bibnamefont {Lauter}}, \bibinfo {author} {\bibfnamefont {R.}~\bibnamefont {Minko}}, \bibinfo {author} {\bibfnamefont {A.}~\bibnamefont {Paetznick}},\ and\ \bibinfo {author} {\bibfnamefont {C.}~\bibnamefont {Petit}},\ }\bibfield  {title} {\bibinfo {title} {Shorter quantum circuits via single-qubit gate approximation},\ }\href {https://doi.org/10.22331/q-2023-12-18-1208} {\bibfield  {journal} {\bibinfo  {journal} {{Quantum}}\ }\textbf {\bibinfo {volume} {7}},\ \bibinfo {pages} {1208} (\bibinfo {year} {2023})}\BibitemShut {NoStop}%
\bibitem [{\citenamefont {Koczor}(2024)}]{Koczor2024}%
  \BibitemOpen
  \bibfield  {author} {\bibinfo {author} {\bibfnamefont {B.}~\bibnamefont {Koczor}},\ }\bibfield  {title} {\bibinfo {title} {Sparse probabilistic synthesis of quantum operations},\ }\href {https://doi.org/10.1103/PRXQuantum.5.040352} {\bibfield  {journal} {\bibinfo  {journal} {PRX Quantum}\ }\textbf {\bibinfo {volume} {5}},\ \bibinfo {pages} {040352} (\bibinfo {year} {2024})}\BibitemShut {NoStop}%
\bibitem [{\citenamefont {Khitrin}\ \emph {et~al.}(2026)\citenamefont {Khitrin}, \citenamefont {Brown},\ and\ \citenamefont {Anand}}]{khitrin2026unbiased}%
  \BibitemOpen
  \bibfield  {author} {\bibinfo {author} {\bibfnamefont {D.}~\bibnamefont {Khitrin}}, \bibinfo {author} {\bibfnamefont {K.~R.}\ \bibnamefont {Brown}},\ and\ \bibinfo {author} {\bibfnamefont {A.}~\bibnamefont {Anand}},\ }\bibfield  {title} {\bibinfo {title} {Unbiased observable estimation with approximate channels in fault-tolerant quantum computation},\ }\href@noop {} {\bibfield  {journal} {\bibinfo  {journal} {Quantum Science and Technology}\ }\textbf {\bibinfo {volume} {11}},\ \bibinfo {pages} {015034} (\bibinfo {year} {2026})}\BibitemShut {NoStop}%
\bibitem [{\citenamefont {Su}\ \emph {et~al.}(2021)\citenamefont {Su}, \citenamefont {Huang},\ and\ \citenamefont {Campbell}}]{Su2021}%
  \BibitemOpen
  \bibfield  {author} {\bibinfo {author} {\bibfnamefont {Y.}~\bibnamefont {Su}}, \bibinfo {author} {\bibfnamefont {H.-Y.}\ \bibnamefont {Huang}},\ and\ \bibinfo {author} {\bibfnamefont {E.~T.}\ \bibnamefont {Campbell}},\ }\bibfield  {title} {\bibinfo {title} {Nearly tight {T}rotterization of interacting electrons},\ }\href {https://doi.org/10.22331/q-2021-07-05-495} {\bibfield  {journal} {\bibinfo  {journal} {{Quantum}}\ }\textbf {\bibinfo {volume} {5}},\ \bibinfo {pages} {495} (\bibinfo {year} {2021})}\BibitemShut {NoStop}%
\bibitem [{\citenamefont {Zhao}\ \emph {et~al.}(2022)\citenamefont {Zhao}, \citenamefont {Zhou}, \citenamefont {Shaw}, \citenamefont {Li},\ and\ \citenamefont {Childs}}]{Zhao2022}%
  \BibitemOpen
  \bibfield  {author} {\bibinfo {author} {\bibfnamefont {Q.}~\bibnamefont {Zhao}}, \bibinfo {author} {\bibfnamefont {Y.}~\bibnamefont {Zhou}}, \bibinfo {author} {\bibfnamefont {A.~F.}\ \bibnamefont {Shaw}}, \bibinfo {author} {\bibfnamefont {T.}~\bibnamefont {Li}},\ and\ \bibinfo {author} {\bibfnamefont {A.~M.}\ \bibnamefont {Childs}},\ }\bibfield  {title} {\bibinfo {title} {Hamiltonian simulation with random inputs},\ }\href {https://doi.org/10.1103/PhysRevLett.129.270502} {\bibfield  {journal} {\bibinfo  {journal} {Phys. Rev. Lett.}\ }\textbf {\bibinfo {volume} {129}},\ \bibinfo {pages} {270502} (\bibinfo {year} {2022})}\BibitemShut {NoStop}%
\bibitem [{\citenamefont {Yi}\ and\ \citenamefont {Crosson}(2022)}]{Yi2022}%
  \BibitemOpen
  \bibfield  {author} {\bibinfo {author} {\bibfnamefont {C.}~\bibnamefont {Yi}}\ and\ \bibinfo {author} {\bibfnamefont {E.}~\bibnamefont {Crosson}},\ }\bibfield  {title} {\bibinfo {title} {Spectral analysis of product formulas for quantum simulation},\ }\href {https://doi.org/10.1038/s41534-022-00548-w} {\bibfield  {journal} {\bibinfo  {journal} {npj Quantum Inf.}\ }\textbf {\bibinfo {volume} {8}},\ \bibinfo {pages} {37} (\bibinfo {year} {2022})}\BibitemShut {NoStop}%
\bibitem [{\citenamefont {Mizuta}\ and\ \citenamefont {Kuwahara}(2025)}]{Mizuta2025}%
  \BibitemOpen
  \bibfield  {author} {\bibinfo {author} {\bibfnamefont {K.}~\bibnamefont {Mizuta}}\ and\ \bibinfo {author} {\bibfnamefont {T.}~\bibnamefont {Kuwahara}},\ }\bibfield  {title} {\bibinfo {title} {Trotterization is substantially efficient for low-energy states},\ }\href {https://doi.org/10.1103/q87n-5xhz} {\bibfield  {journal} {\bibinfo  {journal} {Phys. Rev. Lett.}\ }\textbf {\bibinfo {volume} {135}},\ \bibinfo {pages} {130602} (\bibinfo {year} {2025})}\BibitemShut {NoStop}%
\bibitem [{\citenamefont {Blunt}\ \emph {et~al.}(2025)\citenamefont {Blunt}, \citenamefont {Ivanov},\ and\ \citenamefont {Bay-Smidt}}]{Blunt2025}%
  \BibitemOpen
  \bibfield  {author} {\bibinfo {author} {\bibfnamefont {N.~S.}\ \bibnamefont {Blunt}}, \bibinfo {author} {\bibfnamefont {A.~V.}\ \bibnamefont {Ivanov}},\ and\ \bibinfo {author} {\bibfnamefont {A.~J.}\ \bibnamefont {Bay-Smidt}},\ }\href@noop {} {\bibinfo {title} {A monte carlo approach to bound trotter error}} (\bibinfo {year} {2025}),\ \Eprint {https://arxiv.org/abs/2510.11621} {arXiv:2510.11621 [quant-ph]} \BibitemShut {NoStop}%
\bibitem [{\citenamefont {Bay-Smidt}\ \emph {et~al.}(2026)\citenamefont {Bay-Smidt}, \citenamefont {Glaser}, \citenamefont {Fabian}, \citenamefont {Campbell}, \citenamefont {Blunt},\ and\ \citenamefont {Solomon}}]{baysmidt2026}%
  \BibitemOpen
  \bibfield  {author} {\bibinfo {author} {\bibfnamefont {A.~J.}\ \bibnamefont {Bay-Smidt}}, \bibinfo {author} {\bibfnamefont {N.}~\bibnamefont {Glaser}}, \bibinfo {author} {\bibfnamefont {M.~D.}\ \bibnamefont {Fabian}}, \bibinfo {author} {\bibfnamefont {E.~T.}\ \bibnamefont {Campbell}}, \bibinfo {author} {\bibfnamefont {N.~S.}\ \bibnamefont {Blunt}},\ and\ \bibinfo {author} {\bibfnamefont {G.~C.}\ \bibnamefont {Solomon}},\ }\href {https://arxiv.org/abs/2605.00745} {\bibinfo {title} {Quantum simulation of nanographenes and trotter error cancellation}} (\bibinfo {year} {2026}),\ \Eprint {https://arxiv.org/abs/2605.00745} {arXiv:2605.00745 [quant-ph]} \BibitemShut {NoStop}%
\bibitem [{\citenamefont {Toshio}\ \emph {et~al.}(2025)\citenamefont {Toshio}, \citenamefont {Akahoshi}, \citenamefont {Fujisaki}, \citenamefont {Oshima}, \citenamefont {Sato},\ and\ \citenamefont {Fujii}}]{Toshio2025}%
  \BibitemOpen
  \bibfield  {author} {\bibinfo {author} {\bibfnamefont {R.}~\bibnamefont {Toshio}}, \bibinfo {author} {\bibfnamefont {Y.}~\bibnamefont {Akahoshi}}, \bibinfo {author} {\bibfnamefont {J.}~\bibnamefont {Fujisaki}}, \bibinfo {author} {\bibfnamefont {H.}~\bibnamefont {Oshima}}, \bibinfo {author} {\bibfnamefont {S.}~\bibnamefont {Sato}},\ and\ \bibinfo {author} {\bibfnamefont {K.}~\bibnamefont {Fujii}},\ }\bibfield  {title} {\bibinfo {title} {Practical quantum advantage on partially fault-tolerant quantum computer},\ }\href {https://doi.org/10.1103/PhysRevX.15.021057} {\bibfield  {journal} {\bibinfo  {journal} {Phys. Rev. X}\ }\textbf {\bibinfo {volume} {15}},\ \bibinfo {pages} {021057} (\bibinfo {year} {2025})}\BibitemShut {NoStop}%
\bibitem [{\citenamefont {Akahoshi}\ \emph {et~al.}(2025)\citenamefont {Akahoshi}, \citenamefont {Toshio}, \citenamefont {Fujisaki}, \citenamefont {Oshima}, \citenamefont {Sato},\ and\ \citenamefont {Fujii}}]{Akahoshi2025}%
  \BibitemOpen
  \bibfield  {author} {\bibinfo {author} {\bibfnamefont {Y.}~\bibnamefont {Akahoshi}}, \bibinfo {author} {\bibfnamefont {R.}~\bibnamefont {Toshio}}, \bibinfo {author} {\bibfnamefont {J.}~\bibnamefont {Fujisaki}}, \bibinfo {author} {\bibfnamefont {H.}~\bibnamefont {Oshima}}, \bibinfo {author} {\bibfnamefont {S.}~\bibnamefont {Sato}},\ and\ \bibinfo {author} {\bibfnamefont {K.}~\bibnamefont {Fujii}},\ }\bibfield  {title} {\bibinfo {title} {Compilation of trotter-based time evolution for partially fault-tolerant quantum computing architecture},\ }\href {https://doi.org/10.1103/93zr-1ykb} {\bibfield  {journal} {\bibinfo  {journal} {PRX Quantum}\ }\textbf {\bibinfo {volume} {6}},\ \bibinfo {pages} {040319} (\bibinfo {year} {2025})}\BibitemShut {NoStop}%
\bibitem [{\citenamefont {Chung}\ \emph {et~al.}(2026)\citenamefont {Chung}, \citenamefont {Kavaki}, \citenamefont {Scherer}, \citenamefont {Khalid}, \citenamefont {Kong}, \citenamefont {Kawakubo}, \citenamefont {Anand}, \citenamefont {Dagnew}, \citenamefont {Webb}, \citenamefont {Silva}, \citenamefont {Gyawali}, \citenamefont {Yan}, \citenamefont {Fujii}, \citenamefont {Ho}, \citenamefont {Mohseni}, \citenamefont {Ronagh},\ and\ \citenamefont {Martinis}}]{Chung2026}%
  \BibitemOpen
  \bibfield  {author} {\bibinfo {author} {\bibfnamefont {M.-Z.}\ \bibnamefont {Chung}}, \bibinfo {author} {\bibfnamefont {A.~H.~Z.}\ \bibnamefont {Kavaki}}, \bibinfo {author} {\bibfnamefont {A.}~\bibnamefont {Scherer}}, \bibinfo {author} {\bibfnamefont {A.}~\bibnamefont {Khalid}}, \bibinfo {author} {\bibfnamefont {X.}~\bibnamefont {Kong}}, \bibinfo {author} {\bibfnamefont {T.}~\bibnamefont {Kawakubo}}, \bibinfo {author} {\bibfnamefont {N.}~\bibnamefont {Anand}}, \bibinfo {author} {\bibfnamefont {G.~A.}\ \bibnamefont {Dagnew}}, \bibinfo {author} {\bibfnamefont {Z.}~\bibnamefont {Webb}}, \bibinfo {author} {\bibfnamefont {A.}~\bibnamefont {Silva}}, \bibinfo {author} {\bibfnamefont {G.}~\bibnamefont {Gyawali}}, \bibinfo {author} {\bibfnamefont {T.}~\bibnamefont {Yan}}, \bibinfo {author} {\bibfnamefont {K.}~\bibnamefont {Fujii}}, \bibinfo {author} {\bibfnamefont {A.}~\bibnamefont {Ho}}, \bibinfo {author} {\bibfnamefont {M.}~\bibnamefont {Mohseni}}, \bibinfo {author} {\bibfnamefont {P.}~\bibnamefont {Ronagh}},\
  and\ \bibinfo {author} {\bibfnamefont {J.}~\bibnamefont {Martinis}},\ }\href {https://arxiv.org/abs/2603.13093} {\bibinfo {title} {Partially fault-tolerant quantum computation for megaquop applications}} (\bibinfo {year} {2026}),\ \Eprint {https://arxiv.org/abs/2603.13093} {arXiv:2603.13093 [quant-ph]} \BibitemShut {NoStop}%
\bibitem [{\citenamefont {Kanasugi}\ \emph {et~al.}(2026)\citenamefont {Kanasugi}, \citenamefont {Toshio}, \citenamefont {Maruyama},\ and\ \citenamefont {Oshima}}]{Kanasugi2026}%
  \BibitemOpen
  \bibfield  {author} {\bibinfo {author} {\bibfnamefont {S.}~\bibnamefont {Kanasugi}}, \bibinfo {author} {\bibfnamefont {R.}~\bibnamefont {Toshio}}, \bibinfo {author} {\bibfnamefont {K.}~\bibnamefont {Maruyama}},\ and\ \bibinfo {author} {\bibfnamefont {H.}~\bibnamefont {Oshima}},\ }\href {https://arxiv.org/abs/2603.22778} {\bibinfo {title} {Enabling chemically accurate quantum phase estimation in the early fault-tolerant regime}} (\bibinfo {year} {2026}),\ \Eprint {https://arxiv.org/abs/2603.22778} {arXiv:2603.22778 [quant-ph]} \BibitemShut {NoStop}%
\bibitem [{\citenamefont {Toshio}\ \emph {et~al.}(2026)\citenamefont {Toshio}, \citenamefont {Kanasugi}, \citenamefont {Fujisaki}, \citenamefont {Oshima}, \citenamefont {Sato},\ and\ \citenamefont {Fujii}}]{Toshio2026}%
  \BibitemOpen
  \bibfield  {author} {\bibinfo {author} {\bibfnamefont {R.}~\bibnamefont {Toshio}}, \bibinfo {author} {\bibfnamefont {S.}~\bibnamefont {Kanasugi}}, \bibinfo {author} {\bibfnamefont {J.}~\bibnamefont {Fujisaki}}, \bibinfo {author} {\bibfnamefont {H.}~\bibnamefont {Oshima}}, \bibinfo {author} {\bibfnamefont {S.}~\bibnamefont {Sato}},\ and\ \bibinfo {author} {\bibfnamefont {K.}~\bibnamefont {Fujii}},\ }\href {https://arxiv.org/abs/2603.22891} {\bibinfo {title} {Star-magic mutation: Even more efficient analog rotation gates for early fault-tolerant quantum computer}} (\bibinfo {year} {2026}),\ \Eprint {https://arxiv.org/abs/2603.22891} {arXiv:2603.22891 [quant-ph]} \BibitemShut {NoStop}%
\bibitem [{\citenamefont {Campbell}(2019)}]{Campbell2019}%
  \BibitemOpen
  \bibfield  {author} {\bibinfo {author} {\bibfnamefont {E.}~\bibnamefont {Campbell}},\ }\bibfield  {title} {\bibinfo {title} {Random compiler for fast hamiltonian simulation},\ }\href {https://doi.org/10.1103/PhysRevLett.123.070503} {\bibfield  {journal} {\bibinfo  {journal} {Phys. Rev. Lett.}\ }\textbf {\bibinfo {volume} {123}},\ \bibinfo {pages} {070503} (\bibinfo {year} {2019})}\BibitemShut {NoStop}%
\bibitem [{\citenamefont {G\"unther}\ \emph {et~al.}(2026)\citenamefont {G\"unther}, \citenamefont {Witteveen}, \citenamefont {Schmidhuber}, \citenamefont {Miller}, \citenamefont {Christandl},\ and\ \citenamefont {Harrow}}]{Gunther2025}%
  \BibitemOpen
  \bibfield  {author} {\bibinfo {author} {\bibfnamefont {J.}~\bibnamefont {G\"unther}}, \bibinfo {author} {\bibfnamefont {F.}~\bibnamefont {Witteveen}}, \bibinfo {author} {\bibfnamefont {A.}~\bibnamefont {Schmidhuber}}, \bibinfo {author} {\bibfnamefont {M.}~\bibnamefont {Miller}}, \bibinfo {author} {\bibfnamefont {M.}~\bibnamefont {Christandl}},\ and\ \bibinfo {author} {\bibfnamefont {A.~W.}\ \bibnamefont {Harrow}},\ }\bibfield  {title} {\bibinfo {title} {Phase estimation with partially randomized time evolution},\ }\href {https://doi.org/10.1103/ynxb-p2xq} {\bibfield  {journal} {\bibinfo  {journal} {PRX Quantum}\ }\textbf {\bibinfo {volume} {7}},\ \bibinfo {pages} {020332} (\bibinfo {year} {2026})}\BibitemShut {NoStop}%
\bibitem [{\citenamefont {Kiumi}\ and\ \citenamefont {Koczor}(2025)}]{Kiumi2025}%
  \BibitemOpen
  \bibfield  {author} {\bibinfo {author} {\bibfnamefont {C.}~\bibnamefont {Kiumi}}\ and\ \bibinfo {author} {\bibfnamefont {B.}~\bibnamefont {Koczor}},\ }\bibfield  {title} {\bibinfo {title} {Te-pai: exact time evolution by sampling random circuits},\ }\href {https://doi.org/10.1088/2058-9565/ae1160} {\bibfield  {journal} {\bibinfo  {journal} {Quantum Science and Technology}\ }\textbf {\bibinfo {volume} {10}},\ \bibinfo {pages} {045071} (\bibinfo {year} {2025})}\BibitemShut {NoStop}%
\bibitem [{\citenamefont {Luthra}\ \emph {et~al.}(2025)\citenamefont {Luthra}, \citenamefont {Moylett}, \citenamefont {Browne},\ and\ \citenamefont {Campbell}}]{Luthra2025}%
  \BibitemOpen
  \bibfield  {author} {\bibinfo {author} {\bibfnamefont {S.}~\bibnamefont {Luthra}}, \bibinfo {author} {\bibfnamefont {A.~E.}\ \bibnamefont {Moylett}}, \bibinfo {author} {\bibfnamefont {D.~E.}\ \bibnamefont {Browne}},\ and\ \bibinfo {author} {\bibfnamefont {E.~T.}\ \bibnamefont {Campbell}},\ }\bibfield  {title} {\bibinfo {title} {Unlocking early fault-tolerant quantum computing with mitigated magic dilution},\ }\href {https://doi.org/10.1088/2058-9565/ae0aef} {\bibfield  {journal} {\bibinfo  {journal} {Quantum Science and Technology}\ }\textbf {\bibinfo {volume} {10}},\ \bibinfo {pages} {045066} (\bibinfo {year} {2025})}\BibitemShut {NoStop}%
\bibitem [{\citenamefont {Watrous}(2018)}]{watrous2018theory}%
  \BibitemOpen
  \bibfield  {author} {\bibinfo {author} {\bibfnamefont {J.}~\bibnamefont {Watrous}},\ }\href@noop {} {\emph {\bibinfo {title} {The theory of quantum information}}}\ (\bibinfo  {publisher} {Cambridge university press},\ \bibinfo {year} {2018})\BibitemShut {NoStop}%
\bibitem [{pyg(2026)}]{pygridsynth_github}%
  \BibitemOpen
  \href {https://github.com/quantum-programming/pygridsynth} {\bibinfo {title} {quantum-programming/pygridsynth}} (\bibinfo {year} {2026}),\ \bibinfo {note} {original-date: 2024-11-27T09:42:35Z}\BibitemShut {NoStop}%
\bibitem [{\citenamefont {Yamamoto}\ and\ \citenamefont {Yoshioka}(2026)}]{pygridsynth_paper}%
  \BibitemOpen
  \bibfield  {author} {\bibinfo {author} {\bibfnamefont {S.}~\bibnamefont {Yamamoto}}\ and\ \bibinfo {author} {\bibfnamefont {N.}~\bibnamefont {Yoshioka}},\ }\href {https://arxiv.org/abs/2604.21333} {\bibinfo {title} {pygridsynth: A fast numerical tool for ancilla-free clifford+t synthesis}} (\bibinfo {year} {2026}),\ \Eprint {https://arxiv.org/abs/2604.21333} {arXiv:2604.21333 [quant-ph]} \BibitemShut {NoStop}%
\bibitem [{\citenamefont {Giles}\ and\ \citenamefont {Selinger}(2019)}]{Giles2019}%
  \BibitemOpen
  \bibfield  {author} {\bibinfo {author} {\bibfnamefont {B.}~\bibnamefont {Giles}}\ and\ \bibinfo {author} {\bibfnamefont {P.}~\bibnamefont {Selinger}},\ }\href {https://arxiv.org/abs/1312.6584} {\bibinfo {title} {Remarks on matsumoto and amano's normal form for single-qubit clifford+t operators}} (\bibinfo {year} {2019}),\ \Eprint {https://arxiv.org/abs/1312.6584} {arXiv:1312.6584 [quant-ph]} \BibitemShut {NoStop}%
\bibitem [{\citenamefont {Kliuchnikov}\ \emph {et~al.}(2013)\citenamefont {Kliuchnikov}, \citenamefont {Maslov},\ and\ \citenamefont {Mosca}}]{kliuchnikov_fast_2013}%
  \BibitemOpen
  \bibfield  {author} {\bibinfo {author} {\bibfnamefont {V.}~\bibnamefont {Kliuchnikov}}, \bibinfo {author} {\bibfnamefont {D.}~\bibnamefont {Maslov}},\ and\ \bibinfo {author} {\bibfnamefont {M.}~\bibnamefont {Mosca}},\ }\href {https://doi.org/10.48550/arXiv.1206.5236} {\bibinfo {title} {Fast and efficient exact synthesis of single qubit unitaries generated by {Clifford} and {T} gates}} (\bibinfo {year} {2013}),\ \bibinfo {note} {arXiv:1206.5236 [quant-ph]}\BibitemShut {NoStop}%
\bibitem [{\citenamefont {Kivlichan}\ \emph {et~al.}(2020)\citenamefont {Kivlichan}, \citenamefont {Gidney}, \citenamefont {Berry}, \citenamefont {Wiebe}, \citenamefont {McClean}, \citenamefont {Sun}, \citenamefont {Jiang}, \citenamefont {Rubin}, \citenamefont {Fowler}, \citenamefont {Aspuru-Guzik}, \citenamefont {Neven},\ and\ \citenamefont {Babbush}}]{Kivlichan2020}%
  \BibitemOpen
  \bibfield  {author} {\bibinfo {author} {\bibfnamefont {I.~D.}\ \bibnamefont {Kivlichan}}, \bibinfo {author} {\bibfnamefont {C.}~\bibnamefont {Gidney}}, \bibinfo {author} {\bibfnamefont {D.~W.}\ \bibnamefont {Berry}}, \bibinfo {author} {\bibfnamefont {N.}~\bibnamefont {Wiebe}}, \bibinfo {author} {\bibfnamefont {J.}~\bibnamefont {McClean}}, \bibinfo {author} {\bibfnamefont {W.}~\bibnamefont {Sun}}, \bibinfo {author} {\bibfnamefont {Z.}~\bibnamefont {Jiang}}, \bibinfo {author} {\bibfnamefont {N.}~\bibnamefont {Rubin}}, \bibinfo {author} {\bibfnamefont {A.}~\bibnamefont {Fowler}}, \bibinfo {author} {\bibfnamefont {A.}~\bibnamefont {Aspuru-Guzik}}, \bibinfo {author} {\bibfnamefont {H.}~\bibnamefont {Neven}},\ and\ \bibinfo {author} {\bibfnamefont {R.}~\bibnamefont {Babbush}},\ }\bibfield  {title} {\bibinfo {title} {Improved fault-tolerant quantum simulation of condensed-phase correlated electrons via trotterization},\ }\href {https://doi.org/10.22331/q-2020-07-16-296} {\bibfield  {journal} {\bibinfo  {journal}
  {{Quantum}}\ }\textbf {\bibinfo {volume} {4}},\ \bibinfo {pages} {296} (\bibinfo {year} {2020})}\BibitemShut {NoStop}%
\bibitem [{\citenamefont {Campbell}(2021)}]{Campbell2022}%
  \BibitemOpen
  \bibfield  {author} {\bibinfo {author} {\bibfnamefont {E.~T.}\ \bibnamefont {Campbell}},\ }\bibfield  {title} {\bibinfo {title} {Early fault-tolerant simulations of the hubbard model},\ }\href {https://doi.org/10.1088/2058-9565/ac3110} {\bibfield  {journal} {\bibinfo  {journal} {Quantum Sci. Technol.}\ }\textbf {\bibinfo {volume} {7}},\ \bibinfo {pages} {015007} (\bibinfo {year} {2021})}\BibitemShut {NoStop}%
\bibitem [{\citenamefont {Bay-Smidt}\ \emph {et~al.}(2025)\citenamefont {Bay-Smidt}, \citenamefont {Klausen}, \citenamefont {S\"underhauf}, \citenamefont {Izs\'ak}, \citenamefont {Solomon},\ and\ \citenamefont {Blunt}}]{Bay-Smidt2025}%
  \BibitemOpen
  \bibfield  {author} {\bibinfo {author} {\bibfnamefont {A.~J.}\ \bibnamefont {Bay-Smidt}}, \bibinfo {author} {\bibfnamefont {F.~R.}\ \bibnamefont {Klausen}}, \bibinfo {author} {\bibfnamefont {C.}~\bibnamefont {S\"underhauf}}, \bibinfo {author} {\bibfnamefont {R.}~\bibnamefont {Izs\'ak}}, \bibinfo {author} {\bibfnamefont {G.~C.}\ \bibnamefont {Solomon}},\ and\ \bibinfo {author} {\bibfnamefont {N.~S.}\ \bibnamefont {Blunt}},\ }\bibfield  {title} {\bibinfo {title} {Fault-tolerant quantum simulation of generalized hubbard models},\ }\href {https://doi.org/10.1103/gr4t-b1w5} {\bibfield  {journal} {\bibinfo  {journal} {PRX Quantum}\ }\textbf {\bibinfo {volume} {6}},\ \bibinfo {pages} {030348} (\bibinfo {year} {2025})}\BibitemShut {NoStop}%
\bibitem [{\citenamefont {Blunt}(2021)}]{Blunt2021}%
  \BibitemOpen
  \bibfield  {author} {\bibinfo {author} {\bibfnamefont {N.~S.}\ \bibnamefont {Blunt}},\ }\bibfield  {title} {\bibinfo {title} {Fixed- and partial-node approximations in slater determinant space for molecules},\ }\href {https://doi.org/10.1021/acs.jctc.1c00500} {\bibfield  {journal} {\bibinfo  {journal} {Journal of Chemical Theory and Computation}\ }\textbf {\bibinfo {volume} {17}},\ \bibinfo {pages} {6092} (\bibinfo {year} {2021})},\ \bibinfo {note} {pMID: 34549947},\ \Eprint {https://arxiv.org/abs/https://doi.org/10.1021/acs.jctc.1c00500} {https://doi.org/10.1021/acs.jctc.1c00500} \BibitemShut {NoStop}%
\bibitem [{\citenamefont {Smith}\ \emph {et~al.}(2017)\citenamefont {Smith}, \citenamefont {Mussard}, \citenamefont {Holmes},\ and\ \citenamefont {Sharma}}]{smith2017}%
  \BibitemOpen
  \bibfield  {author} {\bibinfo {author} {\bibfnamefont {J.~E.~T.}\ \bibnamefont {Smith}}, \bibinfo {author} {\bibfnamefont {B.}~\bibnamefont {Mussard}}, \bibinfo {author} {\bibfnamefont {A.~A.}\ \bibnamefont {Holmes}},\ and\ \bibinfo {author} {\bibfnamefont {S.}~\bibnamefont {Sharma}},\ }\bibfield  {title} {\bibinfo {title} {Cheap and near exact casscf with large active spaces},\ }\href {https://doi.org/10.1021/acs.jctc.7b00900} {\bibfield  {journal} {\bibinfo  {journal} {Journal of Chemical Theory and Computation}\ }\textbf {\bibinfo {volume} {13}},\ \bibinfo {pages} {5468} (\bibinfo {year} {2017})},\ \bibinfo {note} {pMID: 28968097},\ \Eprint {https://arxiv.org/abs/https://doi.org/10.1021/acs.jctc.7b00900} {https://doi.org/10.1021/acs.jctc.7b00900} \BibitemShut {NoStop}%
\bibitem [{\citenamefont {Smith}\ \emph {et~al.}(2026)\citenamefont {Smith}, \citenamefont {Mussard}, \citenamefont {Holmes},\ and\ \citenamefont {Sharma}}]{smith_github}%
  \BibitemOpen
  \bibfield  {author} {\bibinfo {author} {\bibfnamefont {J.~E.~T.}\ \bibnamefont {Smith}}, \bibinfo {author} {\bibfnamefont {B.}~\bibnamefont {Mussard}}, \bibinfo {author} {\bibfnamefont {A.~A.}\ \bibnamefont {Holmes}},\ and\ \bibinfo {author} {\bibfnamefont {S.}~\bibnamefont {Sharma}},\ }\href {https://github.com/jamesETsmith/JCTC-11-5468/tree/master} {\bibinfo {title} {Github for ``cheap and near exact casscf with large active spaces'''}} (\bibinfo {year} {2026})\BibitemShut {NoStop}%
\bibitem [{\citenamefont {Li~Manni}\ \emph {et~al.}(2016)\citenamefont {Li~Manni}, \citenamefont {Smart},\ and\ \citenamefont {Alavi}}]{LiManni2016}%
  \BibitemOpen
  \bibfield  {author} {\bibinfo {author} {\bibfnamefont {G.}~\bibnamefont {Li~Manni}}, \bibinfo {author} {\bibfnamefont {S.~D.}\ \bibnamefont {Smart}},\ and\ \bibinfo {author} {\bibfnamefont {A.}~\bibnamefont {Alavi}},\ }\bibfield  {title} {\bibinfo {title} {Combining the complete active space self-consistent field method and the full configuration interaction quantum monte carlo within a super-ci framework, with application to challenging metal-porphyrins},\ }\href {https://doi.org/10.1021/acs.jctc.5b01190} {\bibfield  {journal} {\bibinfo  {journal} {Journal of Chemical Theory and Computation}\ }\textbf {\bibinfo {volume} {12}},\ \bibinfo {pages} {1245} (\bibinfo {year} {2016})},\ \bibinfo {note} {pMID: 26808894},\ \Eprint {https://arxiv.org/abs/https://doi.org/10.1021/acs.jctc.5b01190} {https://doi.org/10.1021/acs.jctc.5b01190} \BibitemShut {NoStop}%
\bibitem [{\citenamefont {Sun}\ \emph {et~al.}(2018)\citenamefont {Sun}, \citenamefont {Berkelbach}, \citenamefont {Blunt}, \citenamefont {Booth}, \citenamefont {Guo}, \citenamefont {Li}, \citenamefont {Liu}, \citenamefont {McClain}, \citenamefont {Sayfutyarova}, \citenamefont {Sharma}, \citenamefont {Wouters},\ and\ \citenamefont {Chan}}]{pyscf}%
  \BibitemOpen
  \bibfield  {author} {\bibinfo {author} {\bibfnamefont {Q.}~\bibnamefont {Sun}}, \bibinfo {author} {\bibfnamefont {T.~C.}\ \bibnamefont {Berkelbach}}, \bibinfo {author} {\bibfnamefont {N.~S.}\ \bibnamefont {Blunt}}, \bibinfo {author} {\bibfnamefont {G.~H.}\ \bibnamefont {Booth}}, \bibinfo {author} {\bibfnamefont {S.}~\bibnamefont {Guo}}, \bibinfo {author} {\bibfnamefont {Z.}~\bibnamefont {Li}}, \bibinfo {author} {\bibfnamefont {J.}~\bibnamefont {Liu}}, \bibinfo {author} {\bibfnamefont {J.~D.}\ \bibnamefont {McClain}}, \bibinfo {author} {\bibfnamefont {E.~R.}\ \bibnamefont {Sayfutyarova}}, \bibinfo {author} {\bibfnamefont {S.}~\bibnamefont {Sharma}}, \bibinfo {author} {\bibfnamefont {S.}~\bibnamefont {Wouters}},\ and\ \bibinfo {author} {\bibfnamefont {G.~K.-L.}\ \bibnamefont {Chan}},\ }\bibfield  {title} {\bibinfo {title} {Pyscf: the python-based simulations of chemistry framework},\ }\href {https://doi.org/https://doi.org/10.1002/wcms.1340} {\bibfield  {journal} {\bibinfo  {journal} {WIREs Computational
  Molecular Science}\ }\textbf {\bibinfo {volume} {8}},\ \bibinfo {pages} {e1340} (\bibinfo {year} {2018})},\ \Eprint {https://arxiv.org/abs/https://wires.onlinelibrary.wiley.com/doi/pdf/10.1002/wcms.1340} {https://wires.onlinelibrary.wiley.com/doi/pdf/10.1002/wcms.1340} \BibitemShut {NoStop}%
\bibitem [{\citenamefont {Sun}\ \emph {et~al.}(2020)\citenamefont {Sun}, \citenamefont {Zhang}, \citenamefont {Banerjee}, \citenamefont {Bao}, \citenamefont {Barbry}, \citenamefont {Blunt}, \citenamefont {Bogdanov}, \citenamefont {Booth}, \citenamefont {Chen}, \citenamefont {Cui}, \citenamefont {Eriksen}, \citenamefont {Gao}, \citenamefont {Guo}, \citenamefont {Hermann}, \citenamefont {Hermes}, \citenamefont {Koh}, \citenamefont {Koval}, \citenamefont {Lehtola}, \citenamefont {Li}, \citenamefont {Liu}, \citenamefont {Mardirossian}, \citenamefont {McClain}, \citenamefont {Motta}, \citenamefont {Mussard}, \citenamefont {Pham}, \citenamefont {Pulkin}, \citenamefont {Purwanto}, \citenamefont {Robinson}, \citenamefont {Ronca}, \citenamefont {Sayfutyarova}, \citenamefont {Scheurer}, \citenamefont {Schurkus}, \citenamefont {Smith}, \citenamefont {Sun}, \citenamefont {Sun}, \citenamefont {Upadhyay}, \citenamefont {Wagner}, \citenamefont {Wang}, \citenamefont {White}, \citenamefont {Whitfield}, \citenamefont
  {Williamson}, \citenamefont {Wouters}, \citenamefont {Yang}, \citenamefont {Yu}, \citenamefont {Zhu}, \citenamefont {Berkelbach}, \citenamefont {Sharma}, \citenamefont {Sokolov},\ and\ \citenamefont {Chan}}]{pyscf_2}%
  \BibitemOpen
  \bibfield  {author} {\bibinfo {author} {\bibfnamefont {Q.}~\bibnamefont {Sun}}, \bibinfo {author} {\bibfnamefont {X.}~\bibnamefont {Zhang}}, \bibinfo {author} {\bibfnamefont {S.}~\bibnamefont {Banerjee}}, \bibinfo {author} {\bibfnamefont {P.}~\bibnamefont {Bao}}, \bibinfo {author} {\bibfnamefont {M.}~\bibnamefont {Barbry}}, \bibinfo {author} {\bibfnamefont {N.~S.}\ \bibnamefont {Blunt}}, \bibinfo {author} {\bibfnamefont {N.~A.}\ \bibnamefont {Bogdanov}}, \bibinfo {author} {\bibfnamefont {G.~H.}\ \bibnamefont {Booth}}, \bibinfo {author} {\bibfnamefont {J.}~\bibnamefont {Chen}}, \bibinfo {author} {\bibfnamefont {Z.-H.}\ \bibnamefont {Cui}}, \bibinfo {author} {\bibfnamefont {J.~J.}\ \bibnamefont {Eriksen}}, \bibinfo {author} {\bibfnamefont {Y.}~\bibnamefont {Gao}}, \bibinfo {author} {\bibfnamefont {S.}~\bibnamefont {Guo}}, \bibinfo {author} {\bibfnamefont {J.}~\bibnamefont {Hermann}}, \bibinfo {author} {\bibfnamefont {M.~R.}\ \bibnamefont {Hermes}}, \bibinfo {author} {\bibfnamefont {K.}~\bibnamefont {Koh}},
  \bibinfo {author} {\bibfnamefont {P.}~\bibnamefont {Koval}}, \bibinfo {author} {\bibfnamefont {S.}~\bibnamefont {Lehtola}}, \bibinfo {author} {\bibfnamefont {Z.}~\bibnamefont {Li}}, \bibinfo {author} {\bibfnamefont {J.}~\bibnamefont {Liu}}, \bibinfo {author} {\bibfnamefont {N.}~\bibnamefont {Mardirossian}}, \bibinfo {author} {\bibfnamefont {J.~D.}\ \bibnamefont {McClain}}, \bibinfo {author} {\bibfnamefont {M.}~\bibnamefont {Motta}}, \bibinfo {author} {\bibfnamefont {B.}~\bibnamefont {Mussard}}, \bibinfo {author} {\bibfnamefont {H.~Q.}\ \bibnamefont {Pham}}, \bibinfo {author} {\bibfnamefont {A.}~\bibnamefont {Pulkin}}, \bibinfo {author} {\bibfnamefont {W.}~\bibnamefont {Purwanto}}, \bibinfo {author} {\bibfnamefont {P.~J.}\ \bibnamefont {Robinson}}, \bibinfo {author} {\bibfnamefont {E.}~\bibnamefont {Ronca}}, \bibinfo {author} {\bibfnamefont {E.~R.}\ \bibnamefont {Sayfutyarova}}, \bibinfo {author} {\bibfnamefont {M.}~\bibnamefont {Scheurer}}, \bibinfo {author} {\bibfnamefont {H.~F.}\ \bibnamefont {Schurkus}},
  \bibinfo {author} {\bibfnamefont {J.~E.~T.}\ \bibnamefont {Smith}}, \bibinfo {author} {\bibfnamefont {C.}~\bibnamefont {Sun}}, \bibinfo {author} {\bibfnamefont {S.-N.}\ \bibnamefont {Sun}}, \bibinfo {author} {\bibfnamefont {S.}~\bibnamefont {Upadhyay}}, \bibinfo {author} {\bibfnamefont {L.~K.}\ \bibnamefont {Wagner}}, \bibinfo {author} {\bibfnamefont {X.}~\bibnamefont {Wang}}, \bibinfo {author} {\bibfnamefont {A.}~\bibnamefont {White}}, \bibinfo {author} {\bibfnamefont {J.~D.}\ \bibnamefont {Whitfield}}, \bibinfo {author} {\bibfnamefont {M.~J.}\ \bibnamefont {Williamson}}, \bibinfo {author} {\bibfnamefont {S.}~\bibnamefont {Wouters}}, \bibinfo {author} {\bibfnamefont {J.}~\bibnamefont {Yang}}, \bibinfo {author} {\bibfnamefont {J.~M.}\ \bibnamefont {Yu}}, \bibinfo {author} {\bibfnamefont {T.}~\bibnamefont {Zhu}}, \bibinfo {author} {\bibfnamefont {T.~C.}\ \bibnamefont {Berkelbach}}, \bibinfo {author} {\bibfnamefont {S.}~\bibnamefont {Sharma}}, \bibinfo {author} {\bibfnamefont {A.~Y.}\ \bibnamefont
  {Sokolov}},\ and\ \bibinfo {author} {\bibfnamefont {G.~K.-L.}\ \bibnamefont {Chan}},\ }\bibfield  {title} {\bibinfo {title} {Recent developments in the pyscf program package},\ }\href {https://doi.org/10.1063/5.0006074} {\bibfield  {journal} {\bibinfo  {journal} {The Journal of Chemical Physics}\ }\textbf {\bibinfo {volume} {153}},\ \bibinfo {pages} {024109} (\bibinfo {year} {2020})}\BibitemShut {NoStop}%
\bibitem [{\citenamefont {McClean}\ \emph {et~al.}(2020)\citenamefont {McClean}, \citenamefont {Rubin}, \citenamefont {Sung}, \citenamefont {Kivlichan}, \citenamefont {Bonet-Monroig}, \citenamefont {Cao}, \citenamefont {Dai}, \citenamefont {Fried}, \citenamefont {Gidney}, \citenamefont {Gimby}, \citenamefont {Gokhale}, \citenamefont {Häner}, \citenamefont {Hardikar}, \citenamefont {Havl{\'{\i}}{\v{c}}ek}, \citenamefont {Higgott}, \citenamefont {Huang}, \citenamefont {Izaac}, \citenamefont {Jiang}, \citenamefont {Liu}, \citenamefont {McArdle}, \citenamefont {Neeley}, \citenamefont {O'Brien}, \citenamefont {O'Gorman}, \citenamefont {Ozfidan}, \citenamefont {Radin}, \citenamefont {Romero}, \citenamefont {Sawaya}, \citenamefont {Senjean}, \citenamefont {Setia}, \citenamefont {Sim}, \citenamefont {Steiger}, \citenamefont {Steudtner}, \citenamefont {Sun}, \citenamefont {Sun}, \citenamefont {Wang}, \citenamefont {Zhang},\ and\ \citenamefont {Babbush}}]{openfermion}%
  \BibitemOpen
  \bibfield  {author} {\bibinfo {author} {\bibfnamefont {J.~R.}\ \bibnamefont {McClean}}, \bibinfo {author} {\bibfnamefont {N.~C.}\ \bibnamefont {Rubin}}, \bibinfo {author} {\bibfnamefont {K.~J.}\ \bibnamefont {Sung}}, \bibinfo {author} {\bibfnamefont {I.~D.}\ \bibnamefont {Kivlichan}}, \bibinfo {author} {\bibfnamefont {X.}~\bibnamefont {Bonet-Monroig}}, \bibinfo {author} {\bibfnamefont {Y.}~\bibnamefont {Cao}}, \bibinfo {author} {\bibfnamefont {C.}~\bibnamefont {Dai}}, \bibinfo {author} {\bibfnamefont {E.~S.}\ \bibnamefont {Fried}}, \bibinfo {author} {\bibfnamefont {C.}~\bibnamefont {Gidney}}, \bibinfo {author} {\bibfnamefont {B.}~\bibnamefont {Gimby}}, \bibinfo {author} {\bibfnamefont {P.}~\bibnamefont {Gokhale}}, \bibinfo {author} {\bibfnamefont {T.}~\bibnamefont {Häner}}, \bibinfo {author} {\bibfnamefont {T.}~\bibnamefont {Hardikar}}, \bibinfo {author} {\bibfnamefont {V.}~\bibnamefont {Havl{\'{\i}}{\v{c}}ek}}, \bibinfo {author} {\bibfnamefont {O.}~\bibnamefont {Higgott}}, \bibinfo {author} {\bibfnamefont
  {C.}~\bibnamefont {Huang}}, \bibinfo {author} {\bibfnamefont {J.}~\bibnamefont {Izaac}}, \bibinfo {author} {\bibfnamefont {Z.}~\bibnamefont {Jiang}}, \bibinfo {author} {\bibfnamefont {X.}~\bibnamefont {Liu}}, \bibinfo {author} {\bibfnamefont {S.}~\bibnamefont {McArdle}}, \bibinfo {author} {\bibfnamefont {M.}~\bibnamefont {Neeley}}, \bibinfo {author} {\bibfnamefont {T.}~\bibnamefont {O'Brien}}, \bibinfo {author} {\bibfnamefont {B.}~\bibnamefont {O'Gorman}}, \bibinfo {author} {\bibfnamefont {I.}~\bibnamefont {Ozfidan}}, \bibinfo {author} {\bibfnamefont {M.~D.}\ \bibnamefont {Radin}}, \bibinfo {author} {\bibfnamefont {J.}~\bibnamefont {Romero}}, \bibinfo {author} {\bibfnamefont {N.~P.~D.}\ \bibnamefont {Sawaya}}, \bibinfo {author} {\bibfnamefont {B.}~\bibnamefont {Senjean}}, \bibinfo {author} {\bibfnamefont {K.}~\bibnamefont {Setia}}, \bibinfo {author} {\bibfnamefont {S.}~\bibnamefont {Sim}}, \bibinfo {author} {\bibfnamefont {D.~S.}\ \bibnamefont {Steiger}}, \bibinfo {author} {\bibfnamefont {M.}~\bibnamefont
  {Steudtner}}, \bibinfo {author} {\bibfnamefont {Q.}~\bibnamefont {Sun}}, \bibinfo {author} {\bibfnamefont {W.}~\bibnamefont {Sun}}, \bibinfo {author} {\bibfnamefont {D.}~\bibnamefont {Wang}}, \bibinfo {author} {\bibfnamefont {F.}~\bibnamefont {Zhang}},\ and\ \bibinfo {author} {\bibfnamefont {R.}~\bibnamefont {Babbush}},\ }\bibfield  {title} {\bibinfo {title} {{OpenFermion}: the electronic structure package for quantum computers},\ }\href {https://doi.org/10.1088/2058-9565/ab8ebc} {\bibfield  {journal} {\bibinfo  {journal} {Quantum Science and Technology}\ }\textbf {\bibinfo {volume} {5}},\ \bibinfo {pages} {034014} (\bibinfo {year} {2020})}\BibitemShut {NoStop}%
\bibitem [{\citenamefont {Lin}\ and\ \citenamefont {Tong}(2022)}]{Lin2022}%
  \BibitemOpen
  \bibfield  {author} {\bibinfo {author} {\bibfnamefont {L.}~\bibnamefont {Lin}}\ and\ \bibinfo {author} {\bibfnamefont {Y.}~\bibnamefont {Tong}},\ }\bibfield  {title} {\bibinfo {title} {Heisenberg-limited ground-state energy estimation for early fault-tolerant quantum computers},\ }\href {https://doi.org/10.1103/PRXQuantum.3.010318} {\bibfield  {journal} {\bibinfo  {journal} {PRX Quantum}\ }\textbf {\bibinfo {volume} {3}},\ \bibinfo {pages} {010318} (\bibinfo {year} {2022})}\BibitemShut {NoStop}%
\bibitem [{\citenamefont {Wan}\ \emph {et~al.}(2022)\citenamefont {Wan}, \citenamefont {Berta},\ and\ \citenamefont {Campbell}}]{Wan2022}%
  \BibitemOpen
  \bibfield  {author} {\bibinfo {author} {\bibfnamefont {K.}~\bibnamefont {Wan}}, \bibinfo {author} {\bibfnamefont {M.}~\bibnamefont {Berta}},\ and\ \bibinfo {author} {\bibfnamefont {E.~T.}\ \bibnamefont {Campbell}},\ }\bibfield  {title} {\bibinfo {title} {Randomized quantum algorithm for statistical phase estimation},\ }\href {https://doi.org/10.1103/PhysRevLett.129.030503} {\bibfield  {journal} {\bibinfo  {journal} {Phys. Rev. Lett.}\ }\textbf {\bibinfo {volume} {129}},\ \bibinfo {pages} {030503} (\bibinfo {year} {2022})}\BibitemShut {NoStop}%
\bibitem [{\citenamefont {Dutkiewicz}\ \emph {et~al.}(2022)\citenamefont {Dutkiewicz}, \citenamefont {Terhal},\ and\ \citenamefont {O'Brien}}]{Dutkiewicz2022}%
  \BibitemOpen
  \bibfield  {author} {\bibinfo {author} {\bibfnamefont {A.}~\bibnamefont {Dutkiewicz}}, \bibinfo {author} {\bibfnamefont {B.~M.}\ \bibnamefont {Terhal}},\ and\ \bibinfo {author} {\bibfnamefont {T.~E.}\ \bibnamefont {O'Brien}},\ }\bibfield  {title} {\bibinfo {title} {Heisenberg-limited quantum phase estimation of multiple eigenvalues with few control qubits},\ }\href {https://doi.org/10.22331/q-2022-10-06-830} {\bibfield  {journal} {\bibinfo  {journal} {{Quantum}}\ }\textbf {\bibinfo {volume} {6}},\ \bibinfo {pages} {830} (\bibinfo {year} {2022})}\BibitemShut {NoStop}%
\bibitem [{\citenamefont {Wang}\ \emph {et~al.}(2023)\citenamefont {Wang}, \citenamefont {Fran{\c{c}}a}, \citenamefont {Zhang}, \citenamefont {Zhu},\ and\ \citenamefont {Johnson}}]{Wang2023}%
  \BibitemOpen
  \bibfield  {author} {\bibinfo {author} {\bibfnamefont {G.}~\bibnamefont {Wang}}, \bibinfo {author} {\bibfnamefont {D.~S.}\ \bibnamefont {Fran{\c{c}}a}}, \bibinfo {author} {\bibfnamefont {R.}~\bibnamefont {Zhang}}, \bibinfo {author} {\bibfnamefont {S.}~\bibnamefont {Zhu}},\ and\ \bibinfo {author} {\bibfnamefont {P.~D.}\ \bibnamefont {Johnson}},\ }\bibfield  {title} {\bibinfo {title} {{Quantum algorithm for ground state energy estimation using circuit depth with exponentially improved dependence on precision}},\ }\href {https://doi.org/10.22331/q-2023-11-06-1167} {\bibfield  {journal} {\bibinfo  {journal} {{Quantum}}\ }\textbf {\bibinfo {volume} {7}},\ \bibinfo {pages} {1167} (\bibinfo {year} {2023})}\BibitemShut {NoStop}%
\bibitem [{\citenamefont {Blunt}\ \emph {et~al.}(2023)\citenamefont {Blunt}, \citenamefont {Caune}, \citenamefont {Izs\'ak}, \citenamefont {Campbell},\ and\ \citenamefont {Holzmann}}]{Blunt2023}%
  \BibitemOpen
  \bibfield  {author} {\bibinfo {author} {\bibfnamefont {N.~S.}\ \bibnamefont {Blunt}}, \bibinfo {author} {\bibfnamefont {L.}~\bibnamefont {Caune}}, \bibinfo {author} {\bibfnamefont {R.}~\bibnamefont {Izs\'ak}}, \bibinfo {author} {\bibfnamefont {E.~T.}\ \bibnamefont {Campbell}},\ and\ \bibinfo {author} {\bibfnamefont {N.}~\bibnamefont {Holzmann}},\ }\bibfield  {title} {\bibinfo {title} {Statistical phase estimation and error mitigation on a superconducting quantum processor},\ }\href {https://doi.org/10.1103/PRXQuantum.4.040341} {\bibfield  {journal} {\bibinfo  {journal} {PRX Quantum}\ }\textbf {\bibinfo {volume} {4}},\ \bibinfo {pages} {040341} (\bibinfo {year} {2023})}\BibitemShut {NoStop}%
\bibitem [{\citenamefont {Ding}\ and\ \citenamefont {Lin}(2023)}]{Ding2023}%
  \BibitemOpen
  \bibfield  {author} {\bibinfo {author} {\bibfnamefont {Z.}~\bibnamefont {Ding}}\ and\ \bibinfo {author} {\bibfnamefont {L.}~\bibnamefont {Lin}},\ }\bibfield  {title} {\bibinfo {title} {Even shorter quantum circuit for phase estimation on early fault-tolerant quantum computers with applications to ground-state energy estimation},\ }\href {https://doi.org/10.1103/PRXQuantum.4.020331} {\bibfield  {journal} {\bibinfo  {journal} {PRX Quantum}\ }\textbf {\bibinfo {volume} {4}},\ \bibinfo {pages} {020331} (\bibinfo {year} {2023})}\BibitemShut {NoStop}%
\bibitem [{\citenamefont {Temme}\ \emph {et~al.}(2017)\citenamefont {Temme}, \citenamefont {Bravyi},\ and\ \citenamefont {Gambetta}}]{Temme2017}%
  \BibitemOpen
  \bibfield  {author} {\bibinfo {author} {\bibfnamefont {K.}~\bibnamefont {Temme}}, \bibinfo {author} {\bibfnamefont {S.}~\bibnamefont {Bravyi}},\ and\ \bibinfo {author} {\bibfnamefont {J.~M.}\ \bibnamefont {Gambetta}},\ }\bibfield  {title} {\bibinfo {title} {Error mitigation for short-depth quantum circuits},\ }\href {https://doi.org/10.1103/PhysRevLett.119.180509} {\bibfield  {journal} {\bibinfo  {journal} {Phys. Rev. Lett.}\ }\textbf {\bibinfo {volume} {119}},\ \bibinfo {pages} {180509} (\bibinfo {year} {2017})}\BibitemShut {NoStop}%
\bibitem [{\citenamefont {Camilo}\ \emph {et~al.}(2026)\citenamefont {Camilo}, \citenamefont {Maciel}, \citenamefont {Tosta}, \citenamefont {Alhajri}, \citenamefont {de~Lima~Silva}, \citenamefont {França},\ and\ \citenamefont {Aolita}}]{Camilo2026}%
  \BibitemOpen
  \bibfield  {author} {\bibinfo {author} {\bibfnamefont {G.}~\bibnamefont {Camilo}}, \bibinfo {author} {\bibfnamefont {T.~O.}\ \bibnamefont {Maciel}}, \bibinfo {author} {\bibfnamefont {A.}~\bibnamefont {Tosta}}, \bibinfo {author} {\bibfnamefont {A.}~\bibnamefont {Alhajri}}, \bibinfo {author} {\bibfnamefont {T.}~\bibnamefont {de~Lima~Silva}}, \bibinfo {author} {\bibfnamefont {D.~S.}\ \bibnamefont {França}},\ and\ \bibinfo {author} {\bibfnamefont {L.}~\bibnamefont {Aolita}},\ }\href {https://arxiv.org/abs/2508.20174} {\bibinfo {title} {Compilation-informed probabilistic logical-error cancellation}} (\bibinfo {year} {2026}),\ \Eprint {https://arxiv.org/abs/2508.20174} {arXiv:2508.20174 [quant-ph]} \BibitemShut {NoStop}%
\bibitem [{\citenamefont {Diamond}\ and\ \citenamefont {Boyd}(2016)}]{cvxpy_paper_1}%
  \BibitemOpen
  \bibfield  {author} {\bibinfo {author} {\bibfnamefont {S.}~\bibnamefont {Diamond}}\ and\ \bibinfo {author} {\bibfnamefont {S.}~\bibnamefont {Boyd}},\ }\bibfield  {title} {\bibinfo {title} {{CVXPY}: {A} {P}ython-embedded modeling language for convex optimization},\ }\href@noop {} {\bibfield  {journal} {\bibinfo  {journal} {Journal of Machine Learning Research}\ }\textbf {\bibinfo {volume} {17}},\ \bibinfo {pages} {1} (\bibinfo {year} {2016})}\BibitemShut {NoStop}%
\bibitem [{\citenamefont {Agrawal}\ \emph {et~al.}(2018)\citenamefont {Agrawal}, \citenamefont {Verschueren}, \citenamefont {Diamond},\ and\ \citenamefont {Boyd}}]{cvxpy_paper_2}%
  \BibitemOpen
  \bibfield  {author} {\bibinfo {author} {\bibfnamefont {A.}~\bibnamefont {Agrawal}}, \bibinfo {author} {\bibfnamefont {R.}~\bibnamefont {Verschueren}}, \bibinfo {author} {\bibfnamefont {S.}~\bibnamefont {Diamond}},\ and\ \bibinfo {author} {\bibfnamefont {S.}~\bibnamefont {Boyd}},\ }\bibfield  {title} {\bibinfo {title} {A rewriting system for convex optimization problems},\ }\href@noop {} {\bibfield  {journal} {\bibinfo  {journal} {Journal of Control and Decision}\ }\textbf {\bibinfo {volume} {5}},\ \bibinfo {pages} {42} (\bibinfo {year} {2018})}\BibitemShut {NoStop}%
\bibitem [{\citenamefont {Granet}\ and\ \citenamefont {Dreyer}(2024)}]{Granet2024}%
  \BibitemOpen
  \bibfield  {author} {\bibinfo {author} {\bibfnamefont {E.}~\bibnamefont {Granet}}\ and\ \bibinfo {author} {\bibfnamefont {H.}~\bibnamefont {Dreyer}},\ }\bibfield  {title} {\bibinfo {title} {Hamiltonian dynamics on digital quantum computers without discretization error},\ }\href {https://doi.org/10.1038/s41534-024-00877-y} {\bibfield  {journal} {\bibinfo  {journal} {npj Quantum Information}\ }\textbf {\bibinfo {volume} {10}},\ \bibinfo {pages} {82} (\bibinfo {year} {2024})}\BibitemShut {NoStop}%
\end{thebibliography}
\end{document}